\documentclass[12pt]{article}
\usepackage[utf8]{inputenc}
\usepackage{physics}

\usepackage{mathrsfs}
\usepackage{graphicx}
\usepackage{authblk}
\usepackage{comment}
\usepackage{dsfont}
\usepackage{amsmath}
\usepackage{amsfonts}
\usepackage{amssymb}
\usepackage{braket}
\usepackage{url}
\usepackage{amsthm}
\usepackage[colorlinks=true,urlcolor=blue,citecolor=blue,linkcolor=blue]{hyperref}
\usepackage{caption}
\usepackage{lipsum}
\usepackage{subcaption}
\usepackage{mwe}
\usepackage{enumerate}
\usepackage{sidecap}
\usepackage[numbers]{natbib}
\usepackage{appendix}
\usepackage{enumitem}
\usepackage{multirow}
\usepackage{array}
\usepackage{physics}
\usepackage{mathtools}
\usepackage{tcolorbox}
\usepackage{thmtools}
\usepackage{hyperref}
\usepackage{booktabs}
\usepackage{cleveref}
\usepackage[paperwidth=215mm,paperheight=297mm,centering,hmargin=2.45cm,vmargin=2.5cm]{geometry}

\setcitestyle{square}
\usepackage{nicematrix}

\newtheorem{theorem}{Theorem}[section] 
\newtheorem{lemma}[theorem]{Lemma}     
\newtheorem{cor}[theorem]{Corollary} 
\newtheorem{prop}[theorem]{Proposition}

\newtheorem{definition}[theorem]{Definition}

\theoremstyle{definition}

\newcommand{\iseq}{\overset{?}{=}}

\newcommand{\Fg}[3]{\operatorname{F}_{#3}\!\left(#1, #2\right)}

\newcommand{\Fu}[2]{\operatorname{F}^\mathrm{U}\!\left(#1, #2\right)}

\newcommand{\Fga}[3]{\operatorname{F}_{#3}^\alpha\left(#1, #2\right)}

\DeclareMathOperator*{\argmin}{argmin}
\DeclareMathOperator*{\argmax}{argmax}
 
\def \Log {\mathrm{Log}} 
 
\DeclareMathSymbol{\hyph}{\mathord}{AMSa}{"39}

\newtcolorbox{definitionbox}[1][]{colback=green!5!white, colframe=white, boxrule=0pt, 
    fonttitle=\bfseries, #1}

\newtcolorbox{theorembox}[1][]{colback=blue!5!white, colframe=white, boxrule=0pt, 
    fonttitle=\bfseries, #1}

\date{}

\AtBeginDocument{
  \addtocontents{toc}{\footnotesize
}
}

\usepackage{xpatch}
\makeatletter
\xpatchcmd{\tableofcontents}{\contentsname \@mkboth}{\small\contentsname \@mkboth}{}{}
\makeatother

\begin{document}

\title{\vspace*{-2cm}Riemannian-geometric generalizations of quantum fidelities and Bures-Wasserstein distance}

\author[1,2]{A. Afham\thanks{Corresponding author: \texttt{afham@student.uts.edu.au}}}
\author[1]{Chris Ferrie}

\affil[1]{\small Centre for Quantum Software and Information, University of Technology Sydney, NSW 2007, Australia}
\affil[2]{\small Sydney Quantum Academy, Sydney, NSW 2000, Australia}

\maketitle
\vspace*{-1cm}
\begin{abstract}
    We introduce a family of fidelities, termed generalized fidelity, which are based on the Riemannian geometry of the Bures-Wasserstein manifold. We show that this family of fidelities generalizes standard quantum fidelities such as Uhlmann-, Holevo-, and Matsumoto-fidelity and demonstrate that it satisfies analogous celebrated properties. The generalized fidelity naturally arises from a \textit{generalized} Bures distance, the natural distance obtained by \textit{linearizing} the Bures-Wasserstein manifold. We prove various invariance and covariance properties of generalized fidelity as the point of linearization moves along geodesic-related paths. We also provide a Block-matrix characterization and prove an Uhlmann-like theorem, as well as provide further extensions to the multivariate setting and to quantum R\'enyi divergences, generalizing Petz-, Sandwich-, Reverse sandwich-, and Geometric-R\'enyi divergences of order $\alpha$.

\end{abstract}

\vspace{1cm}

\renewcommand{\baselinestretch}{.4}\normalsize
\tableofcontents
\renewcommand{\baselinestretch}{1.0}\normalsize

\pagebreak



\begin{figure}[htbp]
    \centering
    \begin{subfigure}{0.49\linewidth}
        \centering
        \includegraphics[width=\linewidth]{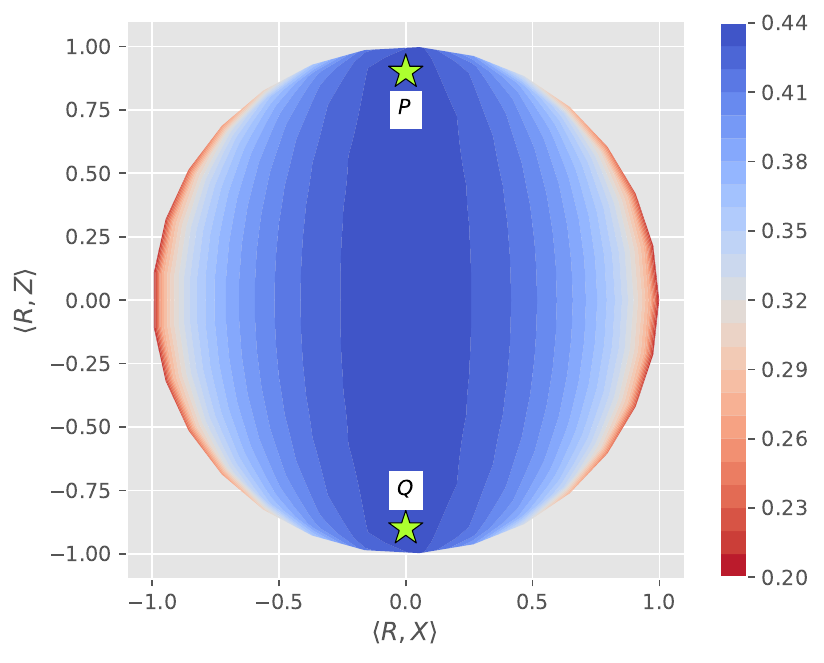}  
        \label{fig:RebitGenFidelity2}
    \end{subfigure}
    \hfill
    \begin{subfigure}{0.49\linewidth}
        \centering
        \includegraphics[width=\linewidth]{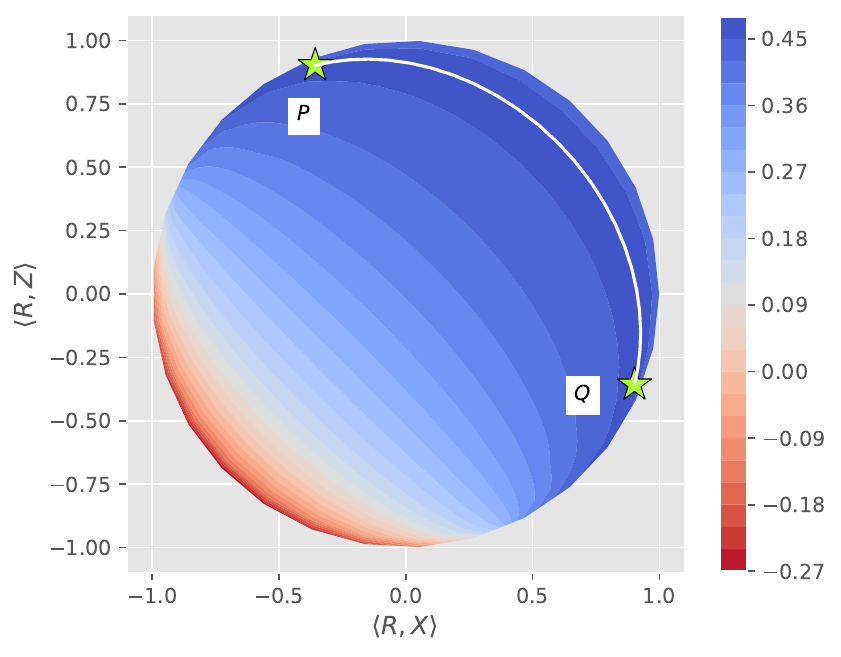}  
        \label{fig:RebitGenFidelity1}
    \end{subfigure}
    \caption{\small Contour plots of the generalized fidelity $\operatorname{F}_R(P,Q)$ as a function of the base $R$. 
    Each point indicates the generalized fidelity between $P$ and $Q$ at $R$, where $R$ is the rebit state with Bloch vector $(\langle R,X \rangle, 0, \langle R,Z \rangle)$. The first subfigure shows that the generalized fidelity between two commuting quantum states can be unequal to the Bhattacharyya coefficient (classical fidelity) for general bases. The second subfigure shows that the generalized fidelity between two states can be negative (generally complex-valued) and that the generalized fidelity attains its maximum value if the base is any point along the Bures-Wasserstein geodesic (white curve) between $P$ and $Q$. Along this geodesic, the value of generalized fidelity $\Fg{P}{Q}{R}$ is equal to the Uhlmann fidelity $\Fu{P}{Q}$.} Code to generate plots is available at~\cite{afhamgithub}.
    \label{fig:RebitGenFidelity}
\end{figure}

\section{Introduction}

Fidelity is a fundamental figure of merit in quantum sciences used to measure the similarity of two quantum states~\cite{uhlmann1976transition, jozsa1994fidelity, nielsen2001quantum, watrous2018theory}. It enjoys many valuable properties, such as achieving its maximum value of 1 if the states are equal and the minimum value of 0 if they have orthogonal support. Fidelity---specifically Uhlmann fidelity---is one of the various non-commutative generalizations of the Bhattacharyya coefficient~\cite{bhattacharyya1943measure} from statistics, which is a comparator for entrywise-positive vectors. It also features prominently in the definition of the Bures distance (a.k.a. Bures-Wasserstein distance), which is the natural distance on the Riemannian manifold of positive definite matrices once it is endowed with the Bures metric (a.k.a. Bures-Wasserstein metric)~\cite{bures1969extension, bhatia2019bures}. This manifold is ubiquitous not only in quantum information~\cite{bengtsson2017geometry, luo2004informational, brahmachari2023fixed, afham2022} but also in the machine learning world via its relation to the 2-Wasserstein distance on the space of probability measures~\cite{otto2001geometry, villani2009optimal, panaretos2019statistical}.

The Bures-Wasserstein (BW) distance is one of the non-commutative generalizations of Hellinger distance between two positive vectors. The (squared) Hellinger distance between two positive vectors $p,q \in [0, \infty)^d$,  is defined as

\begin{equation}
  \mathrm{d}^2_\mathrm{H}(p,q) := \left\| \sqrt{p}  - \sqrt{q} \right\|_2^2 = \sum_{i=1}^{d} p_i+ q_i - 2 \sqrt{p_i q_i},  
\end{equation}
where the square root is taken element-wise. One sees that the Hellinger distance can be interpreted as the difference between the arithmetic mean and the geometric mean of the two vectors:
\begin{equation}
    \mathrm{A}(p,q) := \frac{p+q}{2}, \quad \mathrm{G}(p,q) := \sqrt{pq} \quad  \implies \quad    
    \frac12\operatorname{d}^2_\mathrm H(p,q) = \sum_{i=1}^d \left[\mathrm{A}(p,q) - \mathrm{G}(p,q)\right]_i,
\end{equation} 
where the product and square root of vectors are taken element-wise.
While generalizing to the set of positive semidefinite matrices, (the sum of) the arithmetic mean generalizes straightforwardly. However, there is no unique way of generalizing (the sum of) the geometric mean $\sum_{i=1}^{d} [\mathrm{G}(p,q)]_i$ part, which is the \textit{Bhattacharyya coefficient} or \textit{classical fidelity} between $p$ and $q$. 
\begin{equation}
    \operatorname{F}_\text{cl}(p,q) = \sum_{i=1}^{n} \sqrt{p_i q_i}.
\end{equation}
The non-commutative (quantum) generalization of this inner-product term is usually called a \textit{quantum fidelity}. Various versions of quantum fidelities are central to quantum information, and some prominent ones are defined as follows. 
\begin{enumerate}
    \item Uhlmann fidelity~\cite{uhlmann1976transition, jozsa1994fidelity, nielsen2001quantum, watrous2018theory}: $\operatorname{F}^\mathrm U(P,Q) := \operatorname{Tr}\left[\sqrt{P ^{\frac12} Q P ^{\frac12} }\right]$.
    \item Holevo fidelity~\cite{Holevo1972, WildeRecoverability2018}: $\operatorname{F}^\mathrm H(P,Q) := \operatorname{Tr}\left[P ^{\frac12} Q^{\frac12} \right]$.
    \item Matsumoto fidelity~\cite{matsumoto2010reverse, cree2020fidelity}: $\operatorname{F}^\mathrm U(P,Q) := \operatorname{Tr}\left[ P \# Q\right]$.
    \item Log-Euclidean fidelity: $     \operatorname{F}^{\operatorname{LE}}(P,Q) := \text{Tr}\left[\exp \left( \frac{\ln P + \ln Q}{2} \right)\right].$
    \item $z$-fidelity: $\operatorname{F}^z(P,Q) := \operatorname{Tr}\left[ \left(P^{\frac{1}{2z}} Q^\frac{1}{2z}\right)^z \right]$ for $z \geq 0$. 
    
\end{enumerate}

The $z$-fidelities are obtained by setting $\alpha=\frac12$ in (the fidelity part of) the family of $\alpha\hyph z$ R\'{e}nyi relative entropies~\cite{audenaert2013alpha, Wilde2024multivariate} and satisfies the data processing inequality for $z \geq 1/2$. The limiting case $z \to \infty$ yields the Log-Euclidean fidelity. We remark that Uhlmann~\cite{uhlmann1976transition} initially defined the square of $\operatorname{F}^\mathrm U(P,Q) $ as \textit{transition probability}. 

For each of the above fidelities, one can define the associated (squared) \textit{Hellinger quantity} as
\begin{equation}
    \operatorname{B}(P, Q) := \operatorname{Tr}[P+Q] - 2 \operatorname{F}(P,Q).
\end{equation}
Choosing the Uhlmann fidelity yields the Bures-Wasserstein distance: 
\begin{equation}
\mathrm{d}_\mathrm{BW}^2(P,Q) := \operatorname{Tr}[P+Q] - 2 \operatorname{F}^\mathrm{U}(P,Q) = \min_{U \in \mathbb U_d} \left\| P ^{\frac12} - Q ^{\frac12} U \right\|_2^2,
\end{equation}
which is the natural distance of the (Riemannian) Bures-Wasserstein  manifold. Choosing the Holevo fidelity yields the (matrix) Hellinger distance:
\begin{equation}
d_\mathrm{H}^2(P,Q) := \operatorname{Tr}\left[P+Q - 2P^{\frac 12} Q^{\frac 12}\right] = \operatorname{Tr}[P+Q] - 2 \operatorname{F}^\mathrm H(P,Q) =  \left\|P^{\frac 12} - Q^{\frac 12}\right\|_2^2.
\end{equation}

For Uhlmann fidelity and Holevo fidelity, the corresponding Hellinger quantities define valid distances. However, the corresponding Hellinger quantities are not distances for Matsumoto fidelity and Log-Euclidean fidelity. Instead, their squared versions are divergences~\cite{bhatia2019Hellinger}. Various information-theoretic properties of these quantum fidelities are also subjects of intense study~\cite{Zhang2020, lin2015investigating, matsumoto2014quantum}. 

In this paper, we define generalizations of certain fidelities and Bures-Wasserstein distance motivated by the Riemannian geometry of the Bures-Wasserstein manifold. This approach is based on the linearization of the manifold about an arbitrary \emph{base} point. Such a linearization has been previously studied in the context of the 2-Wasserstein distance on the space of probability measures~\cite{ambrosio2008gradient, altschuler2021averaging, chewi2020gradient}.

Let $P,Q,R \in \mathbb P_d$ be a triple of positive definite matrices. The \textit{generalized fidelity between $P$ and $Q$ at $R$} is defined as
\begin{equation}
    \operatorname{F}_R(P,Q) = \operatorname{Tr}\left[\sqrt{ R^{\frac12} P R^{\frac12}} R^{-1}\sqrt{ R^{\frac12} Q R^{\frac12}}  \right]. 
\end{equation}
See Figure~\ref{fig:RebitGenFidelity} for an illustration of generalized fidelity between fixed rebit states $P$ and $Q$ with the base $R$ varying over the rebit disk. As shown, the generalized fidelity between positive matrices can take negative values depending on the base. Other properties illustrated in the figure include the fact that even if $P$ and $Q$ commute, the generalized fidelity can differ from the classical fidelity. Additionally, it is shown that the generalized fidelity achieves its maximum value when the base is along the Bures-Wasserstein geodesic between $P$ and $Q$. Along this geodesic, the generalized fidelity $\Fg{P}{Q}{R}$ equals the Uhlmann fidelity $\Fu{P}{Q}$.

The corresponding squared \textit{generalized Bures distance} between $P$ and $Q$ at $R$ is defined as
\begin{equation}
    \operatorname{B}_R(P,Q) = \operatorname{Tr}[P+Q] - 2 \Re\operatorname{F}_R(P,Q),  
\end{equation}
where $\Re z$ denotes the real part of a complex number $z$. 

The matrix $R$ is referred to as the \textit{base} of the generalized fidelity ( generalized Bures distance) and, as we later show, is the point such that linearizing the BW manifold at this point yields the generalized Bures distance as the natural distance (induced by the inner-product on the tangent space).

The generalized fidelity exhibits remarkable properties as a function of its base. For example, the Uhlmann fidelity between $P$ and $Q$ can be interpreted as the generalized fidelity when one chooses the base as $P$ or $Q$ (among other points). The Holevo fidelity corresponds to the generalized fidelity when the base is chosen to be the identity matrix $\mathbb I$, and Matsumoto fidelity is the generalized fidelity when one linearizes the manifold at $P^{-1}$ or $Q^{-1}$ (among other points). Additionally, various invariance and covariance properties emerge as the base moves along geodesic-related paths on the Riemannian manifold of positive definite matrices. These results are proven in Section~\ref{Sec:GeometricProperties}. See Figure~\ref{Fig:Geodesic} for an illustrative summary of the geometric results.

This article is structured as follows. In Section~\ref{Sec:MathPre}, we briefly review the necessary concepts from Riemannian geometry. In Section~\ref{Sec:BasicProperties}, we formally define generalized fidelity and generalized Bures distance, and study their basic properties. In Section~\ref{Sec:GeometricProperties}, we explore the invariance and covariance properties of the neralized fidelity between two fixed matrices as the base moves along various geodesic-related paths, corresponding to different metrics: Bures-Wasserstein, Affine-invariant, and Euclidean. 

The rest of the paper is devoted to studying further properties of generalized fidelity and generalized Bures distance. In Section~\ref{Sec:PolarFidelity}, we introduce a different family of fidelities called the $x$-\textit{Polar fidelities}, which is a one-parameter family of fidelities that recovers Uhlmann-, Holevo-, and Matsumoto-fidelity. We also discuss certain results that relate generalized fidelity to the special unitary group $\mathrm{SU}(d)$. This leads to the realization that generalized fidelities can be seen to be (the elements of) the \textit{extreme points} of a larger class of fidelities, which is obtained by taking convex combinations of generalized fidelity over different bases.

In Section~\ref{Sec:BlockMatrix}, we derive a block-matrix representation of generalized fidelity and generalized Bures distance. In Section~\ref{Sec:Multivariate}, we establish results connecting the recently-introduced multivariate fidelity~\cite{Wilde2024multivariate}, average fidelity maximizer (normalized Bures-Wasserstein barycenter)~\cite{afham2022, chewi2020gradient}, and generalized fidelity. In Section~\ref{Sec:UhlmannTheorem} we introduce an Uhlmann-like theorem for generalized fidelity, which characterizes generalized fidelity between states based on the inner-product of their purifications.  

Finally, in Section~\ref{Sec:Divergence}, we introduce an analogous base-dependent formulation of quantum R\'enyi divergences and show that it recovers various $\alpha$-$z$ divergences~\cite{audenaert2013alpha} and $\alpha$-geometric Renyi divergence as we vary the base point. We end the article with a few open problems in Section~\ref{Sec:OpenProb} and discussions in Section~\ref{Sec:Concl}. Code to generate plots and further numerical simulations is available at GitHub~\cite{afhamgithub}.

\subsection*{Related Works}

The BW manifold has been extensively studied in classical machine learning because of its relationship with the optimal transport problem and 2-Wasserstein distance. See \cite{bhatia2019bures} for a thorough analysis of the Riemannian geometry of this manifold and its relation to optimal transport and Wasserstein distance. \cite{modin2016geometry} is a great source to understand the interplay between matrix decomposition, optimal transport, and Riemannian geometry. The relationship between Bures distance and Wasserstein distance can be summarized as follows: the 2-Wasserstein distance between two centered Gaussian probability distributions (supported on $\mathbb C^d$) is equal to the Bures distance between their covariance matrices~\cite{bhatia2019bures}.

Given a metric space and a distribution over it, it is possible to define a \textit{barycenter} (of the distribution) as the minimizer of the average squared distance over this distribution. The Bures-Wasserstein barycenter~\cite{agueh2011barycenters, alvarez2016fixed, bhatia2019bures, chewi2020gradient} appears in many contexts including machine 
learning and inference~\cite{cuturi2014fast, ho2017multilevel, dognin2019wasserstein}, quantum information~\cite{afham2022, brahmachari2023fixed, konig2009operational, Wilde2023estimating}, probability theory~\cite{knott1994generalization, ruschendorf2002n}, computer vision and graphics~\cite{rabin2012wasserstein, solomon2015convolutional}, and signal processing~\cite{elvander2020multi}.

In this article, we primarily study the generalizations of quantum fidelities and Bures distance through the linearization of the BW manifold at an arbitrary point. See \cite{hamm2023manifold, cai2022linearized} which studies the linearization of the Wasserstein manifold of probability measures. A closely related work is~\cite{JawanpuriaGeneralizedBuresDistance}, where a different approach to generalizing the Bures-Wasserstein distance is taken. This generalization of BW distance is defined as 
\begin{equation}
    \mathrm{d}^2_M(P,Q) = \operatorname{Tr}\left[M ^{-1} P\right] + \operatorname{Tr}\left[M ^{-1} Q\right] - 2 \operatorname{Tr} \left[\sqrt{P M ^{-1}  Q M^{-1} }\right],
\end{equation}
for any invertible matrix $M$. One can show that this generalization is equivalent to the (usual) Bures-Wasserstein distance between $M ^{-\frac12} P M^{-\frac12} $ and $ M ^{-\frac12} Q M^{-\frac12} $. We note that this generalization and ours are incompatible: their definition does not recover ours, nor does ours recover theirs.

Other approaches to the generalization of Bures-Wasserstein distance can be found in~\cite{pitrik2020quantum, dinh2021alpha}, where the emphasis is on removing the symmetry on the (arithmetic and geometric) means to obtain weighted variants which are shown to be divergences.

\begin{table}[ht]
    \centering
    \begin{tabular}{|c|c|c|}
        \hline
        \textbf{Symbol} & \textbf{Object} & \textbf{Definition} \\
        \hline
        $ \mathbb H_d $ & Set of $d \times d $ Hermitian matrices & $ \mathbb H_d := \{H \in \mathbb C^{d \times d} : H = H^* \} $ \\
        $ \mathbb P_d $ & Set of $d \times d $ positive definite matrices & $ \mathbb P_d := \{P \in \mathbb H_d : \lambda_i(P) > 0\}$ \\
        $ \mathbb U_d $ & Set of $d \times d $ unitary matrices & $ \mathbb U_d := \{U \in \mathbb C^{d \times d} : U U^* = U^*U = \mathbb I\}$ \\

        $ A \# B $ & Matrix geometric mean of $A, B \in \mathbb P_d$ & $ A \# B := A ^{\frac12} \sqrt{A ^{-\frac12} B A ^{-\frac12} } A ^{\frac12}  $ \\
        $ \operatorname{Pol}(A)$ & Polar factor of an invertible matrix $A$ & $ \operatorname{Pol}(A) = A \sqrt{A^*A}^{-1} \in \mathbb U_d $ \\
        
        $ \frac{A}{B}$ & Symmetrized matrix division & $ \frac AB := B ^{-\frac12} A B ^{-\frac12} \quad \text{for invertible } B $  \\

        \hline
    \end{tabular}
    \caption{Table of notations.}
    \label{tab:notation}
\end{table}

\section{Mathematical preliminaries}\label{Sec:MathPre}

This section provides an overview of essential concepts from Riemannian geometry. Additional mathematical preliminaries, such as matrix geometric mean and Bures-Wasserstein barycenter, are discussed in Appendix~\ref{App:MoreMathPrelim}.

\subsection{Riemannian geometry}
We now give a brief summary of relevant concepts from Riemannian geometry. For detailed expositions, see~\cite{lee2018introduction, do1992riemannian}. For a more recent treatment focused on applications to optimization, see~\cite{boumal2023introduction}, and for an exploration of the Riemannian geometry of the Bures-Wasserstein manifold, see~\cite{chewi2020gradient}. Recall that a Riemannian manifold is defined as a smooth manifold equipped with a metric tensor. In the context of generalized fidelity, we study three distinct metric tensors (and, thereby, three different Riemannian manifolds). These are the \textit{Bures-Wasserstein}, \textit{Affine-invariant} (AI), and Euclidean metric tensors.

Each concept will be illustrated with examples from the Euclidean, BW, and AI metric instances. These concepts (and specific instances) are summarized in Table~\ref{Tab:Metrics}. See~\cite{Mishra2021Compare} for a detailed and comparative study of the BW and AI metrics.

\begin{table}[ht]
  \centering
  \small
  \setlength{\tabcolsep}{8pt} 
  \begin{tabular*}{\textwidth}{@{\hspace{5pt}} >{\raggedright\arraybackslash}p{9em} @{\hspace{20pt}} >{\raggedright\arraybackslash}p{12em} @{\hspace{20pt}} >{\raggedright\arraybackslash}p{9em} >{\raggedright\arraybackslash}p{6em} @{\hspace{5pt}}}
    \toprule
    \textbf{Object} & \textbf{BW metric} & \textbf{AI metric} & \textbf{Euc. metric} \\
    \midrule
    Inner product: $\langle X, Y \rangle_P$ & $\operatorname{Tr}[\mathcal L_P(X) P \mathcal L_P(Y)]$ & $\operatorname{Tr}\left[P^{-1} X P^{-1} Y\right]$ & $\operatorname{Tr}[XY]$ \\
    \midrule
    Exp. map: $ \operatorname{Exp}_P [X]$ & $ (\mathbb I  + \mathcal L_P(X)) P (\mathbb I + \mathcal L_P(X))$ & $P^{\frac12} \exp \left(\dfrac X P\right) P^{\frac12}$ & $P + X$ \\ 
    \midrule
    Log map: \  $ \operatorname{Log}_P [Q]$ & $ \mathcal L_P^{-1}((P ^{-1}  \# Q) - \mathbb I)$ & $P^{\frac12} \log \left(\dfrac Q P \right) P^{\frac12}$ & $Q - P$ \\ 
    \midrule
    Geodesic: $ \gamma_{PQ} (t)$ & $ [(1-t) \mathbb I + t S] P [(1-t) \mathbb I + t S]$  & $P^{\frac12} \left(\dfrac QP\right)^t P^{\frac12}$ & $(1-t)P + t Q$ \\ 
    \midrule
    Squared distance: \ $\operatorname{dist}^2 (P,Q)$ & $\operatorname{Tr}[P+Q ]- 2 \operatorname{F}^\mathrm{U}(P,Q)$ & $\left\|\log\left(\dfrac P Q\right)\right\|^2_2$ & $\|P-Q\|^2_2$ \\ 
    \bottomrule
  \end{tabular*}
  \caption{Different Riemannian metrics on $\mathbb P_d$ and associated objects. Here $S := P ^{-1} \operatorname{\#} Q$, $\mathcal L_P(X)$ is the unique solution to the equation $\mathcal L_P(X)P + P \mathcal L_P(X) = X $,  and $\mathcal L_P^{-1}(Y) := Y P + P Y$.}
  \label{Tab:Metrics}
\end{table}

\textbf{Riemannian manifold and metric tensor.} A Riemannian manifold is a tuple $(\mathcal M, \mathfrak{g})$ of a smooth manifold $\mathcal{M}$ and a \textit{metric tensor} $\mathfrak{g}$. For each point $p \in \mathcal M$ on the manifold, one can define a \textit{tangent space} $T_p \mathcal M$, which is a vector space. For the manifold $\mathbb P_d$ of positive definite matrices, the tangent space at any point is isomorphic to the set of Hermitian matrices $\mathbb H_d$. 

The metric tensor endows an inner product to this vector space as follows:
$\mathfrak{g} (p) \equiv \mathfrak{g}_p  \equiv \langle \cdot, \cdot \rangle _p: T_p \mathcal M \times T_p \mathcal M \to \mathbb C. $ which in turn allows one to define notions such as angle and distance on the manifold. For the three manifolds of interest to us, the metric tensors at some point $P \in \mathbb P_d$ are defined as
\begin{equation}
\begin{aligned}
    \langle U, V \rangle_P^\mathrm{Euc} &:= \mathrm{Tr}[U V], \\
    \langle U, V \rangle_P^\mathrm{BW} &:= \mathrm{Tr}\left[\mathcal L_P (U) P \mathcal L_P(V)\right], \\ 
    \langle U, V \rangle_P^\mathrm{AI} &:= \mathrm{Tr}\left[P ^{-1} U P ^{-1} V\right],
\end{aligned}
\end{equation}
where $U, V \in T_P \mathbb  P_d \cong \mathbb H_d$ and $\mathcal L_P (U) $ is the unique solution to the matrix Lyapunov equation $X P + P X = U$ (with $\mathcal  L_P (U) = X$)~\cite{Mishra2021Compare, Malago2018Wasserstein}. 

\textbf{Distance.} The distance between two points on a Riemannian manifold is defined in terms of the minimization of an integral and need not have a closed form in general. Fortunately, for all the metrics that interest us, closed-form solutions exist for the distance function. The squared distances between arbitrary $P, Q \in \mathbb P_d$ for the three metrics are given by:
\begin{equation}
\begin{aligned}
        \mathrm{d}_\mathrm{Euc}^2(P, Q) &= \|P-Q\|_2^2,  \\
        \mathrm{d}_\mathrm{BW}^2(P, Q) &= \operatorname{Tr}[P+Q] - 2 \operatorname{F}^\mathrm U(P,Q),  \\
        \mathrm{d}_\mathrm{AI}^2(P, Q) &= \left\|\operatorname{log}\left(\frac PQ \right) \right\|^2_2,
\end{aligned} 
\end{equation}
where $\|A\|_2 = \sqrt{\text{Tr}[A^*A]}$ is the Euclidean (Hilbert-Schmidt) norm.

\textbf{Geodesic.} For a formal definition and description of geodesics, see~\cite{lee2018introduction, do1992riemannian, vishnoi2018geodesic}. For our purposes, an informal definition would suffice. Informally, the geodesic between two points is the curve of \textit{minimal length} over all smooth curves joining $p$ and $q$. For the metrics that interest us, the geodesics on $\mathbb P_d$ are as follows. 
\begin{equation}
    \begin{aligned}
            \gamma_{PQ}^\mathrm{Euc} (t) &= (1-t)P +  t Q, \\
            \gamma_{PQ}^\mathrm{BW} (t) &= [(1-t) \mathbb I + t S] P [(1-t) \mathbb I + t S], \\
            \gamma_{PQ}^\mathrm{AI} (t) &=  P^{\frac12} \left(\frac Q P\right)^t P^{\frac12}, \\ 
    \end{aligned}
\end{equation}
where $S:= P^{-1}  \# Q$ is the geometric mean between $P ^{-1} $ and $Q$. 
 
\textbf{Riemannian Exponential map.} For any $p \in \mathcal M$, there always exists an $\epsilon > 0$ such that for any tangent vector $v \in T_p \mathcal M$ with $\|v\|_p \leq \epsilon $, there is a unique constant-speed geodesic $\gamma_{p,v}:[0,1] \to \mathcal M$ with initial conditions
\begin{equation}
    \gamma_{p,v}(0) = p \quad \text{and} \quad \dot{\gamma}_{p,v}(0) = v.
\end{equation}
Call this neighborhood $N_p \subset T_p \mathcal M$. With these tangent vectors as the domain, we can define the Riemannian \textit{Exponential map} $\mathrm{Exp}_p$ as
\begin{equation}
    \mathrm{Exp}_p[v] = \gamma_{p,v}(1) \quad \text{for any } v \in N_p \subset T_p \mathcal M. 
\end{equation}

In words, $\operatorname{Exp}_p$ maps a (sufficiently small) tangent vector $v$ of $p$ to the point on the manifold, which is obtained by `traveling' along the geodesic $\gamma_{p,v}$ for a unit \textit{time}. 

The exponential maps have the following closed forms for the three Riemannian manifolds that interest us.
\begin{equation}
    \begin{aligned}
            \mathrm{Exp}_P^\mathrm{Euc} [V] &= P + V, \\
            \mathrm{Exp}_P^\mathrm{BW} [V] &= (\mathbb I + \mathcal L_P (V))P(\mathbb I + \mathcal L_P (V)), \\
            \mathrm{Exp}_P^\mathrm{AI} [V] &= P^{\frac12}  \exp\left(\frac VP\right) P^{\frac12} . \\ 
    \end{aligned}
\end{equation}
 
As previously mentioned, the exponential map is not usually defined over all of the tangent space. Instead, it can be the case that $ \mathrm{dom}(\mathrm{Exp}_p) \subset T_p \mathcal M$, which is the case for the BW and Euclidean metric. However, for the AI metric, we have $\mathrm{dom}(\mathrm{Exp}_p) = T_p \mathcal M$. 

\textbf{Riemannian Logarithmic map.} The Riemannian exponential map is a diffeomorphism onto its image, and thus we define its inverse --  the \textit{Riemannian logarithmic map}. It has the form 
\begin{equation}
     \mathrm{Log}_p : \mathrm{Exp}_p[N_p] (\subseteq \mathcal M ) \to T_p \mathcal M. 
\end{equation}
One can think of it as the following mapping. Suppose we \textit{linearize} the manifold $\mathcal M$ around some point $p$. The \textit{flattened} space is identified with the tangent space $T_p \mathcal M$. The $\mathrm{Log}_p$ map then maps every \textit{nearby} point $q \in \mathcal M$ in the manifold to a tangent vector $\mathrm{Log}_p[q] \in T_p\mathcal M$, such that if you start from $p$ and move along said tangent vector $\mathrm{Log}_p[q]$ for \textit{unit time}, then we reach $q$. 

The Log map has the following closed form for the Riemannian manifolds that interest us. 
\begin{equation}
    \begin{aligned}
            \mathrm{Log}_P^\mathrm{Euc} [Q] &= Q - P, \\
            \mathrm{Log}_P^\mathrm{BW} [Q] &= \mathcal L_P ^{-1} (P ^{-1} \# Q - \mathbb I), \\
            \mathrm{Log}_P^\mathrm{AI} [Q] &= P^{\frac12}  \log\left(\frac QP\right) P^{\frac12} , \\ 
    \end{aligned}
\end{equation}
where $\mathcal L_P^{-1} (Z) := Z P + P Z$.

\section{Generalized fidelity and generalized Bures distance} \label{Sec:BasicProperties}

We now formally introduce generalized fidelity and generalized Bures distance. We begin with the definition and some basic properties of generalized fidelity. We then do the same for generalized Bures distance and finally conclude the section with the Riemannian-geometric motivation for the definition. 
\subsection{Generalized fidelity}

Let us now define generalized fidelity. 
\begin{definitionbox}
    \begin{definition}     Let $P,Q,R \in \mathbb{P}_d$ be positive definite matrices. The \emph{generalized fidelity between}  $P$ \textit{and} $Q$ \textit{at} $R$  is defined as
    \begin{equation}
         \operatorname{F}_R(P, Q) :=  \operatorname{Tr}  \left[\sqrt{R ^{\frac12} P R ^{\frac12} } R ^{-1} \sqrt{R ^\frac12 Q R ^\frac12}\right] 
    \end{equation} 
\end{definition}
\end{definitionbox}
Here and henceforth, $R$ is referred to as the \textit{base} of the generalized fidelity between $P$ and $Q$. Throughout the analysis, we require the base $R$ to be positive definite while $P$ and $Q$ can be rank-deficient. However, for most of our analysis, we consider $P$ and $Q$ to be invertible as well. The first thing to notice is that the matrix inside the trace is not Hermitian in general. Thus, generalized fidelity is complex-valued in general. We now list some basic properties of generalized fidelity. Proofs are provided in Appendix~\ref{App:BasicProperties}.

\begin{enumerate}
    \item \textbf{Quantization of classical fidelity.} The generalized fidelity $\operatorname{F}_R(P,Q) $ is a quantization of classical fidelity for any triple $P,Q,R \in \mathbb P_d$. That is, if $P,Q,$ and $R$ mutually commute, the generalized fidelity reduces to the classical fidelity (Bhattacharyya coefficient) between $P$ and $Q$. 
    \item \textbf{Conjugate symmetry.} Generalized fidelity is generally complex-valued and conjugate symmetric: 
    \begin{equation}
        \operatorname{F}_R(P, Q) \in \mathbb{C} \quad \text{and} \quad  \operatorname{F}_R(P, Q) = \Fg{Q}{P}{R}^*.
    \end{equation}
    Moreover, $ \Fg{P}{P}{R} = \operatorname{Tr}[P]$ for any $P, R \in \mathbb P_d$. It follows that the generalized fidelity of a state with itself is 1 at any base.
    \item  \textbf{Equivalent forms.} The generalized fidelity has the following equivalent forms.
    \begin{equation}
       \begin{aligned}
            \operatorname{F}_R(P,Q) :&= \operatorname{Tr} \left[\sqrt{R ^{\frac 12} P R ^{\frac 12} } R ^{-1} \sqrt{R ^{\frac 12} Q R ^{\frac 12} } \right]            \\ &= \operatorname{Tr}\left[Q ^{\frac12} U_Q U_P^* P^{\frac12}  \right] \\  
            &= \operatorname{Tr}\left[ (R ^{-1} \# Q) R (R ^{-1}  \#  P) \right],
       \end{aligned} 
    \end{equation}
    where $U_P := \text{Pol}\left(P^{\frac12}  R ^{\frac12} \right)$, $U_Q := \text{Pol}\left(Q^{\frac12} R^{\frac12} \right)$.
    \item \textbf{Pure state simplification.} If both $P = \ketbra{\psi}$ and $Q = \ketbra{\phi}$ are pure states, then
    \begin{equation}
        \operatorname{F}_R(P,Q) =  \Fg{\ketbra{\psi}}{\ketbra{\phi}}{R} = \frac{\langle \psi, \phi \rangle \langle \phi, R \psi \rangle }{\Fu{P}{R} \Fu{Q}{R}}. 
    \end{equation}
    \item \textbf{Commutation with base implies reality.} If the base $R$ commutes with $P$ or $Q$, then $\operatorname{F}_R(P, Q) \in \mathbb{R}$.
    \item \textbf{Multiplicativity.} Let $P_1, Q_1, R_1 \in \mathbb{P}_{d_1}$ and $P_2, Q_2, R_2 \in \mathbb{P}_{d_2}$. Then
    \begin{equation}
    \Fg{P_1 \otimes P_2}{Q_1 \otimes Q_2}{R_1 \otimes R_2} = \Fg{P_1}{Q_1}{R_1} \cdot \Fg{P_2}{Q_2}{R_2}.
    \end{equation}
    \item \textbf{Additivity.} Let $P_1, Q_1, R_1 \in \mathbb{P}_{d_1}$ and $P_2, Q_2, R_2 \in \mathbb{P}_{d_2}$. Then
    \begin{equation}
    \Fg{P_1 \oplus P_2}{Q_1 \oplus Q_2}{R_1 \oplus R_2} = \Fg{P_1}{Q_1}{R_1} + \Fg{P_2}{Q_2}{R_2}.
    \end{equation}
    \item \textbf{Unitary invariance.} For any triple $P, Q, R \in \mathbb{P}_d$ and any unitary $U \in \mathbb{U}_d$,
    \begin{equation}
        \operatorname{F}_R(P, Q) = \Fg{U P U^*}{U Q U^*}{U R U^*}.
    \end{equation}
    \item \textbf{Unitary contravariance.} For any triple $P, Q, R \in \mathbb P_d $ and any unitary $U \in \mathbb U_d$,
    \begin{equation}
        \operatorname{F}_{U R U^*}(P, Q) = \Fg{U^* P U}{U^* Q U}{R}.
    \end{equation}
    \item \textbf{Scaling.} For positive scalars $p, q, r \in \mathbb{R}_+$,
    \begin{equation}
    \Fg{p P}{q Q}{r R} = \sqrt{p q} \operatorname{F}_R(P, Q).
    \end{equation}
    Since the generalized fidelity is invariant with respect to any scaling of the base $R$, one can always choose the base to be a density matrix without loss of generality.
    \item \textbf{Absolute value is bounded above by Uhlmann fidelity.} For any triple $P, Q, R \in \mathbb{P}_d$,
    \begin{equation}
    |\operatorname{F}_R(P, Q)| \leq \Fu{P}{Q}.
    \end{equation}
    \item \textbf{Orthogonal support implies 0.} If any two matrices from $P,Q,R$ have orthogonal support, then $\operatorname{F}_R(P,Q) =0.$
    \item \textbf{Reduction to other fidelities.} For specific choices of the base $R$, one can recover various named fidelities from generalized fidelity.
    \begin{itemize}
        \item \emph{Uhlmann fidelity:} If (but not only if) $R = P$ or $R = Q$,
        \begin{equation}
        \operatorname{F}_R(P, Q) = \operatorname{F}^\mathrm{U}(P, Q) := \operatorname{Tr}\left[\sqrt{P^\frac12 Q P^\frac12}\right].
        \end{equation}
        \item \emph{Holevo fidelity:} If $R = \mathbb{I}$,
        \begin{equation}
        \operatorname{F}_{R}(P,Q) = \operatorname{F}^\mathrm{H}(P, Q) := \operatorname{Tr}\left[P^\frac12 Q^\frac12\right].
        \end{equation}
        \item \emph{Matsumoto fidelity:} If (but not only if) $R = P^{-1}$ or $R = Q^{-1}$,
        \begin{equation}
        \operatorname{F}_R(P, Q) = \operatorname{F}^\mathrm{M}(P, Q) := \operatorname{Tr}[P \# Q] = \operatorname{Tr}\left[P^\frac12 \sqrt{P^{-\frac 12} Q P^{-\frac 12}} P^\frac12\right].
        \end{equation} 
    \end{itemize}
    We note that these are not the only cases where generalized fidelity reduces to the above-mentioned fidelities. Other scenarios where this happens are discussed in Section~\ref{Sec:GeometricProperties}.
\end{enumerate}

Next, we define the generalized Bures distance and discuss some basic properties and the geometric motivation behind the definition. We also see how generalized fidelity naturally appears in this geometric motivation.

\subsection{Generalized Bures distance}
Having defined generalized fidelity, we are in a position to define the \textit{generalized Bures distance}. After defining it, we will show how it can be naturally derived from the geometry of the Bures-Wasserstein manifold.

\begin{definitionbox}
    \begin{definition}     Let $P,Q,R \in \mathbb{P}_d$ be positive definite matrices. The squared \emph{generalized Bures(-Wasserstein) distance between}  $P$ \textit{and} $Q$ \textit{at} $R$  is defined as
    \begin{equation}
         \operatorname{B}_R(P, Q) :=  \operatorname{Tr}[P+Q] - 2 \Re \operatorname{F}_R(P,Q).  
    \end{equation} 
\end{definition}
\end{definitionbox}

We use capital $\mathrm{B}$ to denote squared distance (and certain divergences) and \textit{small} $\mathrm{b}$ to denote distances. For example, we will henceforth use
\begin{equation}
     \operatorname{b}^\mathrm{U}(P,Q) \equiv \sqrt{\operatorname{B}^\mathrm U(P,Q) } = \sqrt{\operatorname{Tr}[P+Q] - 2\operatorname{F}^\mathrm U(P,Q) }
\end{equation}
to denote the Bures-Wasserstein distance. The superscript indicates the specific type of fidelity used. Similarly, we denote the generalized Bures distance as
\begin{equation}
     \operatorname{b}_R(P,Q) \equiv \sqrt{\operatorname{B}_R(P,Q) } = \sqrt{\operatorname{Tr}[P+Q] - 2 \Re \operatorname{F}_R(P,Q)}.
\end{equation}
 
Essentially, the relation between generalized Bures distance and generalized fidelity is analogous to the relation between Bures distance and Uhlmann fidelity. 

We now show that this definition is geometrically motivated. As mentioned before, the generalized Bures distance between $P$ and $Q$ at $R$ is the distance between $P$ and $Q$ if we \textit{flatten the} Bures-Wasserstein manifold at $R$. This is formalized in the next section. 

\subsubsection{Geometric motivation for generalized Bures distance}
We now provide a (Riemannian-)geometrically motivated derivation of the generalized Bures distance. Since the generalized fidelity features in the definition of the generalized Bures distance, this also provides a geometric interpretation for generalized fidelity. 

Let $(\mathcal M, \mathfrak{g})$ be a Riemannian manifold. It is a defining property of Riemannian manifolds~\cite{do1992riemannian} that for any $x \in \mathcal M$, there is a neighborhood $\mathcal N_x \subseteq \mathcal M$ such that for any $y \in \mathcal N_x$, we have the natural distance between the points to be
\begin{equation}
    \mathrm{d}^2(x,y) = || \mathrm{Log}_x[y] ||_x^2,
\end{equation}
where $\Log_x[y] \in T_x \mathcal M$ and where the norm is taken with respect to the inner product $\mathfrak g_x$ on the tangent spce $T_x \mathcal M$. It turns out that for the Bures-Wasserstein manifold, $\mathcal N_P = \mathbb P_d$ for any $P \in \mathbb P_d $. Thus, we can write
\begin{equation}
    \mathrm{b}^\mathrm{U}(P,Q) =  \| \mathrm{Log}_P[Q] \|_P = \| \mathrm{Log}_Q[P] \|_Q, 
\end{equation}
for any $P, Q \in \mathbb P_d$ where the norm is taken with respect to the Bures-Wasserstein metric tensor in the corresponding tangent space.

One can visualize this process as follows. We linearize the \textit{curved} Bures manifold $\mathbb P_d$ \textit{about} the point $P$. This is equivalent to replacing every point $S \in \mathbb P_d$ with its image under the Riemannian log map $\Log_P[S] \in T_P \mathbb P_d$. The tangent space is an inner-product space, and thus, we can define the induced distance between two tangent vectors $X, Y \in T_P \mathbb P_d$ as $\|X - Y\|_P^2$. Therefore the squared distance between the tangent vectors $\Log_P[P]$ and $\Log_P[Q]$ is given by
\begin{equation}
     \|\Log_P[P] - \Log_P[Q]\|_P^2 = \|\Log_P[Q]\|_P^2 = \operatorname{B}^\mathrm{U}(P,Q),
\end{equation}
where we note that $\Log_S[S] = 0$ for any $S \in \mathbb P_d$. 

To obtain generalized Bures distance and thereby generalized fidelity, we flatten our manifold not at $P$ or $Q$, but at an arbitrary third point $R \in \mathbb P_d$. That is, we map the points $P,Q \in \mathbb P_d$ to the tangent space $T_R \mathbb P_d$ via the $\Log_R$ map. We then compute the natural distance between these tangent vectors. It turns out that this is exactly the generalized Bures distance between $P$ and $Q$ at $R$. This is formalized in the following theorem. 

\begin{theorembox}
\begin{theorem} \label{Thm:GBDTangentSpace}
    Let $P,Q,R \in \mathbb P_d$ be chosen arbitrarily. Then
    \begin{equation}
        \operatorname{B}_R(P,Q) = \|\Log_R[P] - \Log_R[Q]\|_R^2.
    \end{equation}
\end{theorem}
\end{theorembox}
\begin{proof}
    We will show that the RHS evaluates to the definition of the squared generalized Bures distance. For brevity, we will use the shorthand $P_R \equiv \Log_R[P] $ and $Q_R \equiv \Log_R[Q] $. Expanding the RHS, we have
\begin{equation}    \label{Eq:21}
\begin{aligned}
    \|P_R - Q_R \|_R^2 &= \langle P_R - Q_R , P_R - Q_R \rangle_R \\ &= \langle P_R, P_R \rangle_R + \langle Q_R, Q_R \rangle_R  - \langle P_R, Q_R \rangle_R - \langle Q_R, P_R \rangle_R,    
\end{aligned}
\end{equation}
where we used the bilinearity of the inner product to get the second line. Let us now analyze a single term, namely $\langle P_R,Q_R \rangle _R$, of the second line. By the definition of the Bures-Wasserstein metric, we have
\begin{equation}
\begin{aligned}
    \langle P_R, Q_R \rangle _R := \operatorname{Tr}\left[\mathcal L_R(P_R) \cdot R \cdot \mathcal L_R(Q_R) \right] 
    = \operatorname{Tr}[\mathcal L_R (\Log_R[P]) \cdot R \cdot \mathcal L_R (\Log_R[Q])]. 
\end{aligned}
\end{equation}
Now recall that the definition of the BW $\Log_R$ map is $\Log_R[Q] := \mathcal L_R ^{-1} (P ^{-1} \# Q - \mathbb I)$, and thus we have
\begin{equation}
\begin{aligned}
    \langle P_R, Q_R \rangle _R &= \operatorname{Tr}\left[\mathcal L_R (\mathcal L_R ^{-1} (R ^{-1} \# P - \mathbb I)) \cdot R  \cdot \mathcal L_R (\mathcal L_R ^{-1} (R ^{-1} \# Q - \mathbb I)) \right] \\
    &= \operatorname{Tr}[(R ^{-1} \# P - \mathbb I) R (R ^{-1} \# Q - \mathbb I)] \\
    &= \operatorname{Tr}[(R ^{-1} \# P) R (R ^{-1} \# Q)] - \operatorname{Tr}[(R ^{-1} \# P) R] -\operatorname{Tr}[(R ^{-1} \# Q) R] + \operatorname{Tr}[R] \\
    &= \operatorname{F}_R(Q,P) - \Fu{P}{R} - \Fu{Q}{R} + \operatorname{Tr}[R],  
\end{aligned}
\end{equation}
where we used the fact that $\text{Tr}[A(A^{-1}  \# B)] =  \text{Tr}\left[\sqrt{A ^{\frac 12} B A ^{\frac 12} }\right] = \operatorname{F}^\mathrm U(A,B)$.
We thus have
\begin{equation}
\begin{aligned}
    \|P_R\|_R^2 = \operatorname{Tr}[P + R] - 2 \operatorname{F}^\mathrm U(P,R) = \operatorname{B}^\mathrm U(P,R), \\   
    \|Q_R\|_R^2 = \operatorname{Tr}[Q + R] - 2 \operatorname{F}^\mathrm U(Q,R) = \operatorname{B}^\mathrm U(Q,R). \\  
\end{aligned}
\end{equation}
Substituting all of these in \Cref{Eq:21}, we get
\begin{equation}
\begin{aligned}
    \|P_R - Q_R\|_R^2 &= \operatorname{B}^\mathrm U(P,R) + \operatorname{B}^\mathrm U(Q,R) - 2 \left(\operatorname{Tr}[R] - \operatorname{F}^\mathrm U(P, R) - \operatorname{F}^\mathrm U(Q, R) + \Re \operatorname{F}_R(P,Q) \right) \\
    &= \operatorname{Tr}[P+Q] - 2 \Re \operatorname{F}_R(P,Q) =: \operatorname{B}_R(P,Q).  
\end{aligned} 
\end{equation}
This concludes the proof.
\end{proof}

Thus, the geometric interpretation of generalized Bures distance is clear. It is the distance between the points $P$ and $Q$ if we \textit{linearize} the manifold at $R$. Analogously, the generalized fidelity is the \textit{fidelity part} of the generalized Bures distance in this scenario. Thus, the various named fidelities have the following geometric interpretations:
\begin{itemize}
    \item Uhlmann fidelity is the generalized fidelity if we linearize the manifold at $P$ or $Q$ (among other points),
    \item Holevo fidelity is the generalized fidelity if we linearize the manifold at $\mathbb I$, and
    \item Matsumoto fidelity is the generalized fidelity if we linearize the manifold at $P ^{-1} $ or $Q ^{-1} $ (among other points).
\end{itemize}
The \textit{other points} are discussed in Section~\ref{Sec:GeometricProperties}. We now discuss some other properties of the generalized Bures distance.

\subsubsection{Further properties of generalized Bures distance}

We begin with an alternative form of generalized Bures distance.
\begin{theorembox}
\begin{prop} \label{Thm:GBRAlternativeForm}
    Let $P,Q,R \in \mathbb P_d$. Then, the squared generalized Bures distance has the form 
    \begin{equation}
        \operatorname{B}_R(P,Q) = \left\| U^*_P P^{\frac12}  - U^*_Q Q ^{\frac12} \right\|_2^2,
    \end{equation}
    where $U_P := \operatorname{Pol}\left(P^{\frac12}  R ^{\frac12} \right)$, $U_Q := \operatorname{Pol}\left(Q ^{\frac12} R ^{\frac12} \right)$, and $\|A\|_2 := \sqrt{\operatorname{Tr}[A^*A]}$ denotes the Frobenius norm of a matrix $A$. 
\end{prop}
\end{theorembox}
\begin{proof}
We have
\begin{equation}
\begin{aligned}
        \left\| U^*_P P^{\frac12}  - U_Q^* Q ^{\frac12} \right\|_2^2 = &\operatorname{Tr}\left[ (U_P^* P^{\frac12}  - U_Q^* Q ^{\frac12})^* (U_P P^{\frac12}  - U_Q Q ^{\frac12}) \right] \\
        = &\operatorname{Tr} \left[P + Q - P^{\frac12} U_P U_Q^* Q^{\frac12}  + Q ^{\frac12} U_Q U_P^* P^{\frac12}  \right] \\ 
        = &\operatorname{Tr}[P+Q]  -2 \Re \operatorname{F}_R(P,Q) =: \operatorname{B}_R(P,Q).   
\end{aligned}
\end{equation}
This concludes the proof.
\end{proof}

\textbf{Remark.} By the adjoint invariance of Frobenius norm, we also have 
\begin{equation}
    \operatorname{B}_R(P,Q) =  \left\| P^{\frac12}  U_P - Q ^{\frac12} U_Q\right\|_2^2.
\end{equation}
We now show that generalized Bures distance is a bona fide distance. A distance function $\operatorname{d}$ must satisfy the following threey properties.
\begin{enumerate}
    \item Symmetry: $\operatorname{d}(x,y) = \text{d}(y,x)$.
    \item Nonnegativity: $\text{d}(x,y) \geq 0 $ with equality if and only if $x =y$.
    \item Triangle inequality: $\text{d}(x,y) \geq \text{d}(x,z) + \text{d}(z,y)$ for any $z$. 
\end{enumerate}
We show that the generalized Bures distance satisfies the above three properties in the following theorem.

\begin{theorembox}
\begin{theorem}
    The generalized Bures distance $\operatorname{b}_R( \cdot, \cdot)$ at any $R \in \mathbb P_d$ satisfies
    \begin{enumerate}
        \item Symmetry: $\operatorname{b}_R(P,Q) = \operatorname{b}_R(Q,P)$, 
        \item Nonnegativity: $\operatorname{b}_R(P,Q) \geq 0$ with equality if and only if $P = Q$. 
        \item Triangle inequality: For any triple $P,Q,S \in \mathbb P_d$, we have
        \begin{equation}
            \operatorname{b}_R(P,Q) \leq \operatorname{b}_R(P,S) + \operatorname{b}_R(S,Q).
        \end{equation}
    \end{enumerate}
\end{theorem}
\end{theorembox}
\begin{proof}
    The symmetry part follows directly from the definition. For the nonnegativity part, either observe that the generalized Bures distance is a norm in the tangent space $T_R \mathbb P_d$ or observe that by Proposition~\ref{Thm:GBRAlternativeForm}, we have 
    \begin{equation}
        \operatorname{b}_R(P,Q) = \left\| U_P^* P^{\frac12}  - U_Q^* Q ^{\frac12} \right\|_2 \geq 0, 
    \end{equation}    
    where $U_P := \operatorname{Pol}\left(P^{\frac12}  R  ^{\frac12} \right)$ and $U_Q := \operatorname{Pol}\left(Q ^{\frac12} R  ^{\frac12} \right)$. We now show that equality is achieved if and only if $P = Q$. The direction $P=Q \implies \operatorname{b}_R(P,Q) = 0$ follows trivially. In the other direction, we want to prove
    \begin{equation}
        \operatorname{b}_R(P,Q) = 0 \implies P = Q.
    \end{equation}
    To this end, observe that
    \begin{equation}
        \operatorname{b}_R(P,Q) = \left\| U_P^* P^{\frac12}  - U_Q^* Q^{\frac12} \right\|_2 = 0 \quad \implies  \quad U_P^* P^{\frac12}  =  U_Q^* Q^{\frac12}. 
    \end{equation}
    Right multiply both sides by $R ^{\frac12} $ to obtain
    \begin{equation}
        U_P^* P^{\frac12}  R ^{\frac12} = U_Q^* Q ^{\frac12} R ^{\frac12}. 
    \end{equation}
    By Lemma~\ref{Lem:PolarFactorLemma}, we have that the above equality is the same as
    \begin{equation}
        \sqrt{R ^{\frac12} P R ^{\frac12} } = \sqrt{R ^{\frac12} Q R ^{\frac12} }.
    \end{equation}
    Squaring both sides and right-left multiplying by $ R^{-\frac12} $, we get $P = Q$.
    
    To prove that the generalized Bures distance satisfies the triangle inequality, choose arbitrary $P,Q,R,S \in \mathbb P_d$. We then have,
    \begin{equation}
    \begin{aligned}
        \operatorname{b}_R(P,Q) =  &\left\| U_P^* P^{\frac12}  - U_Q^* Q^{\frac12}   \right\|_2  \\ 
        = &\left\| U_P^* P^{\frac12}  - U_S^* S ^{\frac12} + U_S^* S ^{\frac12} +  U_Q^* Q^{\frac12} \right\|_2 \\ \leq & \left\| U_P^* P^{\frac12}  - U_S^* S ^{\frac12} \right\|_2 + \left\| U_S^* S ^{\frac12} - U_Q^* Q ^{\frac12} \right\|_2  \\ = &\operatorname{b}_R(P,S) + \operatorname{b}_R(S,Q).   
    \end{aligned}
    \end{equation} 
    Here $U_S := \operatorname{Pol}\left(R ^{\frac12} S ^{\frac12} \right)$ and we have used triangle inequality for the Frobenius norm.
\end{proof}

We now mention a different way of showing triangle inequality and the saturation of the nonnegativity by invoking the tangent space $T_R \mathbb P_d$. By Theorem~\ref{Thm:GBDTangentSpace}, we have
\begin{equation}
    \text{b}_R(P,Q) = \| \Log_R[P] - \Log_R[Q]\|_R = 0 \quad \implies \quad  \Log_R[P] = \Log_R[Q].  
\end{equation}
Now recall that the map $\Log_R$ is a diffeomorphism from $\cal M$ to its image and, therefore, is invertible, which necessarily means $P = Q$. For the triangle inequality we have that for any quadruple $P,Q,R,S \in \mathbb P_d$, we have
\begin{equation}
\begin{aligned}
\text{b}_R(P,Q) &=  \left\|\Log_R[P] - \Log_R[Q]\right\|_R \\ & = \|\Log_R[P] - \Log_R[S] + \Log_R[S] - \Log_R[Q]\|_R \\ &\leq \|\Log_R[P] - \Log_R[S]\|_R + \|\Log_R[S] - \Log_R[Q]\|_R  \\ &= \text{b}_R(P,S) + \text{b}_R(Q,S),
\end{aligned}
\end{equation}
where the inequality comes from the fact that the tangent space is an inner-product space, and therefore, triangle inequality holds.

\section{Geometric properties of generalized fidelity } \label{Sec:GeometricProperties}

\begin{figure}[htbp]
    \centering
   
    \begin{subfigure}[b]{1\textwidth} 
        \centering
        \includegraphics[width=0.555\textwidth]{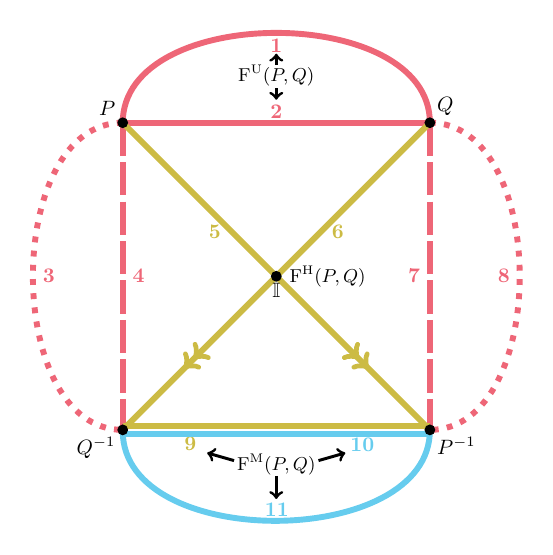} 
        \caption{Generalized fidelity $\operatorname{F}_R(P,Q) $ at different base points $R \in \mathbb P_d$ for fixed $P,Q \in \mathbb P_d$. Red, yellow, and blue curves indicate that the associated metric is Bures-Wasserstein (BW), Affine-invariant (AI), and Euclidean respectively. Straight and curved paths indicate that $R$ and $R ^{-1} $ are elements of geodesics in their respective metrics.  Solid lines indicate that the generalized fidelity is real-valued along these paths, while non-solid lines indicate complex values. Arrows on paths 5 and 6 indicate a conjectured monotonic decrease in $\operatorname{F}_R(P,Q)$ as $R$ moves in the direction of the arrow. Details discussed in Section~\ref{Sec:GeometricProperties}.}
        \label{Fig:GeoFigureDiagram} 
    \end{subfigure}

    \begin{subfigure}[b]{1\textwidth} 
        \centering
        \footnotesize
        \setlength{\tabcolsep}{5pt}
        \renewcommand{\arraystretch}{2}
        \begin{tabular}{|c|c|c|}
            \hline
            \textbf{Path} & \textbf{Path definition} & \textbf{Property of generalized fidelity $\operatorname{F}_{R}(P,Q)  $} \\
            \hline
            1 & $R = \left[\gamma^\text{BW}_{P ^{-1} Q ^{-1} }(t) \right] ^{-1} $  & Invariant. $  \operatorname{F}_R(P,Q) =  \operatorname{F}^\mathrm U(P,Q) $ for all $t \in [0,1].$\\ \hline
                                
            2 & $R = \gamma^\text{BW}_{P Q }(t)  $  & Invariant. Equal to $ \operatorname{F}^\mathrm U(P,Q) $ for all $t \in [0,1].$ \\ \hline

            3 & $R = \left[\gamma^\text{BW}_{P^{-1} Q   }(t)\right]^{-1}   $  & Complex-valued and covaries with Path 8. \\ \hline

            4 & $R = \gamma^\text{BW}_{P Q ^{-1} }(t)  $  & Complex-valued and covaries with Path 7. \\ \hline

            5 & $R = \gamma^\text{AI}_{P P ^{-1} }(t)  $  & Recovers Uhlmann-, Holevo-, and Matsumoto-fidelity. \\ \hline 

            6 & $R = \gamma^\text{AI}_{Q Q^{-1} }(t)  $  & Recovers Uhlmann-, Holevo-, and Matsumoto-fidelity. \\ \hline 

            7 & $R = \gamma^\text{BW}_{Q P ^{-1} }(t)  $  & Complex-valued and covaries with Path 4. \\ \hline

            8 & $R = \left[\gamma^\text{BW}_{Q ^{-1} P }(t)\right]^{-1}   $  & Complex-valued and covaries with Path 3. \\ \hline

            9 & $R = \gamma^\text{AI}_{ P ^{-1} Q ^{-1} }(t)  $  & Invariant. $\operatorname{F}_R(P,Q) = \operatorname{F}^\mathrm M(P,Q) $ for all $t \in [0,1].$ \\ \hline

            10 & $R = \gamma^\text{Euc}_{ P ^{-1} Q ^{-1} }(t)  $  & Invariant. $\operatorname{F}_R(P,Q) = \operatorname{F}^\mathrm M(P,Q) $ for all $t \in [0,1].$ \\ \hline
            
            11 & $R = \left[\gamma^\text{Euc}_{ P Q }(t)\right]^{-1}  $  & Invariant. $\operatorname{F}_R(P,Q) = \operatorname{F}^\mathrm M(P,Q) $ for all $t \in [0,1].$ \\ \hline
        \end{tabular}
        \caption{Definitions and properties of the paths discussed in Figure~\ref{Fig:GeoFigureDiagram}.}
        \label{Fig:GeoFigureTable}
    \end{subfigure}
    \caption{Summary of Geometric properties of generalized fidelity.}
    \label{Fig:Geodesic}
\end{figure}

We now arrive at the central geometric results of this article. We discuss the geometric properties of the generalized fidelity $\operatorname{F}_R(P,Q)$ for fixed $P, Q \in \mathbb P_d$ and variable $R$. In particular, we study the change (or lack thereof) in generalized fidelity $\operatorname{F}_R(P,Q)$ as $R$ varies along certain curves. Each curve discussed is visualized in Figure~\ref{Fig:Geodesic}. In all the below cases, we choose $t \in [0,1]$. The curves we discuss in $\mathbb P_d$ are related to three Riemannian metrics: Bures-Wasserstein, Affine-invariant, and Euclidean. We concisely list important aspects of each metric in Table~\ref{Tab:Metrics}. Each path we study falls into two categories: geodesic paths (represented as straight lines in Figure~\ref{Fig:Geodesic}) and inverse of geodesic paths (represented as curved lines in Figure~\ref{Fig:Geodesic}). By `inverse of geodesic paths' we mean that the inverse $R ^{-1} $ of the base $R$ moves along some geodesic path. We now discuss the properties of generalized fidelity as the base $R$ moves along each specified path, grouped by the metric to which each path is related. For better readability, we defer all proofs of this section to Appendix~\ref{App:GeometricProperties}.

We represent a geodesic path with respect to the Riemannian metric `RM' between points $A, B \in \mathbb P_d$ as \begin{equation}
    \gamma_{A B}^\text{RM} : [0,1] \to \mathbb P_d,
\end{equation}
with $\gamma_{A B}^\text{RM}(0) = A$ and $\gamma_{A B}^\text{RM}(1) = B.$

\subsection{Bures-Wasserstein (red) paths}
We first discuss the paths that are related to the BW metric, which is colored red in Figure~\ref{Fig:Geodesic}. We consider six paths related to this metric. One defining property of the base $R$ as it travels along these paths is that either $R$ or $R ^{-1} $ is an element of a particular BW geodesic. Recall that the BW geodesic between $A,B \in \mathbb P_d$ has the form
\begin{equation}
    C \equiv \gamma_{A B}^{\text{BW}} (t) = [(1-t)\mathbb I + t A ^{-1} \# B] A [(1-t)\mathbb I + t A ^{-1} \# B].
\end{equation}
Moreover, being an element of the BW geodesic (at time $t$) is equivalent to being the BW barycenter of a 2-point distribution with weights $(t, 1-t)$. As such, it uniquely satisfies the following fixed-point equation. 
\begin{equation}
    C = (1-t) \sqrt{C ^{\frac12} A C ^{\frac12} } + t \sqrt{B ^{\frac12} A B ^{\frac12} }.  
\end{equation}
These two properties will be exploited in the proofs. 

\subsubsection*{Path 2: $R = \gamma^{\mathrm{BW}}_{PQ}(t)$}

Here the base $R$ moves along the BW geodesic path between $P$ and $Q$:
\begin{equation}
R = \gamma^{\mathrm{BW}}_{PQ}(t) = \left[(1-t)\mathbb{I} + t S \right] P \left[(1-t)\mathbb{I} + t S \right], \quad \text{for} \; t \in [0,1],
\end{equation}
where $S = P^{-1} \# Q$. As shown in the following theorem, $\operatorname{F}_R(P,Q) $ is constant and equal to Uhlmann fidelity along this curve.

\begin{theorembox}
    \begin{theorem} \label{Thm:Path2}
        Let $P, Q \in \mathbb P_d$ be fixed. Let the base $R = \gamma^{\mathrm{BW}}_{PQ}(t)$ for any $t \in [0,1]$. Then
        \begin{equation}
            \operatorname{F}_R(P,Q) = \operatorname{F}^{\emph U}(P,Q). 
        \end{equation}
    \end{theorem}
\end{theorembox}
\begin{proof}
    See Theorem~\ref{App:ThmPath2}.
\end{proof}

Thus, as the base $R$ varies along the BW geodesic $\gamma_{P Q}^{\text{BW}}$ between $P$ and $Q$, the generalized fidelity is invariant and equal to the Uhlmann fidelity between $P$ and $Q$. This is the first example where the generalized fidelity is real-valued for a non-trivial base ($R \notin \{P^x, Q^y\}$). We next look at another path that exhibits the same property.

\subsubsection*{Path 1: \texorpdfstring{$R = [\gamma^{\mathrm{BW}}_{P^{-1} Q^{-1}}(t)]^{-1}$}{R = [gamma(BW)]^{-1}}}

Each point on this path is the inverse of a point on the BW geodesic between $P^{-1}$ and $Q^{-1}$. 
\begin{equation}
R =  [\gamma^{\mathrm{BW}}_{P^{-1} Q^{-1}}(t)]^{-1} = \left[(1-t)\mathbb{I} + t S \right]^{-1} P \left[(1-t)\mathbb{I} + t S \right]^{-1}, \quad \text{for any} \; t \in [0,1],
\end{equation}
where $S = Q ^{-1} \# P$. As elaborated in the following theorem, the generalized fidelity $\operatorname{F}_R(P,Q) $ is constant and equal to Uhlmann fidelity along this curve.
\begin{theorembox}
    \begin{theorem} \label{Thm:Path1}
        Let $P, Q \in \mathbb P_d$ be fixed. Let the base $R$ be any point along the curve $[\gamma^{\mathrm{BW}}_{P^{-1} Q^{-1}}(t)]^{-1}$. Then
        \begin{equation}
            \operatorname{F}_R(P,Q) = \operatorname{F}^{\emph U}(P,Q). 
        \end{equation}
    \end{theorem}
\end{theorembox}

\begin{proof}
    See Theorem~\ref{AppThm:Path1}. 
\end{proof}

\subsubsection*{Paths 4 and 7: $R_1 = \gamma^{\mathrm{BW}}_{PQ^{-1}}(t)$ and $R_2 = \gamma^{\mathrm{BW}}_{QP^{-1}}(t)$}

These are the BW geodesic paths from $P$ to $Q^{-1}$ and from $Q$ to $P^{-1}$:
\begin{equation}
R_1 = \gamma^{\mathrm{BW}}_{PQ^{-1}}(t) = \left[(1-t)\mathbb{I} + t P ^{-1} \# Q ^{-1}  \right] P \left[(1-t)\mathbb{I} + t P ^{-1} \# Q ^{-1}  \right],
\end{equation}
and
\begin{equation}
R_2 = \gamma^{\mathrm{BW}}_{QP^{-1}}(t) = \left[(1-t)\mathbb{I} + t P ^{-1} \# Q ^{-1}  \right] Q \left[(1-t)\mathbb{I} + t P ^{-1} \# Q ^{-1}  \right],
\end{equation}
The generalized fidelity along these paths is complex-valued. We currently do not know any special closed-form for $\operatorname{F}_R(P,Q) $ when the base moves along either of the paths. However, we can show that the generalized fidelity varies in a \textit{covariant} manner across the two paths. 
\begin{theorembox}
    \begin{theorem}\label{Thm:Paths4and7}
        Let $P,Q \in \mathbb{P}_d$ be fixed. For any fixed $t \in [0,1]$, let 
        \begin{equation}
            R_1 := \gamma_{P Q^{-1}}^{\mathrm{BW}}(t) \quad \text{and} \quad R_2 := \gamma_{Q P^{-1}}^{\mathrm{BW}}(t).
        \end{equation}
        Then $\operatorname{F}_{R_1}(P,Q) = \operatorname{F}_{R_2}(P,Q).$ 
    \end{theorem}
\end{theorembox}

\begin{proof}
    See Theorem~\ref{AppThm:Paths4and7} for proof. 
\end{proof}

\subsubsection*{Paths 3 and 8: $R_1 = [\gamma^\text{BW}_{P^{-1} Q}(t)]^{-1}$ and $R_2 = [\gamma^\text{BW}_{Q^{-1} P}(t)]^{-1}$}

These paths correspond to the inverse of the BW geodesics from $P^{-1}$ to $Q$ and $Q^{-1}$ to $P$:
\begin{equation}
R_1 = [\gamma^\text{BW}_{P^{-1} Q}(t)]^{-1} \quad \text{and} \quad R_2 = [\gamma^\text{BW}_{Q^{-1} P}(t)]^{-1}.
\end{equation}
Similar to the previous result, generalized fidelity covaries along these paths. This is formalized in the following theorem.
\begin{theorembox}
    \begin{theorem}\label{Thm:Paths3and8}
        Let $P,Q \in \mathbb P_d$ be fixed. For any fixed $t \in [0,1]$, let 
        \begin{equation}
            R_1 := [\gamma_{P ^{-1} Q}^{\mathrm{BW}} (t)] ^{-1}  \quad \text{and} \quad R_2 := [\gamma_{Q^{-1} P }^{\mathrm{BW}} (t)] ^{-1} .
        \end{equation}
        Then $\operatorname{F}_{R_1}(P,Q) = \operatorname{F}_{R_2}(P,Q). $ 
    \end{theorem}
\end{theorembox}

\begin{proof}
    See Theorem~\ref{AppThm:Paths3and8} for proof. 
\end{proof}

\subsection{Affine-invariant (yellow) paths}
We now discuss the geometric properties of generalized fidelity as the base moves along geodesic paths related to the Affine-invariant metric. Recall that the AI geodesic path between $A,B \in \mathbb P_d$ is defined as 
\begin{equation}
    \gamma_{AB}^\text{AI}(t) = A ^{\frac12} \left(\frac{B}{A} \right)^t A ^{\frac12}.     
\end{equation}
We now discuss the geodesic properties of generalized fidelity $\operatorname{F}_R(P,Q) $ as the base $R$ moves along three geodesic paths related to this metric.

\subsubsection*{Path 9: $R = \gamma^\mathrm{AI}_{P^{-1} Q^{-1}}(t) = \left[\gamma^\mathrm{AI}_{P Q}(t)\right]^{-1}$.}
Here the base $R$ moves along the AI-geodesic path between $P^{-1}$ and $Q^{-1}$. Note that for any $t \in [0,1]$,
$ \gamma^\mathrm{AI}_{P ^{-1} Q ^{-1} } (t) = [\gamma^\mathrm{AI}_{P Q} (t)]^{-1}.  $

A point $R$ on this curve has the form
\begin{equation}
R = \gamma^\mathrm{AI}_{P^{-1} Q^{-1}}(t) := P^{-\frac 12} \left(P^{\frac 12} Q ^{-1}  P^{\frac 12}\right)^t P^{-\frac 12}.
\end{equation}

For $t = \frac12$, we have $R = P ^{-1} \# Q ^{-1}  = (P \# Q) ^{-1} $. We first provide a short proof for the claim that when $R$ is the midpoint ($t=\frac12$) of the geodesic, the generalized fidelity equals the Matsumoto fidelity $\operatorname{F}^\mathrm M(P,Q) $. Subsequently, we also show that for any point on the AI geodesic between $P ^{-1} $ and $Q ^{-1} $, the generalized fidelity is constantly equal to the Matsumoto fidelity.

\begin{theorembox}
\begin{theorem} \label{Thm:Path9GM}
Let $P,Q \in \mathbb{P}_d$ and choose $R = P^{-1} \# Q^{-1}$. Then $\operatorname{F}_R(P,Q) = \operatorname{F}^\mathrm{M}(P,Q).$
\end{theorem}
\label{Thm:MatsFidelityatGeoMean}
\end{theorembox}
\begin{proof}
    See Theorem~\ref{AppThm:Path9GM} for proof.
\end{proof}

We now show that the generalized fidelity constantly equals the Matsumoto fidelity along any point in the path $\gamma_{ P ^{-1} Q ^{-1} }^{\mathrm{AI}}(t)$. 
\begin{theorembox} \label{Thm:Path9}
    \begin{theorem}
        Let $P, Q \in \mathbb P_d$ and let $R = \gamma_{P ^{-1} Q ^{-1} }^\mathrm{AI}(t)$ for any $t \in [0,1]$. Then 
        \begin{equation}
             \operatorname{F}_R(P,Q) = \operatorname{F}^\mathrm M(P,Q).
        \end{equation}
    \end{theorem}
\end{theorembox}
\begin{proof}
    See Theorem~\ref{AppThm:Path9} for proof.
\end{proof}

\subsubsection*{Paths 5 and 6: $R_1 \gamma^\mathrm{AI}_{P P^{-1}}(t)$ and $R_2 = \gamma^\mathrm{AI}_{Q Q^{-1}}(t)$. } \label{SubSec:PolarPaths}

These are the geodesic paths from $P$ to $P^{-1}$ and $Q$ to $Q^{-1}$ with respect to the AI metric.
\begin{equation}
\begin{aligned}
R_1 &= \gamma^\mathrm{AI}_{P P^{-1}}(t) = P^{\frac 12} \left(P^{-\frac 12} P^{-1} P^{-\frac 12}\right)^t P^{\frac 12}  = P^{1-2t} \equiv P^x, \\
R_2 &= \gamma^\mathrm{AI}_{Q Q^{-1}}(t) = Q^{\frac 12} \left(Q^{-\frac 12} Q^{-1} Q^{-\frac 12}\right)^t Q^{\frac 12}  = Q^{1-2t} \equiv Q^x,
\end{aligned}
\end{equation}
where we have denoted $x \equiv  1 - 2t$ for convenience. As $R_1$ and $R_2$ move along these geodesics, the generalized fidelity takes the form
\begin{equation}
\begin{aligned}
        \operatorname{F}_{R_1}(P,Q) &= \operatorname{Tr}\left[\sqrt{ P^{\frac x2} P P^{\frac x2}} P^{-x}\sqrt{ P^{\frac x2} Q P^{\frac x2}}  \right] \\ &=   \operatorname{Tr} \left[P^{\frac{1-x}2} \sqrt{P^{\frac x2}Q P^{\frac x2}} \right] = \operatorname{Tr}\left[P^\frac{1-x}2 \# P^\frac{1+x}4 Q  P^\frac{1+x}4 \right]. 
\end{aligned}
\end{equation}
A similar calculation shows
\begin{equation}
    \operatorname{F}_{R_2}(P,Q) = \operatorname{Tr}\left[Q^{\frac{1-x}2}\sqrt{Q^\frac{x}{2} P Q^\frac{x}{2}}\right] = \operatorname{Tr}\left[Q^\frac{1-x}2 \# Q^\frac{1+x}4 P Q^\frac{1+x}4\right]. 
\end{equation}

These two paths are special because we recover all the three \textit{named fidelities} along these paths:
\begin{equation}
    \begin{aligned}
        t = 0 &\iff x = 1  \quad &\implies \quad \operatorname{F}_R(P,Q) &= \operatorname{F}^\mathrm U(P,Q), \\
        t = {\textstyle\frac{1}{2}} &\iff x = 0 \quad &\implies \quad \operatorname{F}_R(P,Q) &= \operatorname{F}^\mathrm H(P,Q), \\
        t = 1 &\iff x = -1 \quad &\implies \quad \operatorname{F}_R(P,Q) &= \operatorname{F}^\mathrm M(P,Q).
    \end{aligned}
\end{equation}
Moreover, these are paths along which the generalized fidelity varies while being real-valued throughout. Thus, we have a \textit{one parameter} family of fidelities that recover all the three named fidelities. The one-parameter nature becomes even more apparent with the alternative form of generalized fidelity along these paths, as described in the following theorem. 
\begin{theorembox}
    \begin{theorem}
        Let $P, Q \in \mathbb P_d$ and let $x \in \mathbb R$. Then
        \begin{equation}
            \begin{aligned}
                \Fg{P}{Q}{P^x} = \operatorname{Tr}\left[P^{\frac12}  U_x Q ^{\frac12} \right] \quad \text{and} \quad 
                \Fg{P}{Q}{Q^x} = \operatorname{Tr}\left[P^{\frac12}  V_x Q ^{\frac12} \right], \\
            \end{aligned}
        \end{equation}
        where 
            \begin{equation}
                \begin{aligned}
                    U_x := \operatorname{Pol}\left(P^\frac x2 Q^{\frac12} \right) \quad \text{and} \quad 
                    V_x := \operatorname{Pol}\left(P^{\frac12} Q^\frac x2  \right).
                \end{aligned}
            \end{equation}        
    \end{theorem}
\end{theorembox}
\begin{proof}
    Let $U_x := \operatorname{Pol}\left(P^\frac x2 Q ^{\frac12} \right)$, which implies $U_x^* = \operatorname{Pol}\left(Q ^{\frac12} P^\frac x2  \right)$. Thus we have
    \begin{equation}
        Q ^{\frac12} P^\frac{x}{2} = U_x^* \sqrt{P ^\frac x2 Q  P ^\frac x2} \iff U  _x  = \sqrt{P ^\frac x2 Q  P ^\frac x2  } P^{-\frac{x}{2}} Q ^{-\frac12}. 
    \end{equation}
   It then follows that
   \begin{equation}
    \begin{aligned}
       \operatorname{Tr}\left[P^{\frac12}  U_x Q ^{\frac12} \right] &= \operatorname{Tr}\left[P^{\frac12}  \sqrt{P ^\frac x2 Q  P ^\frac x2  } P^{-\frac{x}{2}} Q ^{-\frac12} Q ^{\frac12} \right] \\ &= \operatorname{Tr}\left[P^{\frac{1-x}2} \sqrt{P ^\frac x2 Q  P ^\frac x2} \right] = \Fg{P}{Q}{P^x},
    \end{aligned}
   \end{equation}
   which proves the first claim. Now let $V_x := \operatorname{Pol}\left(P^{\frac12}  Q^\frac{x}{2}\right)$, which implies $V_x^* = \operatorname{Pol}\left( Q^\frac x2 P^{\frac12} \right)$. Using a similar calculation as above, we have
    \begin{equation}
         V_x =  P ^{-\frac12} Q^{-\frac{x}2} \sqrt{Q^\frac x2 P Q^\frac x2} , 
    \end{equation}
    which implies
    \begin{equation}
    \begin{aligned}
        \operatorname{Tr}\left[P^{\frac12}  V_x Q ^{\frac12} \right] &= \operatorname{Tr}\left[P^{\frac12}  P ^{-\frac12} Q^{-\frac{x}2} \sqrt{Q^\frac x2 P Q^\frac x2} Q ^{\frac12}  \right] \\ 
        &= \operatorname{Tr}\left[ \sqrt{Q^\frac x2 P Q^\frac x2} Q ^{\frac{1-x}2}  \right] = \Fg{P}{Q}{Q^{x}},
    \end{aligned}
    \end{equation}
    as claimed. This completes the proof.
\end{proof}
Note that $\Fg{P}{Q}{P^x} \neq \Fg{P}{Q}{Q^x}$ for general values of $x$, except when $x \in {1, 0, -1}$, where they are equal. In terms of the geodesic paths $\gamma^{\text{AI}}_{P P ^{-1} }(t)$ and $\gamma^{\text{AI}}_{Q Q ^{-1} }(t)$, this corresponds to the values of $t \in \left\{0, 1/2, 1\right\}$ respectively. Moreover, we see that generalized fidelity along these paths is not symmetric under the swap of $P$ and $Q$, as the bases themselves depend on $P$ and $Q$. 
However, we can construct a symmetrized version by taking their average for the same value of $x$:
\begin{equation}
    \overline{\operatorname{F}}_x(P,Q) := \frac{\Fg{P}{Q}{P^x} + \Fg{P}{Q}{Q^x} }{2}.
\end{equation}
We call this parametrized family of fidelities the \textit{$x$-Polar fidelities} and remark that it is a one-parameter family of fidelities that recover Uhlmann-, Holevo-, and Matsumoto-fidelity at $x = 1,0,-1$ respectively. To our knowledge, this is the first such generalization. Further properties, including some remarkable connections to the spectral norm unit ball and the Lie group of $\mathrm{SU}(d)$, are discussed in Section~\ref{Sec:PolarFidelity}.

\begin{figure}[htbp]
    \centering
    \includegraphics[width=0.7\linewidth]{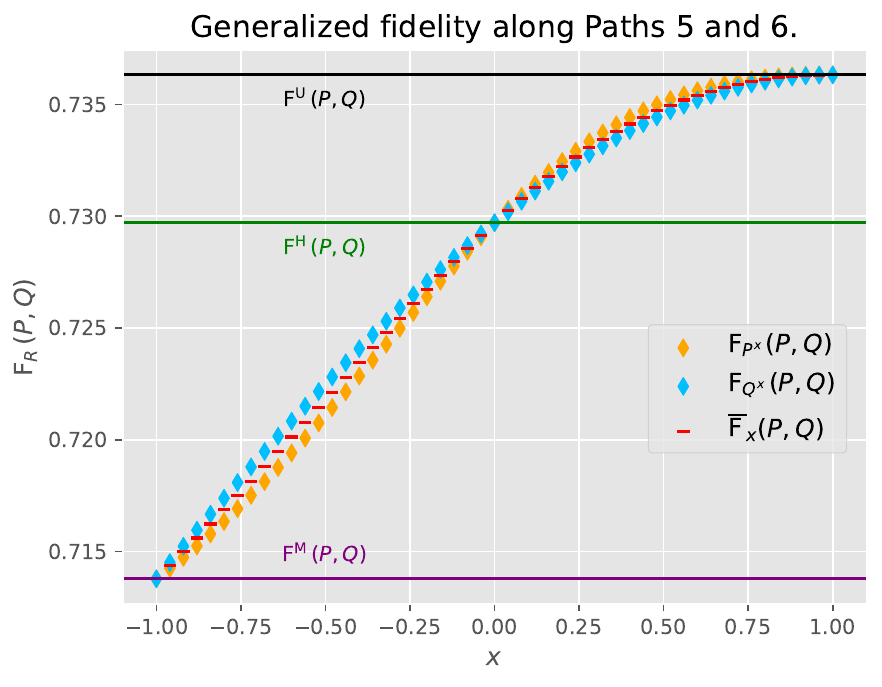}  
    \caption{\small Generalized fidelity along Path 5 $(\Fg{P}{Q}{P^x})$ and Path 6 $(\Fg{P}{Q}{Q^x})$ for two randomly generated density matrices $P$ and $Q$. The plot illustrates a (numerically observed) monotonicity property in the range $x \in [-1,1]$. We also plot the $x$-Polar fidelity $\overline{\operatorname{F}}_x(P,Q)$, which is the average of the generalized fidelities at the same value of $x$. All three quantities recover $\operatorname{F}^\mathrm M(P,Q) , \operatorname{F}^\mathrm H(P,Q) $ and $\operatorname{F}^\mathrm U(P,Q) $ at $x = -1, 0, 1$ respectively. Code to generate plot available at GitHub~\cite{afhamgithub}.}
    \label{fig:example_figure}
\end{figure}

 An interesting numerical observation regarding generalized fidelity along these paths is as follows. For $x, y \in [0,1]$ we observe that
\begin{equation}
    x \geq y \implies 
    \begin{cases}
        \operatorname{F}_{P^x}(P,Q) \geq \operatorname{F}_{P^{y}}(P,Q) \\ 
        \operatorname{F}_{Q^x}(P,Q) \geq \operatorname{F}_{Q^{y}}(P,Q).
    \end{cases}
\end{equation}

This observation is consistent with the known relations~\cite{matsumoto2010reverse, cree2020fidelity} regarding the named fidelities:
\begin{equation}
    \operatorname{F}^\mathrm M(P,Q) \leq \operatorname{F}^\mathrm H(P,Q) \leq \operatorname{F}^\mathrm U(P,Q), 
\end{equation}
for any $P,Q \geq 0$. 

If the numerical observation can be proven to be true, it would also imply that $x$-Polar fidelity is monotonic in $x \in [-1,1]$, which would give us a monotonic family of fidelities that recover the three named fidelities.

\subsection{Euclidean (blue) paths}
Finally, we discuss paths related to the Euclidean metric, where geodesics are simply convex combinations of the endpoints:
\begin{equation}
    \gamma_{AB}^{\text{Euc}}(t) = (1-t) A + t B.
\end{equation}
These paths are colored blue in Figure~\ref{Fig:Geodesic}. We now discuss generalized fidelity along 2 paths related to this metric.

\subsubsection*{Path 10: $R = \gamma^\text{Euc}_{P^{-1} Q^{-1}}(t)$.}
We next study the Euclidean geodesic between $P ^{-1} $ and $Q ^{-1} $, which is simply the convex combination of the end-points: $\gamma^\mathrm{Euc}_{P ^{-1} Q ^{-1} } (t)  = (1-t) P ^{-1} + t Q ^{-1} $. As the base $R$ moves along this path, the generalized fidelity is constant and equal to the Matsumoto Fidelity.
\begin{theorembox}
    \begin{theorem}\label{Thm:Path10}
        Let $P, Q \in \mathbb P_d$ and let $R = \gamma_{P ^{-1} Q ^{-1} }^\mathrm{Euc}(t)$ for any $t \in [0,1]$. Then 
        \begin{equation}
             \operatorname{F}_R(P,Q) = \operatorname{F}^\mathrm M(P,Q). 
        \end{equation}
    \end{theorem}
\end{theorembox}
\begin{proof}
    See Appendix~\ref{App:ThmPath10} for proof.
\end{proof}

\subsubsection{Path 11: $R = [\gamma^\text{Euc}_{PQ}(t)]^{-1}$}
Conversely, we now consider the case where the base moves along the path defined by the inverse of the Euclidean geodesic (straight line) between $P$ and $Q$:
\begin{equation}
R = \left[\gamma^\text{Euc}_{P Q}(t)\right] ^{-1}  = \left[(1-t)P + tQ \right]^{-1}.
\end{equation}
Like in the previous case, the generalized fidelity is constant along this path and equal to the Matsumoto fidelity. 
\begin{theorembox}
    \begin{theorem}\label{Thm:Path11}
       Let $P, Q \in \mathbb P_d$ and let $R = [\gamma_{PQ}^\mathrm{Euc}(t)]^{-1} $ for any $t \in [0,1]$. Then 
        \begin{equation}
             \operatorname{F}_R(P,Q) = \operatorname{F}^\mathrm M(P,Q). 
        \end{equation}    
    \end{theorem}
\end{theorembox}
\begin{proof}
    See Appendix~\ref{App:ThmPath11} for proof.
\end{proof}

\section{Polar fidelities and interior fidelities } \label{Sec:PolarFidelity}

In this section, we introduce a new family of fidelities and show how generalized fidelities can be thought of as elements of the extreme points of this family of fidelities. We begin with recalling one of the equivalent characterizations of generalized fidelity. For any $P,Q,R \in \mathbb P_d$, we have
\begin{equation}
    \operatorname{F}_R(P,Q) = \operatorname{Tr}\left[Q ^{\frac12} U_Q U_P^* P ^{\frac12} \right],
\end{equation}
where $U_P := \operatorname{Pol}\left(P ^{\frac12} R ^{\frac12} \right)$ and $U_Q := \operatorname{Pol}\left(Q ^{\frac12} R ^{\frac12} \right)$. We call the unitary $U_Q U_P^*$ the \textit{unitary factor} of the generalized fidelity $\operatorname{F}_R(P,Q)$. 

We first show that the unitary factor of any generalized fidelity has $+1$ determinant. 
\begin{theorembox}
\begin{prop}
        Let $P,Q,R \in \mathbb P_d$. Then the unitary factor $U_Q U_P^*$ of $\operatorname{F}_R(P,Q) $ has determinant +1. Or equivalently, $U_Q U_P^* \in \mathrm{SU}(d)$.
\end{prop}
\end{theorembox}
\begin{proof}
    The proof will be done in two steps. We first show that the polar factor of any matrix that is the product of two positive definite matrices is a special unitary matrix. To this end, let $A,B \in \mathbb P_d$. We aim to show that $ U:= \operatorname{Pol}\left(AB\right) \in \mathrm{SU}(d)$. 
    By definition, we have
    \begin{equation}
        A B = U |A B| = U \sqrt{B A^2 B}.
    \end{equation}
    Now take determinant across to get
    \begin{equation}
        \det(A B) = \det(U |A B|) = \det(U) \det(|A B| ).
    \end{equation}
    The LHS equals $\det(A) \det(B)$. By properties of determinant, we have
    \begin{equation}
      \det(|AB|) =  \det(\sqrt{A B^2 A}) = \sqrt{\det(A B^2 A)} = \sqrt{\det(A)^2 \det(B)^2} = \det(A) \det(B).
    \end{equation}
    Thus we have $\det(U) = +1$, or $\operatorname{Pol}\left(A B\right) \in \mathrm{SU}(d)$ for any $A,B \in \mathbb P_d$. Thus we have $U_P, U_Q \in \mathrm{SU}(d)$, and by the group structure of special unitaries, we have $U_P U_Q^* \in \mathrm{SU}(d).$    
\end{proof}

Thus, the unitary factor of \textit{any} generalized fidelity is a special unitary.  Let us now recall the definition of $x$-Polar fidelity. For $P,Q \in \mathbb P_d$, we defined
\begin{equation}
    \overline{\operatorname{F}}_x(P,Q) := \frac{\Fg{P}{Q}{P^x} +  \Fg{P}{Q}{Q^x}}{2} = \operatorname{Tr}\left[P ^{\frac12} \left( \frac{U_x + V_x}2 \right) Q ^{\frac12} \right],
\end{equation}
where $U_x := \operatorname{Pol}\left(P^\frac x2 Q ^{\frac12} \right)$ and $V_x := \operatorname{Pol}\left(P^\frac 12 Q ^{\frac x2} \right)$. 
Observe that Polar fidelities are \textit{not} generalized fidelities (except in the cases of $x \in \{1,0,-1\}$). In particular, convex combinations, over bases, of generalized fidelities are not, in general,  generalized fidelities. We call such convex combinations of generalized fidelities over a distribution of bases as \textit{Interior fidelities}. We first formally define it and then show why the name is appropriate.

\begin{definitionbox}
\begin{definition}
       Let $\mu \in \mathcal P_n$ be a probability vector, $\mathcal R = \{R_1, \dots, R_n\} \subset \mathbb P_d$ be a \emph{collection of bases}, and let $P,Q \in \mathbb P_d$. Define the \emph{interior fidelity} between $P$ \textit{and} $Q$ \textit{over} $(\mathcal R, \mu)$ as
   \begin{equation}
    \Fg{P}{Q}{(\mathcal R, \mu)} :=  \sum_{i=1}^{n} \mu_i \Fg{P}{Q}{R_i} = \sum_{i=1}^{n} \mu_i \operatorname{Tr}\left[Q ^{\frac12} V_i P ^{\frac12} \right] = \operatorname{Tr}\left[Q^{\frac12} \overline{V} P^{\frac12}\right], 
\end{equation}
where $V_i$ is the unitary factor of $\operatorname{F}_{R_i}(P,Q)$ and $\overline{V} := \sum_{i=1}^{n} \mu_i V_i$ is the mean of the unitary factors.
\end{definition}   
\end{definitionbox}

One can also construct the corresponding \textit{interior Bures distance}, and geometrically this would correspond to linearizing the manifold at different points and then taking the average distance over these different linearizations. We also note that since any generalized fidelity reduces to classical fidelity in the commuting case, so do interior fidelities. Thus, interior fidelities are valid quantizations of classical fidelity.

We now note a different kind of geometric property of generalized fidelities and interior fidelities. Specifically, generalized fidelities can be seen to be the \textit{extreme points} of a convex family of fidelities, whose interior points are constituted by interior fidelities.

This correspondence is easier to see once we recall that the set of all unitaries from extreme points of the convex and compact \textit{spectral-norm unit ball}~\cite[Theorem 1.10]{watrous2018theory}:
\begin{equation}
    \mathcal B_\infty := \{K : \|K\|_\infty \leq 1\}.
\end{equation}
It then follows from the definition of interior fidelities that the `non-commutative part' is an interior point of $\mathcal B_\infty$. However, we note that the correspondence is not bijective as $\mathcal B_\infty$ also contains unitaries that are not special unitaries, whereas the unitary factor of any generalized fidelity is necessarily a special unitary. 

Thus, we see that the $x$-Polar fidelity is actually an interior fidelity. We showed that the unitary factor of any generalized fidelity is a special unitary. An open question we pose is whether the converse is true. That is, for any $U \in \mathrm{SU}(d)$, does there exist a triple $P,Q,R \in \mathbb P_d$ such that $U = U_Q U_P^*$? In \cite[Remark 5.2]{lawson2008gamma}, it was shown that $\mathrm{SU}(d)$ can be generated by the set of polar factors of the product of two positive definite matrices. However, for our open question to be answered affirmatively, we would require that any special unitary can be written by the product of at most two polar factors of the above kind. It is unclear whether this is true. If the question can be answered affirmatively, then it would shed light on a new collection between polar decomposition, the Bures-Wasserstein manifold of positive definite matrices, and the Lie group $\mathrm{SU}(d)$.

\section{Block matrix characterization of generalized fidelity } \label{Sec:BlockMatrix}

We now show that the generalized fidelity, and thereby the generalized Bures distance, has a Block-matrix representation. This representation is intimately related to the semidefinite program (SDP) for Uhlmann fidelity~\cite{watrous2018theory} and its generalization to SDP for optimal average fidelity~\cite{afham2022}. We will first discuss the SDP for average fidelity (which, in this context, is actually an SDP for \textit{total fidelity}) and then show how one can recover generalized fidelity and generalized Bures distance from it.

\begin{theorembox}
\begin{prop}  \label{Prop:TotalFidelitySDP}
    Let $P,Q,R \in \mathbb P_d$ be arbitrarily chosen. Define the matrices $A, B$ as
    \begin{equation} \label{Eqn:AandB}
A := \frac{1}{2} \begin{pmatrix}
0 & 0 & \mathbb I \\
0 & 0 & \mathbb I \\
\mathbb I & \mathbb I & 0
\end{pmatrix} \in \mathbb H_{3d}, \quad B := \begin{pmatrix}
P & 0 & 0 \\
0 & Q & 0 \\
0 & 0 & R
\end{pmatrix} \in \mathbb P_{3d},
\end{equation}
and $\Phi : \mathbb H_{3d} \to  \mathbb H_{3d}$ to be the Hermitian preserving map which acts as
\begin{equation}
    \begin{pmatrix}
    M_{11} & \cdot & \cdot \\
    \cdot & M_{22} & \cdot \\
    \cdot & \cdot & M_{33}
    \end{pmatrix} \xmapsto{\Phi} \begin{pmatrix}
    M_{11} & 0 & 0 \\
    0 & M_{22} & 0 \\
    0 & 0 & M_{33}
    \end{pmatrix}.
\end{equation}
Consider the SDP $(\Phi, A, B)$ whose primal problem is
\begin{equation}
    \begin{aligned}
        \rm{maximize}: \quad &\langle A,  \mathrm{X} \rangle, \\
        \rm{subject} \,\, \rm{to} : \quad & \mathrm{X} \geq 0, \quad \Phi( \mathrm{X} ) = B.                
    \end{aligned}
\end{equation}

Then the optimal value of this SDP is $ \Fu{P}{R} + \Fu{Q}{R}$, and the unique optimal feasible point that attains this value is of the form
\begin{equation}
    \mathrm{X}_\star := 
    \begin{pmatrix}
    P & P^{\frac12} U_P U_Q^* Q^{\frac12} & P^{\frac12} U_P R^{\frac12} \\
    Q^{\frac12} U_Q U_P^* P^{\frac12} & Q & Q^{\frac12} U_Q R^{\frac12} \\
    R^{\frac12} U_P^* P^{\frac12} & R^{\frac12} U_Q^* Q^{\frac12} & R 
    \end{pmatrix}
\end{equation}
\textit{where $U_P := \operatorname{Pol}\left(R ^{\frac12} P ^{\frac12} \right)$ and $U_Q := \operatorname{Pol}\left(R ^{\frac12} Q ^{\frac12} \right)$}.

\end{prop}
\end{theorembox}
\begin{proof}
    The proof is a straightforward application of results in \cite{watrous2018theory} and \cite{afham2022}. The fact that the value of the objective function cannot exceed $\Fu{P}{R} + \Fu{Q}{R}$ is shown in \cite[Lemma 4 (Equation 37)]{afham2022}. That is, for any feasible point $\mathrm{X}$,
    \begin{equation}
        \langle A,\mathrm{X} \rangle \leq \Fu{P}{R} + \Fu{Q}{R}. 
    \end{equation}
     To show that the inequality can be saturated, we construct a feasible point that attains the optimal value $\Fu{P}{R} + \Fu{Q}{R}$. Later, we will show that it is also the unique optimal feasible point. For $U_P := \operatorname{Pol}\left(P ^{\frac12} R ^{\frac12}  \right)$ and $U_Q := \operatorname{Pol}\left(Q ^{\frac12} R ^{\frac12} \right)$, define the $3d \times d$ block matrix $T$ as follows.
\begin{equation}
        T := 
    \begin{pmatrix}
    P ^{\frac12} U_P \\
    Q ^{\frac12} U_Q \\
    R ^{\frac12} 
    \end{pmatrix}
    \end{equation}
Now consider $\mathrm{X}_\star:= TT^*$:
\begin{equation}
\begin{aligned}
        \mathrm{X}_\star = T T^* &= 
    \begin{pmatrix}
    P ^{\frac12} U_P \\
    Q ^{\frac12} U_Q \\
    R ^{\frac12} 
    \end{pmatrix} 
    \begin{pmatrix}
     U_P^* P ^{\frac12} &
    U_Q^* Q ^{\frac12}  &
    R ^{\frac12} 
    \end{pmatrix} \\ &= 
    \begin{pmatrix}
    P & P^{\frac12} U_P U_Q^* Q^{\frac12} & P^{\frac12} U_P R^{\frac12} \\
    Q^{\frac12} U_Q U_P^* P^{\frac12} & Q & Q^{\frac12} U_Q R^{\frac12} \\
    R^{\frac12} U_P^* P^{\frac12} & R^{\frac12} U_Q^* Q^{\frac12} & R 
    \end{pmatrix} \\ &\equiv    
    \begin{pmatrix}
        P & Z & Y_P \\
        Z^* & Q & Y_Q \\
        Y_P^* & Y_Q^* & R 
    \end{pmatrix} \geq 0.
\end{aligned}      
\end{equation}
In the last equality we renamed some quantities for brevity. We then see that
\begin{equation}
    \langle A, \mathrm{X}_\star \rangle = \Re\operatorname{Tr}[Y_P] + \Re \operatorname{Tr}[Y_Q] = \Fu{P}{R} + \Fu{Q}{R}, 
\end{equation}
which implies that the SDP achieves optimality.  

We now prove the uniqueness of the optimal feasible point. We first note that the positivity of $\mathrm{X}_\star$ necessarily implies the positivity of the principle sub-block matrices
\begin{equation}
    M_P := \begin{pmatrix}
        P & Y_P \\
        Y_P^* & R 
    \end{pmatrix} \geq 0
\quad \text{and} \quad 
    M_Q := \begin{pmatrix}
        Q & Y_Q \\
        Y_Q^* & R 
    \end{pmatrix} \geq 0.
\end{equation}
By \cite[Lemma 3.18]{watrous2018theory}, we have $M_P \geq 0$ if and only if $Y_P = P ^{\frac12} K R ^{\frac12}$ for some contraction $K : \|K\|_\infty \leq 1$. Note that $\Re \operatorname{Tr}[Y_P]$ attains the maximum necessarily for some unitary $K$, which is an extreme point of the set of contractions. To see this, observe that
\begin{equation}
    \Re \operatorname{Tr}[Y_P] = \Re \operatorname{Tr}\left[P ^{\frac12} K R ^{\frac12}\right] = \Re \operatorname{Tr}\left[ K R ^{\frac12}P ^{\frac12}\right] \leq \left\| K R ^{\frac12}P ^{\frac12} \right\|_1,
\end{equation}
where the last equality comes from the variational characterization of trace norm. Now note that for any contraction $K$ we have $K^* K \leq \mathbb I$ and thus
\begin{equation}
    P ^{\frac12} R ^{\frac12} K^* K R ^{\frac12} P ^{\frac12}  \leq P ^{\frac12} R P ^{\frac12}, 
\end{equation}
with equality if and only if $K$ is a unitary. Since the operator square root function over positive semidefinite matrices is monotonic, we have
\begin{equation}
   \sqrt{P ^{\frac12} R ^{\frac12} K^* K R ^{\frac12} P ^{\frac12}}  \leq \sqrt{P ^{\frac12} R P ^{\frac12}},
\end{equation}
and taking trace across, we have
 \begin{equation}
     \left\| K R ^{\frac12} P^{\frac12} \right\|_1=  \operatorname{Tr}\left[\sqrt{P ^{\frac12} R ^{\frac12} K^* K R ^{\frac12} P ^{\frac12}}\right] \leq \operatorname{Tr}\left[\sqrt{P ^{\frac12} R P ^{\frac12}} \right] = \operatorname{F}^\mathrm{U}(P,R). 
 \end{equation}
Thus we see that the inequality is saturated at some unitary. Due to the linearity of the objective function $\operatorname{Tr}[Y_P]$, this also necessarily means that the optimum is attained at a unique unitary. This is because if there are two distinct optimal unitaries, by linearity of the objective function, any convex combination of these unitaries would also attain the optimum. However, such a (non-trivial) convex combination will not be a unitary, which would contradict the previous claim. Thus the optimal unitary is unique. 

Indeed, $U_P$ is the unique optimal feasible point that achieves this value for $M_P$, and similar reasoning shows that $U_Q$ is the unique optimal unitary for $M_Q$. Thus, $\mathrm{X}_\star$ is the unique optimal feasible point of the SDP $(\Phi,A,B)$. This concludes the proof.
\end{proof}

Given the above SDP for a triple $P,Q,R \in \mathbb P_d$ one can \textit{extract} the generalized fidelity $\operatorname{F}_R(P,Q) $ and the squared generalized Bures distance $\operatorname{B}_R(P,Q)$ from the \textit{optimal primal feasible} of SDP. This is formalized in the following theorem.

\begin{theorembox}
\begin{theorem}
    Define the SDP $(\Phi, A, B)$ as above for an arbitrary triple $P,Q,R \in \mathbb P_d$. Let $\mathrm{X}_\star$ be the optimal primal feasible:
    \begin{equation}
        \langle \mathrm{X_\star}, A \rangle = \Fu{P}{R} + \Fu{Q}{R}.
    \end{equation}
\end{theorem}
Then,
\begin{enumerate}
    \item $\operatorname{F}_R(P,Q) = \langle K, \mathrm{X}_\star \rangle, $
    \item $\Re \operatorname{F}_R(P,Q) = \left\langle \frac{K + K^*}2, \mathrm{X}_\star \right\rangle $
    \item $\operatorname{B}_R(P,Q) = \langle J, \mathrm{X}_\star \rangle, $
\end{enumerate}
where
\begin{equation} \label{Eqn:KandJ}
K :=  \begin{pmatrix}
0 & \mathbb I & 0 \\
0 & 0 & 0 \\
0 & 0 & 0
\end{pmatrix}, \quad J := \begin{pmatrix}
\mathbb I & -\mathbb I & 0 \\
-\mathbb I & \mathbb I & 0 \\
0 & 0 & 0
\end{pmatrix},
\end{equation}

\end{theorembox}
\begin{proof}
    From Proposition~\ref{Prop:TotalFidelitySDP}, we have that the optimal feasible points $\mathrm{X}_\star$ is of the form
   \begin{equation}
        \mathrm{X}_\star = \begin{pmatrix}
        P & Z & Y_P \\
        Z^* & Q & Y_Q \\
        Y_P^* & Y_Q^* & R 
    \end{pmatrix} = 
    \begin{pmatrix}
    P & P^{\frac12} U_P U_Q^* Q^{\frac12} & P^{\frac12} U_P R^{\frac12} \\
    Q^{\frac12} U_Q U_P^* P^{\frac12} & Q & Q^{\frac12} U_Q R^{\frac12} \\
    R^{\frac12} U_P^* P^{\frac12} & R^{\frac12} U_Q^* Q^{\frac12} & R 
    \end{pmatrix} \geq 0.
    \end{equation}

Taking trace-inner product of $\langle K, X_\star \rangle = \operatorname{Tr}[K^*\mathrm{X}_\star]$ then gives
\begin{equation}
    \operatorname{Tr}\left[Q^{\frac12} U_Q U_P^* P ^{\frac12}\right] = \operatorname{F}_R(P,Q).
\end{equation}
A similar calculation yields 
\begin{equation}
\Re \operatorname{F}_R(P,Q) = \left\langle \frac{K + K^*}2, \mathrm{X}_\star \right\rangle. 
\end{equation}
Finally, we have
\begin{equation}
    \langle J, \mathrm{X}_\star \rangle = \operatorname{Tr}[J \mathrm{X}_\star] = \operatorname{Tr}[P+Q] - 2 \Re \operatorname{F}_R(P,Q) =: \operatorname{B}_R(P,Q). 
\end{equation}
This concludes the proof.
\end{proof}

Thus, we have shown a block-matrix characterization of generalized fidelity and squared generalized Bures distance. The SDP used in the characterization is closely related to the SDP for \textit{optimal average fidelity}~\cite{afham2022}, which finds a problem that is equivalent to finding the Bures-Wasserstein barycenter of a finite collection of states. Further implications of this relation are explored in the next section.

\section{Generalized multivariate fidelities and relation to optimal average fidelity } \label{Sec:Multivariate}
Let us now consider a slightly different scenario. Let ${\rho_1, \dots, \rho_n}$ be a collection of quantum states. In \cite{Wilde2024multivariate}, the authors introduced and studied various notions of multivariate fidelities—fidelities between a collection of $n \geq 2$ states. One particular version of multivariate fidelity they define and study is the normalized sum of distinct pairwise fidelities:
\begin{equation} 
  \operatorname{F}(\rho_1, \ldots, \rho_n) := \frac1{n(n-1)} \sum_{i,j=1; i \neq j}^n \Fu{\rho_i}{\rho_j}.   
\end{equation}

For a tuple of states $(\rho_1, \ldots \rho_n)$ and a base $\sigma$, one can then define the analogous generalized multivariate fidelity at a state $\sigma \in \mathbb D_d$ as
\begin{equation}
    \operatorname{F}_\sigma(\rho_1, \ldots, \rho_n) := \frac1{n(n-1)} \sum_{i,j=1; i \neq j}^n \Fg{\rho_i}{\rho_j}{\sigma}.   
\end{equation}
Now consider the problem of finding the state that maximizes the average fidelity over a fixed ensemble, as studied in \cite{afham2022}. It is shown there that the \textit{fidelity maximzier}, defined as the state that maximizes the total fidelity $f(\rho):= \sum_{i=1}^{n} \Fu{\rho_i}{\rho}$, uniquely satisfies the fixed-point equation:
\begin{equation} \label{Eq:FPEqn}
    \sigma = \frac{1}{f(\sigma)} \sum_{i=1}^{n} \sqrt{\sigma ^{\frac12} \rho_i \sigma^{\frac12} }.
\end{equation}
We note that due to the relation between Uhlmann fidelity and BW barycenter, finding the average fidelity maximize is equivalent to finding the BW barycenter over the same distribution (up to a trace-normalization). One can relate the total fidelity of this maximizer and the generalized multivariate fidelity as shown in the following theorem.

\begin{theorembox}
    \begin{theorem}
        Let $\{\rho_1, \dots, \rho_n\}  \subseteq \mathbb D_d$ be a collection of states and $\sigma := \argmax_{\rho \in \mathbb D_d} f(\rho)$ be the barycenter of the collection. Then
        \begin{equation}
            f(\sigma)^2 = \sum_{i,j=1}^{n} \Fg{\rho_i}{\rho_j}{\sigma}. 
        \end{equation}
        Moreover, the generalized multivariate fidelity is related to the average fidelity of the maximizer as follows:
        \begin{equation}
            \operatorname{F}_\sigma (\rho_1, \dots, \rho_n) = \frac{f(\sigma)^2-n}{n^2 - n}.
        \end{equation}
    \end{theorem}
\end{theorembox}
\begin{proof}
Observe that
\begin{equation}
    \begin{aligned}
              \sigma = \sigma \sigma ^{-1} \sigma  &= \left(\frac{1}{f(\sigma)} \sum_{i=1}^{n} \sqrt{\sigma ^{\frac12} \rho_i \sigma^{\frac12} }\right) \sigma^{-1} \left(\frac{1}{f(\sigma)} \sum_{j=1}^{n} \sqrt{\sigma ^{\frac12} \rho_j \sigma^{\frac12} } \right) \\
        &= \frac{1}{f(\sigma)^2} \sum_{i,j=1}^n \sqrt{\sigma ^{\frac12} \rho_i \sigma ^{\frac12}}  \sigma ^{-1} \sqrt{\sigma ^{\frac12} \rho_j \sigma ^{\frac12}},
    \end{aligned}
\end{equation}
where the second equality is by \Cref{Eq:FPEqn}. Take trace across and rearrange to obtain
\begin{equation}
    \begin{aligned}
            f(\sigma)^2 &= \sum_{i,j=1}^{n} \operatorname{Tr}\left[ \sqrt{\sigma ^{\frac12} \rho_i \sigma ^{\frac12}}  \sigma ^{-1} \sqrt{\sigma ^{\frac12} \rho_j \sigma ^{\frac12}}\right] = \sum_{i,j=1}^{n} \Fg{\rho_i}{\rho_j}{\sigma}, 
    \end{aligned}
     \end{equation}
which proves the first claim. Note that $n = \sum_{i=1}^{n} \Fg{\rho_i}{\rho_i}{\sigma}$ and subtract this quantity from both sides (respectively) to obtain   
\begin{equation}
     f(\sigma)^2 - n = \sum_{i,j=1; i \neq j}^{n} \Fg{\rho_i}{\rho_j}{\sigma}. 
\end{equation}
Divide throughout by $n(n-1)$ to obtain
\begin{equation} 
    \frac{f(\sigma)^2 - n}{n^2 - n} = \frac1{n (n-1)} \sum_{i,j=1; i \neq j}^{n} \Fg{\rho_i}{\rho_j}{\sigma} =: \operatorname{F}_\sigma (\rho_1, \dots , \rho_n),
\end{equation}
which concludes the proof.
\end{proof}

\section{Uhlmann-(like) theorem for generalized fidelity } \label{Sec:UhlmannTheorem}

We first recall the definition of a purification of a state.

\begin{definitionbox}
    \begin{definition}
Let $P \in \mathbb P_d$. The \emph{purifications} of $P$ is the set of vectors
\begin{equation}
\emph{Pur}(P) := \left\{ |v\rangle \in \mathbb{C}^d_A \otimes \mathbb{C}^d_B : \operatorname{Tr}_B \left[ |v\rangle\langle v| \right] = P \right\}.
\end{equation}
This set is identical to
\begin{equation}
\emph{Pur}(P) = \left\{ \sum_{i=1}^d \lambda_i^{\frac 12} |p_i\rangle \otimes |u_i\rangle : \{ |u_i\rangle \}_{i=1}^d \text{ form an orthonormal set} \right\},
\end{equation}
where $P = \sum_{i=1}^d \lambda_i |p_i\rangle \langle p_i|$ is the eigendecomposition of $P$.
\end{definition}
\end{definitionbox}

Let $P \in \mathbb P_d$. The \textit{canonical purification} of $P$ is given by
\begin{equation}
|P\rangle = (P^{\frac12}   \otimes \mathbb I) |\Omega \rangle ,
\end{equation}
where $|\Omega\rangle := \sum_{i=1}^{d} |i, i\rangle  $ is the unnormalized canonical Bell state. Since there is a unitary degree of freedom in the second space, we may write an arbitrary purification as
\begin{equation}
|P_U\rangle \equiv \left(P^{\frac12}   \otimes U\right) |\Omega \rangle ,
\end{equation}
for an arbitrary $U \in \mathbb U_d$. An arbitrary purification of $Q \in \mathbb P_d$ can similarly be written as
\begin{equation}
|Q_V\rangle \equiv \left(Q^{\frac12}  \otimes V\right) |\Omega \rangle ,
\end{equation}
for an arbitrary $V \in \mathbb U_d$. Their overlap is given by
\begin{equation}
\langle Q_V , P_U \rangle = \left\langle \Omega , \left(Q^{\frac12}  P^{\frac12}   \otimes V^* U\right) \Omega \right\rangle = \operatorname{Tr}\left[Q^{\frac12}  P^{\frac12}   (V^*U)^\top\right] = \operatorname{Tr}\left[Q^{\frac12}  P^{\frac12}   U^\top \overline{V} \right],
\end{equation}
where $\overline{V}$ denotes the complex conjugate of $V$. Uhlmann's theorem states that the maximum (absolute-valued) overlap is achieved when $U^\top \overline{V} = \operatorname{Pol}\left(P^{\frac 12} Q^{\frac 12}\right)$. This result also allows us to state an \emph{Uhlmann's theorem} for generalized fidelity.

\begin{theorembox}
    \begin{theorem}
Let $P, Q, R \in \mathbb P_d$. Consider the purifications of $P$ and $Q$ defined as
\begin{equation}
|P_{U_P^\top}\rangle := \left(P^{\frac12}   \otimes U_P^\top\right) |\Omega \rangle \quad \text{and} \quad |Q_{U_Q^\top}\rangle := \left(Q^{\frac12}  \otimes U_Q^\top\right) |\Omega \rangle ,
\end{equation}
where $U_P := \operatorname{Pol}\left(P^{\frac 12} R^{\frac 12}\right)$ and $U_Q = \operatorname{Pol}\left(Q^{\frac 12} R^{\frac 12}\right)$. Then
\begin{equation}
\operatorname{F}_R(P,Q) = \left\langle P_{U_P^\top}, Q_{U_Q^\top}\right \rangle.
\end{equation}
\end{theorem}
\end{theorembox}

\begin{proof}
Let $U_P := \operatorname{Pol}\left(P^{\frac 12} R^{\frac 12}\right)$ and $U_Q := \operatorname{Pol}\left(Q^{\frac 12} R^{\frac 12}\right)$. Consider the purifications
\begin{equation}
|P_{U_P^\top}\rangle = \left(P^{\frac12}   \otimes U_P^\top\right) |\Omega \rangle \quad \text{and} \quad |Q_{U_Q^\top}\rangle = \left(Q^{\frac12}  \otimes U_Q^\top\right) |\Omega \rangle .
\end{equation}
Now consider the overlap
\begin{equation}
\begin{aligned}
    \langle P_{U_P^\top}, Q_{U_Q^\top} \rangle &= \langle \Omega , \left(P^{\frac12}   Q^{\frac12}  \otimes \overline{U}_P U_Q^\top \right) \Omega \rangle = \operatorname{Tr}\left[P^{\frac12}   Q^{\frac12}  \otimes  \left(\overline{U}_P U_Q^\top\right)^\top\right] 
    \\ &= \operatorname{Tr}\left[P^{\frac12}   Q^{\frac12}  U_Q U_P^*\right]  = \operatorname{Tr}\left[Q^{\frac12}  U_Q U_P^* P^{\frac12} \right] = \operatorname{F}_R(P,Q).
\end{aligned}
\end{equation}
This completes the proof.
\end{proof}

Thus, the generalized fidelity $\operatorname{F}_R(P,Q) $ can be seen as the overlap of a pair of specific purifications of $P$ and $Q$, with the choice of purification depending on $R$.

\section{An analogous generalization of some R\'enyi divergences } \label{Sec:Divergence}
We now briefly extend the formalism of generalized fidelities to quantum R\'enyi divergences. We refer to \cite{audenaert2013alpha, tomamichel2015quantum, muller2013quantum} for a detailed treatment. We restrict our treatment to normalized states (probability vectors and density matrices), keeping in mind that generalization to non-normalized states is straightforward and can be seen in the previously mentioned references. 

Quantum R\'enyi divergences are \textit{quantizations} of \textit{classical} R\'enyi divergences~\cite{renyi1961entropy, van2014renyi}, which, for probability vectors $p, q \in \mathcal P_n$, are defined as
\begin{equation}
    D_{\alpha} (p\|q) := \begin{cases}
        \frac1{\alpha-1} \log \sum_{i=1}^{d} p_i^\alpha q_i^{1-\alpha}  & \text{ if } p \ll q \text{ or } \alpha < 1,    \\
        \infty & \text{otherwise},
    \end{cases}
\end{equation}
for $\alpha \in (0, 1) \cup (1, \infty)$.  In subsequent discussions, we also restrict our attention to the case $\text{supp } P \subset \text{supp } Q$ (and thus conveniently ignore the cases where the divergence diverges to $\infty$).

As with classical fidelity, there is no unique way of generalizing this to positive semidefinite matrices. For an axiomatic approach see~\cite{audenaert2013alpha, tomamichel2015quantum, muller2013quantum}. Here, we will mention the definitions of various well-studied quantum R\'enyi divergences and show how some of them can be unified by defining a quantity inspired by generalized fidelity. Every quantum R\'enyi divergence we study is of the form
\begin{equation}
    D_{\alpha}(P \| Q) := \frac1{\alpha - 1} \log f_\alpha(P, Q),
\end{equation}
where $f_\alpha(\rho, \sigma)$ is a trace functional can be thought of as the (asymmetric) \textit{fidelity part} of the divergence. Since the relation between $D_\alpha(P\|Q)$ and $f_\alpha(P,Q)$ follows directly, we will restrict our attention to the trace functional $f_\alpha(P,Q)$.

The first family of divergences we discuss is the family of $\alpha$-$z$ divergences~\cite{audenaert2013alpha}, which is defined as
\begin{equation}
    D_{\alpha,z}(P\|Q) = \frac{1}{\alpha - 1} \log \operatorname{Tr}\left[ \left(P^\frac{\alpha}{2z} Q^\frac{1-\alpha}z P^\frac{\alpha}{2z} \right)^z \right],
\end{equation}
for $P \ll Q$, $\alpha \in \mathbb R \backslash \{1\}$ (with the limit being taken for $\alpha \to 1$) and $z \in \mathbb R_+$ (with limit being taken for $z \to 0$). This family unifies various various quantum R\'enyi divergences. In particular, set $\alpha = z$ to obtain the \textit{sandwich relative entropy} of order $\alpha$:
\begin{equation}
 D_{\alpha, \alpha} (P\|Q) =  D_\alpha^\text{S} :=  \frac{1}{\alpha-1} \log \operatorname{Tr}\left[ \left( Q^{\frac{1-\alpha}{2\alpha}} P Q^{\frac{1-\alpha}{2\alpha}} \right)^\alpha  \right].
\end{equation}
Setting $z=1$ recovers the \textit{Petz-R\'enyi divergence of order $\alpha$}:

\begin{equation}
  D_{\alpha,1} (P \| Q) = D^\text{PR}_\alpha (P\|Q) := \frac{1}{\alpha-1} \log \operatorname{Tr}[P^\alpha Q^{1-\alpha}]. 
\end{equation}
Setting $z = 1-\alpha$ recovers the \textit{reverse sandwich relative entropy} of order $\alpha$:
\begin{equation}
  D_{\alpha, 1-\alpha} (P \| Q) = D^\text{RS}_\alpha (P\|Q) := \frac{1}{\alpha-1} \log \operatorname{Tr}\left[ \left(P^{\frac\alpha{2(1-\alpha)}} Q P^{\frac\alpha{2(1-\alpha)}}\right)\right]. 
\end{equation}
From the sandwiched relative entropy, one can recover other divergences such as the \textit{min-relative entropy}, \textit{Umegaki quantum relative entropy}, and the \textit{max-relative entropy}:
\begin{equation}
\begin{aligned}
D_{\frac12, \frac12} (P\|Q) =  D^\text{min}(P\|Q) &:= - 2 \log \Fu{P}{Q}, \\ 
\lim_{\alpha \to 1} D_{\alpha, \alpha} (P\|Q) =  D^\text{Um}(P\|Q) &:= \operatorname{Tr}\left[P \left(\log P - \log Q \right)\right], \\ 
\lim_{\alpha \to \infty} D_{\alpha, \alpha} (P\|Q) =  D^\text{max}(P\|Q) &:= \inf\{\lambda : P \leq 2^\lambda Q\}. 
\end{aligned}
\end{equation}

Another quantum generalization of R\'enyi divergence is based on the AI geodesic between $P$ and $Q$. Specifically, the \textit{$\alpha$-geometric R\'enyi divergence}~\cite{matsumoto2015new, fang2021geometric, katariya2021geometric} between $P$ and $Q$ is defined as
\begin{equation}
    D_\alpha^\text{G}(P \|Q) = \frac{1}{\alpha - 1} \log \operatorname{Tr}[Q \#_\alpha P] = \frac{1}{\alpha-1} \log \operatorname{Tr}\left[Q ^{\frac12} \left( Q ^{-\frac12} P Q ^{-\frac12} \right)^\alpha Q ^{\frac12} \right].
\end{equation}
Thus, the quantity was first introduced in~\cite{matsumoto2015new} and is the largest quantum R\'enyi divergence that satisfies data processing inequality. It is also known~\cite{matsumoto2015new, tomamichel2015quantum, katariya2021geometric} that for $P \in \mathbb D_d$ and $Q \in \mathbb P_d$
\begin{equation}
    \lim_{\alpha \to 1} D_\alpha^\text{G}(P\|Q) = D^\text{BS}(P \|Q) := \operatorname{Tr}\left[P \log \left( P ^{\frac12} Q ^{-1} P ^{\frac12} \right)\right],
\end{equation}
where $D^\text{BS}(P\|Q)$ is the \textit{Belavkin-Staszewski} relative entropy~\cite{belavkin1982c}.

We will now define a quantity inspired by the definition of generalized fidelity, which can recover many of the above-mentioned divergences. In particular, the quantity we define will recover the following divergences
\begin{equation}
    D^\text{S}_\alpha, D^\text{PR}_\alpha, D^\text{RS}_\alpha, \text{ and }D^\text{G}_\alpha. 
\end{equation}
Since the sandwich relative entropy, in turn, can recover the min-relative entropy $D^\text{min}$, Umegaki relative entropy $D^\text{Um}$, and max-relative entropy $D^\text{max}$ and the geometric R\'enyi divergence has the Belavkin-Staszewski $D^\text{BS}$ as its $\alpha \to 1$ limit, the quantity we define would recover these divergences too. We note that this section is meant for introductory and illustrative purposes alone, and further information-theoretic properties are deferred to a future article.

The key idea is to define a \textit{base-dependent} quantity that generalizes the trace functional term. Similar to generalized fidelity, this term is complex in general, and therefore, we will only use its real part. 

\begin{definitionbox}
\begin{definition}
Let $P,Q,R \in \mathbb P_d$ and $\alpha \in (0,1) \cup (1, \infty)$. Define
\begin{equation}
    \hat{D}_{\alpha, R} (P \| Q) = \frac1{\alpha-1} \log \Re \operatorname{F}_{R}^\alpha(P,Q),    
\end{equation}
where the trace functional $F_R^\alpha$ is defined as
\begin{equation}
    \Fga{P}{Q}{R} := \operatorname{Tr}\left[\left( R ^{\frac12} P R ^{\frac12} \right)^\alpha R ^{-1} \left( R ^{\frac12} Q R ^{\frac12} \right)^{1-\alpha}  \right]. 
\end{equation}
\end{definition}
\end{definitionbox}

We firstly note that in the \textit{classical scenario}, where $P,Q$ and $R$ mutually commute, the above quantity reduces to the (classical) R\'enyi divergence. 

We now show how the above quantity recovers the previously mentioned divergences. This is formalized in the following theorem.
\begin{theorembox}
 \begin{theorem}
    For $P \in \mathbb D_d$ and $Q \in \mathbb P_d$, define $\hat{D}_{\alpha, R}(P \| Q)$ as above. Then we have
    \begin{alignat}{3}
    R &= \mathbb I  \quad & \implies \quad \hat{D}_{\alpha, R}(P \| Q) &= D_\alpha^\mathrm{PR}(P \| Q)  \quad &\text{(Petz-R\'enyi)}  \\
    R &= Q^\frac{1-\alpha}\alpha \quad  & \implies \quad \hat{D}_{\alpha, R}(P \| Q) &= D_\alpha^\mathrm{S}(P \| Q) \quad &\text{(Sandwich)} \\
    R &= P^\frac\alpha{1-\alpha} \quad  & \implies \quad \hat{D}_{\alpha, R}(P \| Q) &= D_\alpha^\mathrm{RS}(P \| Q)  \quad &\text{(Reverse sandwich)} \\
    R &= Q^{-1}  \quad  & \implies \quad \hat{D}_{\alpha, R}(P \| Q) &= D_\alpha^\mathrm{G}(P \| Q) \quad &\text{(Geometric)}
\end{alignat}
\end{theorem}
\end{theorembox}
\begin{proof}
We only need to work with the trace functional $\Fga{P}{Q}{R}$ as its relation to $\hat D_{\alpha, R}$ follows directly. Recall the form of the trace functional:
\begin{equation}
    \Fga{P}{Q}{R} := \operatorname{Tr}\left[\left( R ^{\frac12} P R ^{\frac12} \right)^\alpha R ^{-1} \left( R ^{\frac12} Q R ^{\frac12} \right)^{1-\alpha}  \right]. 
\end{equation}
Choosing $R = \mathbb I$, we easily see that 
\begin{equation}
    \Fga{P}{Q}{\mathbb I} =  \operatorname{Tr}[P^\alpha Q ^{1-\alpha}],
\end{equation}
which leads to the Petz-R\'enyi divergence.
For the Sandwich R\'enyi relative entropy, we choose $R = Q^\frac{1-\alpha} \alpha$. Substituting in the trace functional, we have

\begin{equation}
\begin{aligned}
\Fga{P}{Q}{R} :&= \operatorname{Tr} \left[ \left( R^\frac 12 P R^\frac 12 \right)^\alpha R^{-1} \left( R^\frac 12 Q R^\frac 12 \right)^{1-\alpha} \right]\\
&= \operatorname{Tr} \left[ \left( Q^{\frac{1-\alpha}{2\alpha}} P Q^{\frac{1-\alpha}{2\alpha}} \right)^\alpha Q^{\frac{\alpha-1}{\alpha}} \left( Q^{\frac{1-\alpha}{2\alpha}} Q Q^{\frac{1-\alpha}{2\alpha}} \right)^{1-\alpha} \right] \\
&= \operatorname{Tr} \left[ \left( Q^{\frac{1-\alpha}{2\alpha}} P Q^{\frac{1-\alpha}{2\alpha}} \right)^\alpha Q^{\frac{\alpha-1}{\alpha}} Q^{\frac{1-\alpha}{\alpha}} \right] \\
&= \operatorname{Tr} \left[ \left( Q^{\frac{1-\alpha}{2\alpha}} P Q^{\frac{1-\alpha}{2\alpha}} \right)^\alpha \right], 
\end{aligned}
\end{equation}
which corresponds to the sandwich R\'enyi relative entropy. For the reverse sandwich R\'enyi relative entropy, set $R = P^\frac{\alpha}{1-\alpha}$, which leads to
\begin{equation}
\begin{aligned}
    \Fga{P}{Q}{R} :&=   \operatorname{Tr} \left[ \left( R^\frac 12 P R^\frac 12 \right)^\alpha R^{-1} \left( R^\frac 12 Q R^\frac 12 \right)^{1-\alpha} \right] \\
&= \operatorname{Tr} \left[ \left( P^{\frac{\alpha}{2(1-\alpha)}} P P^{\frac{\alpha}{2(1-\alpha)}} \right)^\alpha P^{\frac{-\alpha}{1-\alpha}} \left( P^{\frac{\alpha}{2(1-\alpha)}} Q P^{\frac{\alpha}{2(1-\alpha)}} \right)^{1-\alpha} \right] \\
&= \operatorname{Tr} \left[ P^{\frac{\alpha}{1-\alpha}} P^{\frac{-\alpha}{1-\alpha}} \left( P^{\frac{\alpha}{2(1-\alpha)}} Q P^{\frac{\alpha}{2(1-\alpha)}} \right)^{1-\alpha} \right] \\
&= \operatorname{Tr} \left[ \left( P^{\frac{\alpha}{2(1-\alpha)}} Q P^{\frac{\alpha}{2(1-\alpha)}} \right)^{1-\alpha} \right], 
\end{aligned}
\end{equation}
which is the trace functional that defines the reverse sandwich R\'enyi relative entropy. Finally, choose $R = Q ^{-1} $ to obtain the $\alpha$-geometric R\'enyi divergence:
\begin{equation}
        \begin{aligned}
        \Fga{P}{Q}{Q ^{-1} } :&= \operatorname{Tr} \left[ \left( Q^{-\frac 12} P Q^{-\frac 12} \right)^\alpha Q \left( Q^{-\frac 12} Q Q^{-\frac 12} \right)^{1-\alpha} \right] \\
        &= \operatorname{Tr} \left[ \left( Q^{-\frac 12} P Q^{-\frac 12} \right)^\alpha Q   \cdot \mathbb I^{1-\alpha}  \right] \\
        &= \operatorname{Tr} \left[ Q ^{\frac12}  \left( Q^{-\frac 12} P Q^{-\frac 12} \right)^\alpha Q ^{\frac12}  \right] \\
&= \operatorname{Tr}\left[Q \#_\alpha P \right], 
    \end{aligned}
    \end{equation}
which is the trace functional in the definition of $\alpha$-geometric R\'enyi divergence. This concludes the proof.
\end{proof}

\section{Open problems } \label{Sec:OpenProb}

We now discuss some related open problems. 

\begin{enumerate}
    \item \textbf{Data Processing Inequality.} For a given pair $P,Q \in \mathbb P_d$, for what values of the base $R$ does the squared generalized Bures distance satisfy the data processing inequality (DPI)? Equivalently, for what values of $R$ does the following inequality hold for any quantum channel $\Phi$? 
    \begin{equation}
    \operatorname{B}_R(P,Q) \overset{?} {\geq}\operatorname{B}_{\Phi(R)}\left((\Phi(P), \Phi(Q) \right). 
    \end{equation} 
    Preliminary numerical experiments have not identified any instances that violate DPI.   
    \item \textbf{Convexity.} A related open question is regarding (joint) convexity of generalized Bures distance (and/or its squared version) in $P, Q$, and $R$. These appear to be difficult problems, and perhaps it would be easier to tackle these questions for $x$-Polar fidelities (for $x \in [-1,1]$).
    \item \textbf{Recovering $z$-fidelities from generalized fidelities.} Can the $z$-fidelities and/or the Log-Euclidean fidelity be written as the generalized fidelity (or interior fidelity) for some choice of base (or a distribution over bases)? An affirmative answer to the above problem might also lead to the analogous generalization for more members of the $\alpha$-$z$ divergences. 
    \item \textbf{Monotonicity of polar fidelities.} In the discussion of Paths 5 and 6, we remarked on the numerical observation that the generalized fidelity was monotonic along these curves. Does this observation always hold? That is, for any $P,Q \in \mathbb P_d$ and any pair $x, y \in [-1,1]$, does the following statement hold true?
    \begin{equation}
        x \geq y \implies \begin{cases}
            \Fg{P}{Q}{P^x}  \geq \Fg{P}{Q}{P^y}, \\
            \Fg{P}{Q}{Q^x}  \geq \Fg{P}{Q}{Q^y}. \\
         \end{cases}
    \end{equation}
    An affirmative answer would also imply that the $x$-Polar fidelity is monotonic in $x$ in the range $[-1,1]$. This numerically observed monotonic is also in line with the known relation regarding the named fidelities:
    \begin{equation}
        \operatorname{F}^\mathrm M(P,Q) \leq \operatorname{F}^\mathrm H(P,Q) \leq \operatorname{F}^\mathrm U(P,Q), 
    \end{equation}
    for any $P,Q \geq 0$. Moreover, it is also known $\operatorname{F}^\mathrm M$ is the smallest and $\operatorname{F}^\mathrm{U}$ is the largest quantization of classical fidelity that satisfies DPI~\cite{matsumoto2010reverse}. Thus, the family of $x$-Polar fidelities (in the range $x \in [-1,1]$) might be helpful in studying fidelities that satisfy the data processing inequality. 
    \item \textbf{Other bases for Holevo fidelity.} 
Another open problem we pose is whether other non-trivial bases exist where generalized fidelity recovers Holevo fidelity. Recall that we showed that $R = \mathbb I$ implies $\operatorname{F}_R(P,Q) = \operatorname{F}^\mathrm H(P,Q)$. In fact, this is the only choice of $R$ where we showed this equality for a general (non-commuting) pair of matrices $P,Q$. Are there other bases on which the generalized fidelity reduces to Holevo fidelity? To find such a non-trivial base, it would suffice to find $R \in \mathbb P_d$ such that
\begin{equation}
    \operatorname{Pol}\left(R ^{\frac12} P ^{\frac12} \right) = \operatorname{Pol}\left(R ^{\frac12} Q ^{\frac12} \right). 
\end{equation}
We believe that results from \cite{lawson2008gamma} would be helpful in this endeavor.
\item \textbf{SDP representation.} The next question we ask is whether (the real part of) generalized fidelity or squared generalized Bures distance has a true semidefinite program representation. Though we have presented a block-matrix characterization of generalized fidelity, it is not an SDP. An SDP formulation could be vastly beneficial in optimization problems involving generalized Bures distance, such as in a potential formulation of metric learning~\cite{yang2006distance, zadeh2016geometric}. 
\item \textbf{Unitary factor of generalized fidelity and $\mathrm{SU}(d)$.} The final open problem we pose is asking whether the set of unitary factors of generalized fidelity and the set of special unitaries are the same. We have shown that the set of unitary factors forms a subset of $\mathrm{SU}(d)$, but it is unclear whether the reverse inclusion holds. To prove this, one would have to show that any special unitary $U \in \mathrm{SU}_d$ has the form
\begin{equation}
    U = U_Q U_P^*
\end{equation}
for a triple $P,Q,R \in \mathbb P_d$, where $U_P := \operatorname{Pol}\left(P ^{\frac12} R ^{\frac12} \right)$ and $U_Q = \operatorname{Pol}\left(Q ^{\frac12} R ^{\frac12} \right)$.
An affirmative answer would illuminate a connection between the Bures-Wasserstein manifold, generalized fidelity, and the Lie group $\mathrm{SU}(d)$. 
\end{enumerate}

\section{Conclusion and discussions }\label{Sec:Concl}
In this paper, we introduce a family of fidelities between positive definite matrices that generalize and unify various existing quantum fidelities. The definition is motivated by the Riemannian geometry of the Bures-Wasserstein manifold, and it endows the existing fidelities with novel geometric interpretations. We also define and study the generalization of the closely related Bures-Wasserstein distance. 

After studying the basic properties of these objects, we proved several remarkable geometric properties of generalized fidelity including invariance and covariance properties of generalized fidelity along geodesic-related paths related to the Bures-Wasserstein, Affine-invariant, and Euclideain Riemannian metrics on the manifold of positive definite matrices. 

We then showed how convex combinations of generalized fidelity define a new family of fidelities. One such family, the \textit{Polar fidelity}, is shown to be a family of fidelities parametrized by a single real number that recovers Uhlmann-, Holevo-, and Matsumoto fidelity. We then derived a block-matrix characterization of generalized fidelity and squared generalized Bures-Wasserstein distance and showed an interesting relation between generalized fidelity, multivariate fidelities, and Bures-Wasserstein barycenters. 

We also discussed an \textit{Uhlmann-like} theorem for generalized fidelity, and finally, we demonstrated how our formalism can also be extended to generalized certain quantum R\'enyi divergences. We then discussed various open problems.

We conclude this section by discussing potential applications of generalized Bures distance in quantum and classical machine learning. Distances are crucial in machine learning as we often embed data points in high-dimensional metric spaces, and often, we find that different distance metrics (and divergences) are suitable for different tasks. In particular, consider the problem of \textit{Metric learning}~\cite{yang2006distance, zadeh2016geometric, kulis2013metric, davis2007information}. The problem can be succinctly stated as follows: find the distance metric that is \textit{best} suited to represent the data. For example, one could consider the case where we are given pairwise distances between many data points (represented by vectors in $\mathbb  R^d$), and we are interested in finding the positive definite matrix that defines the Mahalanobis distance which would be most consistent with the given distances~\cite{zadeh2016geometric}. Metric learning has applications in various machine learning fields, including classification algorithms like the K Nearest Neighbours algorithm~\cite{peterson2009k} and dimensionality reduction problems~\cite{wang2015survey, harandi2017joint}. 

Generalized Bures distance could extend the problem to the setting where the data points are positive definite matrices or quantum states. Specifically, given a collection of states and associated data (such as pairwise distances or labels for each state), find a suitable base for the given task. For example, this could be a classification problem where the states are given labels indicating membership in disjoint classes. Then, the task would be to find a base such that the generalized Bures distance at this base increases \textit{interclass} distance while decreasing \textit{intraclass} distance. Thus, metric learning based on generalized Bures distance could benefit various classical ML problems involving the BW manifold or even quantum variants of the aforementioned (classical) ML problems \cite{wiebe2015quantum, basheer2020quantum, schuld2014quantum, duan2019quantum, liang2020variational}.

\section{Acknowledgements }
AA is funded by the Sydney Quantum Academy supplementary scholarship. We thank Mark M. Wilde for going through an earlier version of the draft and providing helpful comments. AA also thanks Roberto Rubboli, Ian George, Marco Tomamichel, Bamdev Mishra, and Pratik Jawanpuria for insightful discussions.

\bibliographystyle{unsrt}
\bibliography{my.bib}

\appendix
\newpage

\section{More mathematical preliminaries} \label{App:MoreMathPrelim}

\subsection{Geometric mean} \label{App:GeoMean}
Here, we discuss some basic properties of the geometric mean of two positive definite matrices. For an excellent treatment of this topic, see~\cite{bhatia2009positive}. See~\cite{Wilde2024MatrixGeoMean} for a detailed exploration of quantum algorithms for matrix geometric means, along with a discussion on the role of matrix geometric means in various quantum information problems.
\begin{definitionbox}
    \begin{definition}[Geometric mean]
        Let $A, B \in \mathbb P_d$ be positive definite matrices. The geometric mean between $A$ and $B$ is defined as
        \begin{equation}
            A \# B := A ^{\frac12} \sqrt{A ^{-\frac12} B A ^{-\frac12} } A ^{\frac12}.
        \end{equation}
    \end{definition}
\end{definitionbox}
Though it may not be obvious from the definition, the geometric mean is symmetric: $A \# B = B \# A$. It also enjoys various other properties, some of which we list here. 
\begin{theorembox}
    \begin{prop}[Properties of Geometric mean] \label{Prop:GeoMeanPropertiesApp}
        Let $A \# B$ be the geometric mean of $A, B \in \mathbb P_d$. Then, the following holds true.
        \begin{enumerate}
            \item $A \# B$ is the unique positive definite solution to the matrix Riccati equation
            \begin{equation}
                B = X A ^{-1} X.
            \end{equation}
            \item $A\#B$ is the midpoint of the Affine-invariant geodesic $\gamma_{A B}^{\mathrm{AI}}(t)$ between $A,B$ defined as
            \begin{equation}
                \gamma_{A B}^\mathrm{AI}(t) := A ^{\frac12} \left(A ^{-\frac12} B A ^{-\frac12} \right)^t A ^{\frac12},
            \end{equation}
            for $t \in [0,1]$.
        \end{enumerate}
        
    \end{prop}
\end{theorembox}
The proofs are available in \cite{bhatia2009positive}.



\subsection{Bures-Wasserstein barycenters}

Given a distribution $\mu$ over $\mathbb P_d$, the Bures-Wasserstein barycenter~\cite{agueh2011barycenters, alvarez2016fixed, chewi2020gradient, bhatia2019bures, zemel2019frechet} is the positive definite matrix that minimizes the average squared Bures-Wasserstein distance over this distribution. Hence, this point can be thought of as the BW analog of the mean, which minimizes the average squared Euclidean distance. 

For our purposes, it will suffice to consider finite distributions. Let $\mu$ be an $n$-dimensional probability vector and ${P_1, \ldots, P_n} \subset \mathbb P_d$ be a collection of positive definite matrices. Then, the BW barycenter is defined as
\begin{equation}
    P_\mu := \argmin_{Q \in \mathbb P_d} \underset{P \sim \mu}{\mathbb E} [\mathrm{B}^\text{U}(P, Q)] =  \argmin_{Q \in \mathbb P_d} \sum_{i=1}^{n} \mu_i \mathrm{B}^\text{U}(P_i, Q). 
\end{equation}
This convex problem of finding the BW barycenter of a distribution appears in a variety of fields. In quantum information, it appears in Bayesian quantum tomography as the Bayes estimator for fidelity~\cite{afham2022}, as a secrecy measure in quantum information~\cite{konig2009operational}, and as the square root of the maximum success probability of certain quantum algorithms~\cite{Wilde2023estimating}. In \cite{brahmachari2023fixed}, the authors generalize the barycenter problem to exploit a group structure that captures even more problems relevant to quantum information. Other settings where BW barycenter find applications include machine learning and inference, computer vision and graphics, probability theory and signal processing. See~\cite{chewi2020gradient, altschuler2021averaging, chewi2024statistical} and references within.

In general, $P_\mu$ does not have a closed-form solution. However, it uniquely satisfies the following fixed-point equation:
\begin{equation} \label{Eq:BWBaryFPEqn}
    P_\mu = \sum_{i=1}^{n} \mu_i \sqrt{P_\mu^{\frac12} P_i  P_\mu^{\frac12}. }
\end{equation}
This relation can be derived in various ways, including from first-order convexity conditions~\cite{bhatia2019bures} or from the complementary slackness relation of a certain semidefinite program~\cite{afham2022}. 

Consider the case where the probability distribution is supported on two points $P_0$ and $P_1$ weights $\mu = (1-t,t)$ for $t \in [0,1]$. Then, the BW barycenter is simply the point on the BW geodesic between $P_0$ and $P_1$. In this case, there exists a closed-form solution for the barycenter $P_t$:
\begin{equation}
\begin{aligned}        P_t &= \left((1-t) \mathbb I +  t P_0 ^{-1} \# P_1 \right) P_0 \left((1-t) \mathbb I +  t P_0 ^{-1} \# P_1 \right) \\
        &= (1-t)^2 P_0 + t^2 P_1 + t(1-t) \left[\sqrt{P_0 P_1} + \sqrt{P_1 P_0}\right] \\
        &= \gamma^\mathrm{BW}_{PQ}(t).
        \end{aligned}
\end{equation}

\section{Proofs of basic properties} \label{App:BasicProperties}
In this section, we prove the basic properties of generalized fidelity stated in Section~\ref{Sec:BasicProperties}. For the rest of the section we arbitrarily choose and fix $P, Q, R \in \mathbb P_d$ and define $U_P := \operatorname{Pol}\left(P^{\frac12}  R ^{\frac12} \right)$ and $U_Q = \operatorname{Pol}\left(Q ^{\frac12} R ^{\frac12} \right)$. 

\begin{enumerate}[noitemsep, left=0pt]
    \item \textbf{Conjugate Symmetry}. $\operatorname{F}_R(P,Q) \iseq \operatorname{F}_R(Q,P)^*$. The claim easily follows from the definition:
    \begin{equation}
    \begin{aligned}
              \operatorname {F}_R(P,Q) :&=  \operatorname{Tr}\left[\sqrt{ R^{\frac12} P R^{\frac12}} R^{-1}\sqrt{ R^{\frac12} Q R^{\frac12}}  \right] \\ &= \operatorname{Tr}\left[\sqrt{ R^{\frac12} Q R^{\frac12}} R^{-1}\sqrt{ R^{\frac12} P R^{\frac12}}  \right]^* =: \operatorname{F}_R(Q,P)^*.  
    \end{aligned} 
    \end{equation}
    \item \textbf{Equivalent forms}:
    \begin{equation}
    \begin{aligned}
            \operatorname{F}_R(P,Q) := &\operatorname{Tr}\left[\sqrt{ R^{\frac12} P R^{\frac12}} R^{-1}\sqrt{ R^{\frac12} Q R^{\frac12}}  \right] \\ 
            \iseq  &\operatorname{Tr}\left[ Q ^{\frac12} U_Q U_P^* P^{\frac12}   \right] \\
            \iseq &\operatorname{Tr}\left[(R ^{-1} \# Q) R (R ^{-1} \# P)\right],
    \end{aligned}
    \end{equation}
    To show the first equality, we simply take the polar decompositions $\sqrt{R ^{\frac12} P R ^{\frac12} } = U_P^* P^{\frac12}  R ^{\frac12} $ and $\sqrt{R ^{\frac12} Q R^{\frac12} } = R ^{\frac12} Q ^{\frac12} U_Q$ and then substitute in the definition of generalized fidelity and then use cyclicity of trace. To prove the third equivalent form
    \begin{equation}
            \operatorname{Tr}\left[\sqrt{ R^{\frac12} P R^{\frac12}} R^{-1}\sqrt{ R^{\frac12} Q R^{\frac12}}  \right] \iseq \operatorname{Tr}\left[(R ^{-1} \# Q) R (R ^{-1} \# P)\right],
    \end{equation}
    substitute the formula for geometric mean. The relation immediately follows.

    \item \textbf{Simplified form for pure states.} Assume $P = \ketbra{\psi}$ and $Q = \ketbra{\phi}$. Consider the term
    \begin{equation}
        \sqrt{R^{\frac12}  P R ^{\frac12} } = \sqrt{|u \rangle \langle u|} = \frac{|u \rangle \langle u|}{\sqrt{\langle u, u \rangle }},
    \end{equation}
     where we have defined $|u \rangle := R^{\frac12} |\psi \rangle $. A similar calculation reveals
     \begin{equation}
        \sqrt{R ^{\frac12} Q R ^{\frac12} } = \frac{|v \rangle \langle v|}{\sqrt{\langle v, v \rangle }},
     \end{equation} 
    for $|v \rangle := R ^{\frac12} | \phi \rangle$. Thus we have
    \begin{equation}
    \begin{aligned}
        \operatorname{F}_R(P,Q) & =
        \operatorname{Tr}\left[\sqrt{ R^{\frac12} P R^{\frac12}} R^{-1}\sqrt{ R^{\frac12} Q R^{\frac12}}  \right] \\  &= \frac{1}{\|u\| \|v\|} \operatorname{Tr}\left[\ketbra{u} R ^{-1} \ketbra{v} \right] \\ &= 
        \frac{1}{\|u\| \|v\|} \operatorname{Tr}\left[R ^{\frac12} \ketbra{\psi} R ^{\frac12}  R ^{-1} R ^{\frac12} \ketbra{\phi} R ^{\frac12} \right] \\
        &= \frac{\langle \psi, \phi \rangle \langle \phi, R \psi \rangle }{ \Fu{P}{R} \Fu{Q}{R}}, 
    \end{aligned}
    \end{equation}
    where in the last equality, we used $\|u\| = \sqrt{\langle \psi, R \psi \rangle} = \Fu{P}{R}$ and the analogous identity for $\|v\|$.

    \item \textbf{Commutation with base implies reality.} Without loss of generality, assume $[R,P] = 0$. Then we have $\sqrt{R ^{\frac12} P R ^{\frac12} } = P^{\frac12}  R ^{\frac12} $, which implies
    \begin{equation}
        \operatorname{F}_R(P,Q) = \operatorname{Tr}\left[P^{\frac12}  R ^{\frac12} R^{-\frac12} \sqrt{R ^{\frac12} Q R ^{\frac12} } \right] = \operatorname{Tr}\left[P^{\frac12}  \sqrt{R ^{\frac12} Q R ^{\frac12} } \right],
    \end{equation}
    which is the inner product of two positive matrices, and thus the generalized fidelity is real (and positive).

    \item \textbf{Multiplicativity.} For $P_1,Q_1,R_1 \in \mathbb P_{d_1}$ and $P_2,Q_2,R_2 \in \mathbb P_{d_2}$, the claim is
    \begin{equation}
        \Fg{P_1 \otimes P_2}{Q_1 \otimes Q_2}{R_1 \otimes R_2} \iseq \Fg{P_1}{Q_1}{R_1} \cdot \Fg{P_2}{Q_2}{R_2}. 
    \end{equation}
    Denote $P \equiv P_1 \otimes P_2, Q \equiv Q_1 \otimes Q_2$, and $R \equiv R_1 \otimes R_2$. Then,
    \begin{equation}
        \Fg{P_1 \otimes P_2}{Q_1 \otimes Q_2}{R_1 \otimes R_2} = \operatorname{Tr}\left[\sqrt{ R^{\frac12} P R^{\frac12}} R^{-1}\sqrt{ R^{\frac12} Q R^{\frac12}}\right]. 
    \end{equation}
    Since the matrix square root, product, and inverse factors out with respect to the tensor product, we have
    \begin{equation}
    \begin{aligned}
        \sqrt{ R^{\frac12} P R^{\frac12}} R^{-1}\sqrt{ R^{\frac12} Q R^{\frac12}} = 
       \left [\sqrt{ R_1^{\frac12} P_1 R_1^{\frac12}}   R_1^{-1}\sqrt{ R_1^{\frac12} Q_1 R_1^{\frac12}}  \right] \otimes \left [\sqrt{ R_2^{\frac12} P_2 R_2^{\frac12}} R_2^{-1}\sqrt{ R_2^{\frac12} Q_2 R_2^{\frac12}}  \right].
    \end{aligned}
    \end{equation}
    Take trace across and using the identity $\operatorname{Tr}[A \otimes B] = \operatorname{Tr}[A] \operatorname{Tr}[B]$ to obtain
    $\operatorname{F}_R(P,Q) = \operatorname{F}_{R_1}(P_1,Q_1) \cdot \operatorname{F}_{R_2}(P_2,Q_2). $

    \item \textbf{Additivity.} Here the claim is
    \begin{equation}
        \Fg{P_1 \oplus P_2}{Q_1 \oplus Q_2}{R_1 \oplus R_2} \iseq \Fg{P_1}{Q_1}{R_1} + \Fg{P_2}{Q_2}{R_2}. 
    \end{equation}
    Denote $P \equiv P_1 \oplus P_2, Q \equiv Q_1 \oplus Q_2$, and $R \equiv R_1 \oplus R_2$. From the properties of direct sum we have,
    \begin{equation}
    \begin{aligned}
        \sqrt{ R^{\frac12} P R^{\frac12}} R^{-1}\sqrt{ R^{\frac12} Q R^{\frac12}} = 
       \left [\sqrt{ R_1^{\frac12} P_1 R_1^{\frac12}}   R_1^{-1}\sqrt{ R_1^{\frac12} Q_1 R_1^{\frac12}}  \right] \oplus \left [\sqrt{ R_2^{\frac12} P_2 R_2^{\frac12}} R_2^{-1}\sqrt{ R_2^{\frac12} Q_2 R_2^{\frac12}}  \right].
    \end{aligned}
    \end{equation}
    Take trace across and use the identity that $ \operatorname{Tr}[A \oplus B] = \operatorname{Tr}[A]+\operatorname{Tr}[B]$, we have $\operatorname{F}_R(P,Q) = \Fg{P_1}{Q_1}{R_1} +  \Fg{P_2}{Q_2}{R_2}$.

    \item \textbf{Unitary invariance.} Given $P,Q,R \in \mathbb P_d$ and $U \in \mathbb U_d$, we want to show that
    \begin{equation}
        \Fg{UPU^*}{UQU^*}{URU^*} = \operatorname{F}_R(P,Q). 
    \end{equation}
    Observe that $ \sqrt{VAV^*} = V \sqrt{A} V^*$ for any $ V \in \mathbb U_d$ and $A \in \mathbb P_d$. Moreover, $(U R U^*) ^{-1} = U R ^{-1} U^* $. Substituting this in the definition, we get the required relation.

    \item \textbf{Unitary contravariance.} Given $P,Q,R \in \mathbb P_d$ and $U \in \mathbb U_d$, we want to show that
        \begin{equation}
            \Fg{P}{Q}{URU^*} = \operatorname{F}_R(U^*P U,U^* Q U). 
        \end{equation}
    By definition, we have
    \begin{equation}
    \begin{aligned}
        \operatorname{F}_{U R U^*}(P,Q) := &\operatorname{Tr}\left[\sqrt{(U R ^{\frac12} U^*) P (U R ^{\frac12} U^*)} (U R ^{-1} U^*) \sqrt{(U R ^{\frac12} U^*) Q (U R ^{\frac12} U^*)} \right] \\
        = &\operatorname{Tr} \left[U \sqrt{R ^{\frac12} U^* P^{\frac12}   U R ^{\frac12} } R ^{-1} \sqrt{R ^{\frac12} U^* Q ^{\frac12}  U R ^{\frac12} } U^* \right] \\ 
        = &\operatorname{Tr} \left[\sqrt{R ^{\frac12} \left(U^* P^{\frac12}   U\right) R ^{\frac12} } R ^{-1} \sqrt{R ^{\frac12} \left(U^* Q ^{\frac12}  U\right) R ^{\frac12} } \right]
        = \Fg{U^* P U}{U^* Q U}{R}.
    \end{aligned}
    \end{equation}

    \item \textbf{Scaling}. For positive scalars $p,q,r \in \mathbb R_+$, 
    we want to show that
    \begin{equation}
        \Fg{pP}{qQ}{rR} = \sqrt{pq} \operatorname{F}_R(P,Q). 
    \end{equation}
    The result follows directly from substitution.

    \item \textbf{Upper bound on absolute value.} For any triple $P,Q,R \in \mathbb P_d$, we want to show that
    \begin{equation}
        |\operatorname{F}_R(P,Q)| \leq \operatorname{F}^\mathrm U(P,Q). 
    \end{equation}
    We have
    \begin{equation}
       |\operatorname{F}_R(P,Q)| = \left|\operatorname{Tr}\left[P^{\frac12}  U_P U_Q^* Q ^{\frac12} \right] \right| \leq \max_{V \in \mathbb U_d}\left| \operatorname{Tr} \left[P^{\frac12}  V Q ^{\frac12} \right] \right| = \operatorname{F}^\mathrm U(P,Q),
    \end{equation}
    where the first equality comes from the alternative characterization of generalized fidelity, and the last equality comes from the variational characterization of Uhlmann fidelity. 

    \item \textbf{Reduction to named fidelities} We now show that for specific choices of the base $R$, we can recover the Uhlmann-, Holevo-, and Matsumoto fidelities. 
    \begin{enumerate}
        \item \textbf{Uhlmann fidelity: $R \in \{P,Q\}$}. Without loss of generality, choose $R = P$. We then have
            \begin{equation}
            \begin{aligned}
                \operatorname{F}_R(P,Q) := &\operatorname{Tr}\left[\sqrt{ R^{\frac12} P R^{\frac12}} R^{-1}\sqrt{ R^{\frac12} Q R^{\frac12}}  \right] \\ = &\operatorname{Tr}\left[P \dot P ^{-1} \sqrt{P^{\frac12}  Q P^{\frac12}  }\right] = \operatorname{F}^\mathrm U(P,Q).
            \end{aligned}
            \end{equation}
            The case where $R = Q \implies \operatorname{F}_R(P,Q) = \operatorname{F}^\mathrm U(P,Q) $ is derived similarly. 

        \item Choose \textbf{Holevo fidelity}. Choose $R = \mathbb I$. We then have
            \begin{equation}
            \begin{aligned}
                \operatorname{F}_R(P,Q) := &\operatorname{Tr}\left[\sqrt{ R^{\frac12} P R^{\frac12}} R^{-1}\sqrt{ R^{\frac12} Q R^{\frac12}}  \right] \\ = &\operatorname{Tr}\left[ P^{\frac12}   Q^{\frac12} \right] = \operatorname{F}^\mathrm H(P,Q).
            \end{aligned}
            \end{equation}

        \item \textbf{Matsumoto fidelity: $R \in \{P ^{-1} ,Q ^{-1} \}$}. Without loss of generality, choose $R = P^{-1} $. We then have
            \begin{equation}
            \begin{aligned}
                \operatorname{F}_R(P,Q) := &\operatorname{Tr}\left[\sqrt{ R^{\frac12} P R^{\frac12}} R^{-1}\sqrt{ R^{\frac12} Q R^{\frac12}}  \right] \\ = &\operatorname{Tr} \left[\mathbb I \cdot P \sqrt{P ^{-\frac12} Q P ^{-\frac12} } \right] = \operatorname{Tr}[P\#Q] = \operatorname{F}^\mathrm M(P,Q).
            \end{aligned}
            \end{equation}
            The case where $R = Q^{-1}  \implies \operatorname{F}_R(P,Q) = \operatorname{F}^\mathrm M(P,Q) $ is derived similarly. 
    \end{enumerate}

\end{enumerate}

\section{Useful lemmas}~\label{Sec:UsefulLeamm}
In this section, we list various supporting results used in the main proofs. The first lemma characterizes the unitary factor of $P^{\frac 12} Q^{\frac 12}$ for $P, Q \in \mathbb{P}_d$.

\begin{theorembox}
    \begin{lemma}\label{Lem:PolarFactorLemma}
Let $P, Q \in \mathbb{P}_d$ and let $U := \operatorname{Pol}\left(P^{\frac 12} Q^{\frac 12}\right)$. Then,
\begin{align}
    U^* P^{\frac 12} Q^{\frac 12} &= \sqrt{Q^{\frac 12} P Q^{\frac 12}} = Q^{\frac 12} P^{\frac 12} U, \label{Eq:PolFacLemma1} \\
    U \sqrt{Q^{\frac 12} P Q^{\frac 12}} U^* &= \sqrt{P^{\frac 12} Q P^{\frac 12}}, \label{Eq:PolFacLemma2} \\
    U^* &= \operatorname{Pol}\left(Q^{\frac 12} P^{\frac 12}\right), \\
    U &= \operatorname{Pol}\left(P^{-\frac 12} Q^{-\frac 12}\right), \\
    P \cdot (P^{-1} \# Q) &= \sqrt{PQ} = P^{\frac 12} U Q^{\frac 12} \\  
     (P^{-1} \# Q) \cdot P &= \sqrt{QP} = Q^{\frac 12} U^* P^{\frac 12} .
\end{align}
\end{lemma}
\end{theorembox}
\begin{proof}
The first statement is a direct consequence of the definition of polar decomposition. To prove the second statement, we begin with the first statement. Multiply the first and last sides together and square the middle side to get 
\begin{equation} U^* P^{\frac 12} Q P^{\frac 12} U = Q^{\frac 12} P Q^{\frac 12}. \end{equation}
Conjugate both sides with $U$ and take the square root to obtain the required relation.

For the third statement, we begin with the polar decomposition of $Q^{\frac 12} P^{\frac 12}$ to get
\begin{equation} Q^{\frac 12} P^{\frac 12} = V \sqrt{P^{\frac 12} Q P^{\frac 12}} = \sqrt{Q^{\frac 12} P Q^{\frac 12}} U^*, \end{equation}
where we denote $V = \operatorname{Pol}\left(Q^{\frac 12} P^{\frac 12}\right)$ and the last equality comes from \Cref{Eq:PolFacLemma1}. Multiply by $V^*$ on the left to get
\begin{equation} \sqrt{P^{\frac 12} Q P^{\frac 12}} = V^* \sqrt{Q^{\frac 12} P Q^{\frac 12}} U^*. \end{equation}
Comparing the above equation with \Cref{Eq:PolFacLemma2}, we get $V = U^*$ as claimed. 

To prove the fourth statement, begin with the polar decomposition of $P^{\frac 12} Q^{\frac 12}$ and take inverse across to get
\begin{equation} Q^{-\frac 12} P^{-\frac 12} = \sqrt{Q^{-\frac 12} P^{-1} Q^{-\frac 12}} U^*. \end{equation}
Now take the adjoint to get the desired result. 

For the fifth statement, we will use the fact that for any matrix $A$ with positive eigenvalues (not necessarily Hermitian), there exists a unique matrix $B$ such that $B^2 = B B = A$. Thus, we may denote $B \equiv \sqrt{A}$. See \cite[Excercise 4.5.2]{bhatiamatrixanalysis} for further details. To prove the first part, observe that
\begin{equation}
\begin{aligned}
        (P \cdot P ^{-1} \# Q) \left(P \cdot P ^{-1} \# Q\right) &= \left(P \cdot P ^{-\frac 12}  \sqrt{P ^{\frac 12} Q P ^{\frac 12} } P ^{-\frac 12} \right)  \left( P \cdot P ^{-\frac 12}  \sqrt{P ^{\frac 12} Q P ^{\frac 12} } P ^{-\frac 12} \right) \\ &=  PQ. 
\end{aligned}
\end{equation}
To prove the second equality, it is sufficient to prove the squared version as we have established the uniqueness of the square root. 
\begin{equation} PQ \stackrel{?}{=} P^{\frac 12} U Q^{\frac 12} P^{\frac 12} U Q^{\frac 12}, \end{equation}
which can be proven by noting that $Q^{\frac 12} P^{\frac 12} U = U^* P^{\frac 12} Q^{\frac 12}$ by Statement 1. Substituting, the RHS reduces to $PQ$, thus completing the proof. The final equality is proven in the same manner as the previous equality, and hence, we skip the proof. 
\end{proof}

The following proposition characterizes the \textit{named fidelities} between $P$ and $Q$ in terms of $P^{\frac 12}, Q^{\frac 12}$ and a unitary matrix.

\begin{theorembox}
\begin{prop}\label{Prop:NamedFidelitiesUnitary}
Let $P, Q \in \mathbb{P}_d$. Then we have
\begin{equation}
\operatorname{F}^\mathrm{U}(P, Q) = \operatorname{Tr}\left[P^{\frac 12} U Q^{\frac 12}\right],
\end{equation}
\begin{equation}
\operatorname{F}_\mathrm{M}(P, Q) = \operatorname{Tr}\left[P^{\frac 12} V Q^{\frac 12}\right],
\end{equation}
\begin{equation}
\operatorname{F}_\mathrm{H}(P, Q) = \operatorname{Tr}\left[P^{\frac 12} \mathbb I Q^{\frac12} \right],
\end{equation}
where $U, V \in \mathbb{U}_d$ such that $U = \operatorname{Pol}\left(P^{\frac 12} Q^{\frac 12}\right)$ and $V = \operatorname{Pol}\left(P^{-\frac 12} Q^{\frac 12}\right)$. Moreover, $V$ is the unique unitary such that $P^{\frac 12} V Q^{\frac 12}$ is Hermitian.
\end{prop}
\end{theorembox}
\begin{proof}
The proof of the first statement directly follows from the definition of Uhlmann fidelity. Indeed by Lemma~\ref{Lem:PolarFactorLemma}, we have
\begin{equation} Q^{\frac 12} P^{\frac 12} U = \sqrt{Q^{\frac 12} P Q^{\frac 12}}, \end{equation}
whence it follows
\begin{equation} P^{\frac 12} U Q^{\frac 12} = Q^{-\frac 12} \sqrt{Q^{\frac 12} P Q^{\frac 12}} Q^{\frac 12}. \end{equation}
Take trace and use cyclicity to obtain 
    \begin{equation} 
    \operatorname{F}^\mathrm{U}(P, Q) = \operatorname{Tr}\left[P^{\frac 12} U Q^{\frac12} \right]. 
    \end{equation}

For the second statement, recall that $\operatorname{F}_\mathrm{M}(P, Q) = \operatorname{Tr}[P \# Q].$ The fact that there exists a unique unitary $V$ such that $P \# Q = P^{\frac 12} V Q^{\frac 12}$ is proven in multiple sources including~\cite[Proposition 4.1.8]{bhatia2009positive} and \cite{cree2020fidelity}. Thus, we omit the proof of this part and instead prove that $V = \operatorname{Pol}(P^{-\frac 12} Q^{\frac 12}).$ To this end consider the geometric mean of $P$ and $Q$:
\begin{equation} P \# Q = P^{\frac 12} \sqrt{P^{-\frac 12} Q P^{-\frac 12}} P^{\frac 12} = P^{\frac 12} V Q^{\frac 12} \in \mathbb{P}_d \end{equation}
for some unitary $V \in \mathbb{U}_d$. Now left and right multiply by $P^{-\frac 12}$ to obtain
\begin{equation} P^{-\frac 12} (P \# Q) P^{-\frac 12} = \sqrt{P^{-\frac 12} Q P^{-\frac 12}} = V Q^{\frac 12} P^{-\frac 12}. \end{equation}
By definition, the polar decomposition of $Q^{\frac 12} P^{-\frac 12}$ is
\begin{equation} Q^{\frac 12} P^{-\frac 12} = W \left|Q^{\frac 12} P^{-\frac 12}\right| = W \sqrt{P^{-\frac 12} Q P^{-\frac 12}}, \end{equation}
where $W = \operatorname{Pol}\left(Q^{\frac 12} P^{-\frac 12}\right).$ Comparing with the previous equation, we get $V^* = W$. This completes the proof.
\end{proof}

One can use the above Proposition to provide sufficient conditions when the generalized fidelity will equal these fidelities. 
\begin{theorembox}
    \begin{cor}
Let $P, Q, R \in \mathbb{P}_d$. Then 
\begin{equation} 
\operatorname{F}_R(Q, P) = \operatorname{Tr}\left[P ^{\frac 12} U_P U_Q^* Q ^{\frac 12}\right] =
\begin{cases} 
\operatorname{F}^\mathrm{U}(P, Q) &\text{ if } \quad U_P U_Q^* = U \\
\operatorname{F}^\mathrm{M}(P, Q) &\text{ if } \quad U_P U_Q^* = V \\
\operatorname{F}^\mathrm{H}(P, Q) &\text{ if }  \quad  U_P U_Q^* = \mathbb I \text{ or } U_P = U_Q 
\end{cases}
\end{equation}
where 
\begin{equation}
\begin{aligned}
    U := \operatorname{Pol}\left(P^{\frac 12} Q^{\frac 12}\right)&, \quad V := \operatorname{Pol}\left(P^{-\frac 12} Q^{\frac 12}\right), \\ U_P := \operatorname{Pol}\left(P^{\frac 12} R^{\frac 12}\right)&, \quad U_Q := \operatorname{Pol}\left(Q^{\frac 12} R^{\frac 12}\right).
\end{aligned}
\end{equation}
\end{cor}
\end{theorembox}

We shall use the above corollary to prove certain geometric results of generalized fidelity.



The next lemma lists some defining properties of the Bures-Wasserstein geodesic.

\begin{theorembox}
    \begin{lemma} \label{Lem:BWGeodesicProperties}
        Let $A,B \in \mathbb P_d$ and $C = \gamma_{AB}^\mathrm{BW}(t)$ for any $t \in [0,1]$. Then, the following statements hold true.
        \begin{align}
               C &= [(1-t) \mathbb I + t (A^{-1} \# B)]A[(1-t) \mathbb I + t (A ^{-1} \# B)], \\ \label{Eq:161}
             C &= (1-t) \sqrt{C ^{\frac 12} A C ^{\frac 12} } + t \sqrt{C ^{\frac 12} B C ^{\frac 12} }, \\ 
             \mathbb I &= (1-t) C ^{-1} \# A + t C ^{-1} \# B, \label{Eq:162}
        \end{align}
         where the fixed-point equation (\Cref{Eq:161}) is uniquely satisfied by $C$.
    \end{lemma}
\end{theorembox}
\begin{proof}
    The first equation has been proven in multiple sources such as \cite[Equation 39]{bhatia2019bures}. 
    The second equation can be seen as the $n=2$ version of the fixed-point equation satisfied by the BW barycenter (see \cite{afham2022, altschuler2023}). To obtain the third equation, conjugate the LHS and RHS of the second statement with $C ^{-\frac 12} $. \end{proof}

Next, we present a well-known result regarding Affine-invariant geodesics which states. This, along with the following lemmas, will be useful in proving results related to the geometric mean.

\begin{theorembox}
\begin{lemma} \label{Lem:ConjugationInvarianceAI}
Let $A,B \in \mathbb{P}_d$. Then, for any invertible $X$ with matching dimensions,
\begin{equation}
X (\gamma^{\mathrm{AI}}_{A, B} (t)) X^* = \gamma^{\mathrm{AI}}_{X A X^*, X B X^*} (t),
\end{equation}
for any $t \in [0,1].$
\end{lemma}
\end{theorembox}
\begin{proof}
See~\cite[Theorem 5.1 and Remark 5.2]{lawson2024expanding} for proof.    
\end{proof}

The following lemma is useful to state a sufficient condition for generalized fidelity to reduce to Matsumoto fidelity.

\begin{theorembox}
    \begin{lemma} \label{Lem:GenFidMatFidSuffCondition}
Let $P,Q,R \in \mathbb{P}_d$. Then, the following statements are equivalent.
\begin{enumerate}
    \item $R^{-\frac 12} \sqrt{R^{\frac 12} P R^{\frac 12}} \sqrt{R^{\frac 12} Q R^{\frac 12}} R^{-\frac 12} = P \# Q.$
    \item $\left[R ^{\frac 12} P R ^{\frac 12}, R ^{\frac 12} Q R ^{\frac 12}\right]  = 0.$
    \item $ P R Q = Q R P.$

\end{enumerate}
Here $[A,B] := AB - BA$ denotes the commutator of $A,B$.
\end{lemma}
\end{theorembox}
\begin{proof}
We will establish the equivalence by showing $(1) \iff (2)$ and $(2) \iff (3)$. We begin with $(1) \implies (2)$. Assume
\begin{equation}
R^{-\frac 12} \sqrt{R^{\frac 12} P R^{\frac 12}} \sqrt{R^{\frac 12} Q R^{\frac 12}} R^{-\frac 12} = P \# Q,
\end{equation}
which implies 
\begin{equation}
\sqrt{R^{\frac 12} P R^{\frac 12}} \sqrt{R^{\frac 12} Q R^{\frac 12}} = R^{\frac 12} P \# Q R^{\frac 12} = \sqrt{R^{\frac 12} Q R^{\frac 12}} \sqrt{R^{\frac 12} P R^{\frac 12}},
\end{equation}
where the last equality comes from the fact that $P \# Q$ is positive definite (and thus Hermitian). Hence we have
\begin{equation}
\left[\sqrt{R^{\frac 12} Q R^{\frac 12}}, \sqrt{R^{\frac 12} P R^{\frac 12}}\right] = 0,
\end{equation}
which implies their squares also commute, thereby completing the proof.

Now for the reverse implication $(2) \implies (1)$, assume $\left[ R ^{\frac12}  P R ^{\frac12}  , R ^{\frac12}  Q R ^{\frac12} \right] = 0$. Then their square roots also commute, and thus we have
\begin{equation}
\sqrt{R^{\frac 12} P R^{\frac 12}} \sqrt{R^{\frac 12} Q R^{\frac 12}} > 0 \iff R^{-\frac 12} \sqrt{R^{\frac 12} P R^{\frac 12}} \sqrt{R^{\frac 12} Q R^{\frac 12}} R^{-\frac 12} > 0.
\end{equation}
By polar decomposition, we can write the RHS as
\begin{equation}
0 < R^{-\frac 12} \cdot R^{\frac 12} P^{\frac 12} U_P \cdot U_Q^* Q^{\frac 12} R^{\frac 12} \cdot R^{-\frac 12} = P^{\frac 12} U_P U_Q^* Q^{\frac 12},
\end{equation}
where $U_P := \operatorname{Pol}\left(P ^{\frac 12} R ^{\frac 12} \right)$ and $U_Q := \operatorname{Pol}\left(Q ^{\frac 12} R ^{\frac 12} \right)$. By Proposition~\ref{Prop:NamedFidelitiesUnitary}, if $A, B > 0$ and $V$ is a unitary matrix such that $A^{\frac 12} V B^{\frac 12} > 0$, then $A^{\frac 12} V B^{\frac 12} = A \# B$. Thus we have
\begin{equation}
R^{-\frac 12} \sqrt{R^{\frac 12} P R^{\frac 12}} \sqrt{R^{\frac 12} Q R^{\frac 12}} R^{-\frac 12} = P^{\frac 12} U_P U_Q^* Q^{\frac 12} = P \# Q,
\end{equation}
which proves the reverse implication. Now we prove $(2) \Longleftrightarrow (3)$. This is easily seen as
\begin{equation}
    \left[R ^{\frac 12} P R ^{\frac 12} , R ^{\frac 12} Q R ^{\frac 12} \right] = 0 \quad \Longleftrightarrow \quad   R ^{\frac 12} P R Q R ^{\frac 12}  = R ^{\frac 12} Q R P R ^{\frac 12}  \quad \Longleftrightarrow \quad P R Q = Q R P.
\end{equation}
 This concludes the proof. \end{proof}

We note that triples of matrices of these forms have been studied as \textit{$\Gamma$-commuting matrices} in~\cite{lawson2008gamma}. Observe that the first statement of this lemma is a sufficient condition for the generalized fidelity to be equal to the Matsumoto fidelity (take trace across and use cyclicity on the LHS). Thus, we have the following corollary.

\begin{theorembox}
\begin{cor} \label{Cor:GenFid2MatsFid}
        Let $P,Q,R \in \mathbb P_d$. If $P R Q = Q R P$, then $\operatorname{F}_R(P,Q) = \operatorname{F}^\mathrm M(P,Q). $
\end{cor}
\end{theorembox}
\begin{proof}
    The proof follows directly from Statements 1 and 3 of Lemma~\ref{Lem:GenFidMatFidSuffCondition}.
\end{proof}

\section{Proofs of geometric properties} \label{App:GeometricProperties}
In this section, we provide the proofs of the main results of the article. 

\begin{theorembox}
    \begin{theorem}[Path 2; restated from Theorem~\ref{Thm:Path2}]\label{App:ThmPath2}
        Let $P, Q \in \mathbb P_d$ be fixed. Let the base $R = \gamma^{\mathrm{BW}}_{PQ}(t)$ for any $t \in [0,1]$. Then
        \begin{equation}
            \operatorname{F}_R(P,Q) = \operatorname{F}^{\emph U}(P,Q). 
        \end{equation}
    \end{theorem}
\end{theorembox}
\begin{proof}
Let
\begin{equation}
    U_P := \operatorname{Pol}\left(P ^{\frac 12} R ^{\frac 12}\right), \quad U_Q := \operatorname{Pol}\left(Q ^{\frac 12} R ^{\frac 12}\right), \quad \text{and} \quad U := \operatorname{Pol}\left(P ^{\frac 12} Q ^{\frac 12}\right)
\end{equation}
be polar factors. Using the alternate representation of generalized fidelity, we have
\begin{equation}
     \operatorname{F}_R(P,Q) = \text{Tr} \left[Q ^{\frac 12} U_Q U_P^* P ^{\frac 12} \right]
\end{equation}
and noting that $\operatorname{F}^\mathrm U(P,Q) = \text{Tr}\left[P ^{\frac 12} U Q ^{\frac 12} \right]$, it suffices to show that $U_P U_Q^* = U$. By \Cref{Eq:162}, we have
\begin{equation}
\mathbb{I} = (1-t) R ^{-1} \# P + t R^{-1} \# Q \equiv M + N,
\end{equation}
where, for simplicity, we have denoted $M \equiv (1-t) R^{-1} \# P$ and $N \equiv t R ^{-1} \# Q $. We thus have $M, N \in \mathbb{P}_d$ and $M+N = \mathbb{I}$. It then follows that
\begin{equation}
[M,N] = [M, \mathbb{I} - M] = 0,
\end{equation}
which implies $R^{-1} \# P$ commutes with $R^{-1} \# Q$. Thus, the product of these positive matrices is also positive:
\begin{equation}
0 < (R ^{-1} \# P)(R^{-1} \# Q) = R^{-\frac 12} \sqrt{R^{\frac 12} P R^{\frac 12}} R^{-1} \sqrt{R^{\frac 12} Q R^{\frac 12}} R^{-\frac 12},
\end{equation}
which implies 
\begin{equation}
\sqrt{R^{\frac 12} Q R^{\frac 12}} R^{-1} \sqrt{R^{\frac 12} P R^{\frac 12}} = U^*_Q Q^{\frac 12} P^{\frac 12} U_P = U^*_P P^{\frac 12} Q^{\frac 12} U_Q > 0,
\end{equation}
where the first equality follows from polar decomposition. Conjugating with $U_P$ we get
\begin{equation}
U_P U_Q^* Q^{\frac 12} P^{\frac 12} = P^{\frac 12} Q^{\frac 12} U_Q U_P^*   > 0,
\end{equation}
whence it follows that
\begin{equation}
\left(U_P U_Q^* Q^{\frac 12} P^{\frac 12}\right)^2 = \left(P^{\frac 12} Q^{\frac 12} U_Q U_P^* \right) \left(U_P U_Q^* Q^{\frac 12} P^{\frac 12}\right) = P^{\frac 12} Q P^{\frac 12}.
\end{equation}
Since the positive semidefinite matrices have unique positive semidefinite square roots, it follows that
\begin{equation}
U_P U_Q^* Q^{\frac 12} P^{\frac 12} = \sqrt{P^{\frac 12} Q P^{\frac 12}},
\end{equation}
which implies $U_P U_Q^* = U = \operatorname{Pol}\left(P^{\frac 12} Q^{\frac 12}\right)$, which then further implies
\begin{equation}
\operatorname{F}_R(Q,P) 
= \operatorname{Tr}\left[U_P U_Q^* Q^{\frac 12} P^{\frac 12}\right] = \operatorname{Tr}\left[P^{\frac 12} U Q^{\frac 12}\right] = \operatorname{F}^\mathrm{U}(P,Q),
\end{equation}
where $R = R_t$ for any $t \in [0,1]$. This completes the proof.
\end{proof}

\begin{theorembox}
    \begin{theorem}[Path 1, restated from~\ref{Thm:Path1}] \label{AppThm:Path1}
        Let $P, Q \in \mathbb P_d$ be fixed. Let the base $R$ be any point along the path $[\gamma^{\mathrm{BW}}_{P^{-1} Q^{-1}}(t)]^{-1}$. Then
        \begin{equation}
            \operatorname{F}_R(P,Q) = \operatorname{F}^{\emph U}(P,Q). 
        \end{equation}
    \end{theorem}
\end{theorembox}

\begin{proof}
This theorem will be proved similarly to the previous one. Let us begin with the form of $R$:
\begin{equation}
    R = [\gamma^{\mathrm{BW}}_{P^{-1} Q^{-1}}(t)]^{-1} \Longleftrightarrow R ^{-1} = \gamma^{\mathrm{BW}}_{P^{-1} Q^{-1}}(t).
\end{equation}

Similar to the previous proof, let
\begin{equation}
        U_P := \operatorname{Pol}\left(P ^{\frac 12} R ^{\frac 12}\right), \quad U_Q := \operatorname{Pol}\left(Q ^{\frac 12} R ^{\frac 12}\right), \quad \text{and} \quad U := \operatorname{Pol}\left(P ^{\frac 12} Q^{\frac 12}\right).
\end{equation}
We will show that $U_P U_Q^* = U$.  By \Cref{Eq:162}  (with $(A,B, C) = (P ^{-1} , Q ^{-1} , R ^{-1} )$) we have
\begin{equation}
\mathbb{I} = (1-t) R \# P ^{-1} + t R \# Q ^{-1} = M + N,
\end{equation}
where we wrote $M \equiv (1-t) R \# P ^{-1} $ and $ N = (1-t) R \# Q ^{-1}$ for simplicity. We thus have $M, N \in \mathbb{P}_d$ with $M+ N = \mathbb{I}$. It follows that
\begin{equation}
[M, N] = [M, \mathbb{I} - M] = 0,
\end{equation}
which implies $P ^{-1}  \#R $ commutes with $Q ^{-1} \# R$. We thus have their product to be Hermitian and positive definite:
\begin{equation}
R ^{\frac 12} \sqrt{R ^{-\frac 12} P ^{-1} R ^{-\frac 12}} R ^{\frac 12} R ^{\frac 12} \sqrt{R ^{-\frac 12} Q ^{-1} R ^{-\frac 12}} R ^{\frac 12} > 0,
\end{equation}
which implies
\begin{equation}
\sqrt{R ^{-\frac 12} P ^{-1} R ^{-\frac 12}} R  \sqrt{R ^{-\frac 12} Q ^{-1} R ^{-\frac 12}} > 0.
\end{equation}
Now take inverse across to get
\begin{equation}
\sqrt{R ^{\frac 12} P  R ^{\frac 12}} R ^{-1}   \sqrt{R ^{\frac 12} Q R ^{\frac 12}} = U_P^* P ^{\frac 12} Q ^{\frac 12} U_Q = U_Q^* Q ^{\frac 12} P ^{\frac 12} U_P > 0,
\end{equation}
where the first equality follows from polar decomposition. The rest of the proof mirrors the last part of the previous proof, which we write for the sake of completion.
Conjugating $U_Q^* Q ^{\frac 12} P ^{\frac 12} U_P > 0$ with $U_Q$ we get
\begin{equation}
Q ^{\frac 12} P ^{\frac 12} U_P U_Q^* = U_Q  U_P^* P ^{\frac 12} Q ^{\frac 12}  > 0,
\end{equation}
whence it follows that
\begin{equation}
\left(Q ^{\frac 12} P ^{\frac 12} U_P U_Q^*\right)^2 = \left(Q ^{\frac 12} P ^{\frac 12} U_P U_Q^*\right) \left(U_Q  U_P^* P ^{\frac 12} Q ^{\frac 12}\right) = Q ^{\frac 12} P Q ^{\frac 12}.
\end{equation}
Since the positive semidefinite matrices have unique positive semidefinite square roots, we have
\begin{equation}
Q ^{\frac 12} P ^{\frac 12} U_P U_Q^* = \sqrt{Q ^{\frac 12} P Q ^{\frac 12}},
\end{equation}
which implies $U_P U_Q^* = U = \operatorname{Pol}\left(P ^{\frac 12} Q ^{\frac 12}\right)$. Thus we have
\begin{equation}
\begin{aligned}
    \operatorname{F}_R(P,Q) &= \operatorname{Tr}\left[\sqrt{R ^{\frac 12} P ^{1} R ^{\frac 12}} R ^{-1} \sqrt{R ^{\frac 12} Q ^{1} R ^{\frac 12}}\right] \\
&= \operatorname{Tr}\left[U_P U_Q^* Q ^{\frac 12} P ^{\frac 12}\right] = \operatorname{Tr}\left[P^{\frac 12} U Q ^{\frac 12}\right] = \operatorname{F}^\mathrm{U}(P,Q),
\end{aligned}
\end{equation}
where $R = [\gamma^{\mathrm{BW}}_{P^{-1} Q^{-1}}(t)]^{-1}$ for any $t \in [0,1]$. This completes the proof.
\end{proof}

\begin{theorembox}
\begin{theorem} [Restated from Theorem~\ref{Thm:Paths4and7}] \label{AppThm:Paths4and7}
        Let $P, Q \in \mathbb{P}_d$ be fixed. For any fixed $t \in [0,1]$, let 
        \begin{equation}
            R_1 := \gamma_{P Q^{-1}}^{\mathrm{BW}}(t) \quad \text{and} \quad R_2 := \gamma_{Q P^{-1}}^{\mathrm{BW}}(t).
        \end{equation}
        Then \begin{equation}
            \operatorname{F}_{R_1}(P,Q) = \operatorname{F}_{R_2}(P,Q).
        \end{equation} 
\end{theorem}
\end{theorembox}

\begin{proof}
Fix arbitrary $P, Q \in \mathbb{P}_d$, $t \in [0,1]$ and define
\begin{equation}
R_1 := \gamma^{\mathrm{BW}}_{PQ^{-1}}(t) \quad \text{and} \quad R_2 := \gamma^{\mathrm{BW}}_{QP^{-1}}(t) \implies \operatorname{F}_{R_1}(P,Q)  = \operatorname{F}_{R_2}(P,Q).
\end{equation}
We aim to show that the generalized fidelity between $P$ and $Q$ at these bases are equal. Recall that the generalized fidelity can also be written as
\begin{equation}
\operatorname{F}_{R_1}(P,Q) = \operatorname{Tr}\left[(R_1^{-1} \# Q)  R_1 (R_1^{-1} \# P)\right],
\quad \operatorname{F}_{R_2}(P,Q) = \operatorname{Tr}\left[(R_2^{-1} \# Q) R_2 ( R_2^{-1} \# P)\right].
\end{equation}
Thus, to prove the theorem, it suffices to show that the terms inside the trace are equal, which will be done in two steps. We first prove that 
\begin{align}
    (R_1^{-1} \# Q)  R_1 (R_1^{-1} \# P) = \sqrt{Q M P M} M ^{-1}, \\
    (R_2^{-1} \# Q)  R_2 (R_2^{-1} \# P) = M ^{-1} \sqrt{M Q M P},
\end{align}
for a particular $M \in \mathbb P_d$, whose form is described later. In the second step, we show that the two RHS terms are equal. 

Let us perform the first step. To this end, we first write down the explicit forms of $R_1$ and $R_2$. For an arbitrary fixed $t \in [0,1]$ we have
\begin{equation}
R_1 = \gamma_{P Q^{-1}}^{\mathrm{BW}}(t) = \left[(1-t)\mathbb{I} + t P^{-1} \# Q^{-1}\right] P \left[(1-t)\mathbb{I} + t P^{-1} \# Q^{-1}\right],
\end{equation}
\begin{equation}
R_2 = \gamma_{Q P^{-1}}^{\mathrm{BW}}(t) = \left[(1-t)\mathbb{I} + t Q^{-1} \# P^{-1}\right] Q \left[(1-t)\mathbb{I} + t Q^{-1} \# P^{-1}\right].
\end{equation}
Denote $M \equiv \left[(1-t)\mathbb{I} + t P^{-1} \# Q^{-1}\right]$. Thus, and noting that the geometric mean is symmetric, the above relations can be written as
\begin{equation}
R_1 = M P M \quad \text{and} \quad R_2 = M Q M.
\end{equation}
Proposition~\ref{Prop:GeoMeanPropertiesApp} then implies
\begin{equation}
R_1 \# P^{-1} = M = R_2 \# Q^{-1}.  \label{Eq:50}
 \end{equation}
Moreover by Lemma~\ref{Lem:PolarFactorLemma}, we have that $(A^{-1} \# B)A = \sqrt{BA}$ and $A (A^{-1} \# B)= \sqrt{A B}$ for any $A, B \in \mathbb P_d$. Thus, we can write
\begin{align}
(R_1^{-1} \# Q) R_1 = \sqrt{Q R_1} \quad \text{and} \quad R_2 (R_2 ^{-1} \# P) = \sqrt{R_2 P}.
\end{align}
By inverting \Cref{Eq:50}, and using the above relation, we have
\begin{align}
    (R_1^{-1} \# Q)  R_1 (R_1^{-1} \# P) &= \sqrt{Q R_1} M ^{-1}  = \sqrt{Q M P M} M ^{-1}, \\
    (R_2^{-1} \# Q)  R_2 (R_2^{-1} \# P) &= M^{-1} \sqrt{R_2 P} = M ^{-1} \sqrt{M Q M P},
\end{align}
as claimed, where we also used the relations $R_1 = M P M$ and $R_2 = M Q M$. Now we prove that the RHS terms are equal, which amounts to proving,
\begin{equation}
    \sqrt{Q M P M} M ^{-1} \iseq M ^{-1} \sqrt{M Q M P}.
\end{equation}
This is equivalent to proving
\begin{equation}
\sqrt{Q M P M} \iseq M^{-1} \sqrt{M Q M P} M.
\end{equation}
To prove this, it suffices to prove the squared version as matrices with positive eigenvalues have a unique square root (see~\cite[Excercise 4.5.2]{bhatiamatrixanalysis}). Square the LHS to obtain $Q M P M$. Now square the RHS to obtain
\begin{equation}
\left(M^{-1} \sqrt{M Q M P} M\right) \left(M^{-1} \sqrt{M Q M P} M\right) = M^{-1} (M Q M P) M = Q M P M,
\end{equation}
which is equal to the square of the LHS. By uniqueness of square root, we have
\begin{equation}
(R_1^{-1} \# Q) R_1 (R_1^{-1} \# P) = (R_2^{-1} \# Q) R_2 (R_2^{-1} \# P).
\end{equation}
Trace both sides to obtain $\operatorname{F}_{R_1}(P,Q) = \operatorname{F}_{R_2}(P,Q)$ as claimed.
\end{proof}

\begin{theorembox}
    \begin{theorem}[Restated from Theorem~\ref{Thm:Paths3and8}]\label{AppThm:Paths3and8}
        Let $P,Q \in \mathbb P_d$ be fixed. For any fixed $t \in [0,1]$, let 
        \begin{equation}
            R_1 := [\gamma_{P ^{-1} Q}^{\mathrm{BW}} (t)] ^{-1}  \quad \text{and} \quad R_2 := [\gamma_{Q^{-1} P }^{\mathrm{BW}} (t)] ^{-1} .
        \end{equation}
        Then $\operatorname{F}_{R_1}(P,Q) = \operatorname{F}_{R_2}(P,Q). $ 
    \end{theorem}
\end{theorembox}

\begin{proof}
    The proof works similarly to the previous proof. We aim to show that for arbitrary $P, Q \in \mathbb{P}_d$ and any $t \in [0,1]$,
\begin{equation}
R_1 ^{-1}  = \gamma^{\mathrm{BW}}_{P ^{-1} Q}(t) \quad \text{and} \quad R_2 ^{-1}  = \gamma^{\mathrm{BW}}_{Q ^{-1} P}(t) \implies \operatorname{F}_{R_1}(P,Q)  = \operatorname{F}_{R_2}(P,Q).
\end{equation}
We first recall the alternative form of generalized fidelity:
\begin{equation}
\operatorname{F}_{R_1}(P,Q) = \operatorname{Tr}\left[(R_1^{-1} \# Q)  R_1 (R_1^{-1} \# P)\right],
\quad \operatorname{F}_{R_2}(P,Q) = \operatorname{Tr}\left[(R_2^{-1} \# Q) R_2 ( R_2^{-1} \# P)\right].
\end{equation}
We will show that the terms inside the trace are equal, which will be done in two steps. We first prove that 
\begin{align}
    (R_1^{-1} \# Q)  R_1 (R_1^{-1} \# P) = \sqrt{Q N ^{-1}  P  N ^{-1} } N, \\
    (R_2^{-1} \# Q)  R_2 (R_2^{-1} \# P) = N \sqrt{N ^{-1} Q N ^{-1}  P},
\end{align}
for a particular choice of $N \in \mathbb P_d$, whose form is described later. In the second step, we show that the two RHS terms are equal. 

To this end, we first write down the explicit forms of $R_1 ^{-1} $ and $R_2 ^{-1} $. For an arbitrary fixed $t \in [0,1]$ we have
\begin{equation}
R_1 ^{-1}  = \gamma_{P ^{-1}  Q}^{\mathrm{BW}}(t) := \left[(1-t)\mathbb{I} + t P \# Q\right] P ^{-1}  \left[(1-t)\mathbb{I} + t P \# Q\right],
\end{equation}
\begin{equation}
R_2 = \gamma_{Q ^{-1} P}^{\mathrm{BW}}(t) := \left[(1-t)\mathbb{I} + t Q \# P\right] Q ^{-1}  \left[(1-t)\mathbb{I} + t Q\# P\right].
\end{equation}
Denote $N \equiv \left[(1-t)\mathbb{I} + t P \# Q\right]$. Thus, the above relations can be written as
\begin{equation}
R_1 ^{-1}  = N P^{-1}  N \quad \text{and} \quad R_2 ^{-1}  = N Q ^{-1} N.
\end{equation}
which implies
\begin{equation}
P \# R_1 ^{-1} = N = Q \# R_2 ^{-1}.  \label{Eq:69}
 \end{equation}
Using Lemma~\ref{Lem:PolarFactorLemma} as before, we have that $(A^{-1} \# B)A = \sqrt{BA}$ and $A (A^{-1} \# B)= \sqrt{A B}$ for any $A, B \in \mathbb P_d$. Thus, we can write
\begin{align}
(R_1^{-1} \# Q) R_1 = \sqrt{Q R_1} = \sqrt{Q N ^{-1} P N ^{-1} } \quad \text{and} \quad R_2 (R_2 ^{-1} \# P) = \sqrt{R_2 P} = \sqrt{N ^{-1} Q N ^{-1}  P}.
\end{align}
Using \Cref{Eq:69} and the above relation, we have
\begin{align}
    (R_1^{-1} \# Q)  R_1 (R_1^{-1} \# P) &= \sqrt{Q N ^{-1}  P  N ^{-1} } N, \\
    (R_2^{-1} \# Q)  R_2 (R_2^{-1} \# P) &= N \sqrt{N ^{-1}  Q N ^{-1}  P}
\end{align}
as claimed. Now we prove that the RHS terms are equal:
\begin{equation}
    \sqrt{Q N ^{-1}  P N ^{-1} } N  \iseq N  \sqrt{N ^{-1}  Q N ^{-1}  P},
\end{equation}
which is equivalent to proving
\begin{equation}
    \sqrt{Q N ^{-1}  P N ^{-1} }   \iseq N  \sqrt{N ^{-1}  Q N ^{-1}  P} N ^{-1} ,
\end{equation}
As in the previous proof, it suffices to prove the squared version of the above equation. First, square the LHS to obtain $Q N ^{-1}  P N ^{-1} $. Now square the RHS to obtain
\begin{equation}
\left(N  \sqrt{N ^{-1}  Q N ^{-1}  P} N ^{-1}\right) \left(N  \sqrt{N ^{-1}  Q N ^{-1}  P} N ^{-1} \right) = N (N ^{-1}  Q N ^{-1}  P) N = Q N ^{-1}  P N ^{-1} ,
\end{equation}
which is equal to the square of LHS. We have thus shown, by the uniqueness of the positive square root,
\begin{equation}
(R_1^{-1} \# Q) R_1 (R_1^{-1} \# P) = (R_2^{-1} \# Q) R_2 (R_2^{-1} \# P).
\end{equation}
Taking the trace across, we get $\operatorname{F}_{R_1}(P,Q) = \operatorname{F}_{R_2}(P,Q)$ as claimed.
\end{proof}

\begin{theorembox}
\begin{theorem} [Restated from Theorem~\ref{Thm:Path9GM}] \label{AppThm:Path9GM}
Let $P,Q \in \mathbb{P}_d$ and choose $R = P^{-1} \# Q^{-1}$. Then $\operatorname{F}_R(P,Q) = \operatorname{F}_\mathrm{M}(P,Q).$
\end{theorem}
\label{AppThm:MatsFidelityatGeoMean}
\end{theorembox}

\begin{proof}
By definition, we have
\begin{equation}
\operatorname{F}_R(P,Q) = \operatorname{F}_R(P,Q) = \operatorname{Tr}\left(\sqrt{R^{\frac 12} P R^{\frac 12}} (P \# Q) \sqrt{R^{\frac 12} Q R^{\frac 12}}\right), 
\label{Eq:58}
\end{equation}
where we used the property of the geometric mean that $(A \# B)^{-1} = A^{-1} \# B^{-1}$ for any $A,B \in \mathbb{P}_d$. 

Recall that $\operatorname{F}_\mathrm{M}(P,Q) = \operatorname{Tr}[P \# Q]$ by definition. Thus it suffices to show that $\sqrt{R^{\frac 12} P R^{\frac 12}} \stackrel{?}{=} \left(\sqrt{R^{\frac 12} Q R^{\frac 12}}\right)^{-1}$. Equivalently, it suffices to show that
\begin{equation}
R^{\frac 12} P R^{\frac 12} \stackrel{?}{=} R^{-\frac 12} Q^{-1} R^{-\frac 12} \Longleftrightarrow R P R \stackrel{?}{=} Q^{-1}.
\end{equation} 
By Proposition~\ref{Prop:GeoMeanPropertiesApp}, we have that the geometric mean $R = P^{-1} \# Q^{-1}$ uniquely satisfies the second condition, and thus we have $R^{\frac 12} P R^{\frac 12} = R^{-\frac 12} Q^{-1} R^{-\frac 12}$, which implies their matrix square roots equal, which in turn implies  
\begin{equation}
\operatorname{F}_R(P,Q) = \operatorname{Tr}[P \# Q] = \operatorname{F}^\mathrm{M}(P,Q).
\end{equation} 
This completes the proof.
\end{proof}

Now we prove it for the whole curve $\gamma_{P ^{-1} Q ^{-1} }^{\mathrm{AI}} $.                                                 

\begin{theorembox}
    \begin{theorem}[Path 9, restated from Theorem~\ref{Thm:Path9}]\label{AppThm:Path9}
        Let $P, Q \in \mathbb P_d$ and let $R = \gamma_{P ^{-1} Q ^{-1} }^\mathrm{AI}(t)$ for any $t \in [0,1]$. Then 
        \begin{equation}
             \operatorname{F}_R(P,Q) = \operatorname{F}^\mathrm M(P,Q).
        \end{equation}
    \end{theorem}
\end{theorembox}

\begin{proof}
Let
\begin{equation} R = \gamma_{P ^{-1} Q ^{-1}}^{\mathrm{AI}}(t) = P ^{-\frac 12} \left( P ^{\frac 12} Q ^{-1} P ^{\frac 12} \right)^t P ^{-\frac 12} \end{equation}
for some $t \in [0,1]$.
By Corollary~\ref{Cor:GenFid2MatsFid}, it is sufficient to show that 
\begin{equation}
    P R Q \stackrel{?}{=} Q R P     
\end{equation}
Let us begin with the LHS. We first left and right multiply by $P ^{-\frac 12}$ to obtain
\begin{equation}
    \begin{aligned}
    P ^{-\frac 12} (P R Q) P ^{-\frac 12} &= P ^{\frac 12} R P ^{\frac 12} \left(\frac Q P \right) \\ &\stackrel{1}{=} (P ^{\frac 12} \gamma_{P ^{-1} Q ^{-1}}^{\mathrm{AI}}(t) P ^{\frac 12}) \left(\frac Q P \right) \\
    &\stackrel{2}{=} \gamma_{ \mathbb I, P ^{\frac 12} Q ^{-1} P ^{\frac 12}}^{\mathrm{AI}}(t) \left(\frac Q P \right) \\  &\stackrel{3}{=} \left(\frac Q P \right)^{-t} \left(\frac Q P \right) \\
    &= \left(\frac Q P \right) ^{1-t} = \left(P ^{-\frac 12} Q P ^{-\frac 12}\right)^{1-t}
    \end{aligned}    
\end{equation}
where we use the shorthand $\frac{A}{B} := B ^{-\frac 12} A B ^{-\frac 12}$. Here in (1) we have used the fact that $R$ is an element of the geodesic $\gamma^\mathrm{AI}_{P ^{-1} Q ^{-1}}$, in (2) we have used the Affine invariance property (Lemma~\ref{Lem:ConjugationInvarianceAI}), and in (3) we used the definition of the AI geodesic.

A similar calculation on the RHS reveals $ P ^{-\frac 12} (Q R P ) P ^{-\frac 12}  = \left( \frac Q P\right)^{1-t} = (P ^{-\frac 12} Q P ^{-\frac 12} )^{1-t}$, which implies $PRQ = QRP$. In then follows from Corollary~\ref{Cor:GenFid2MatsFid} that $ \operatorname{F}_R(P,Q) = \operatorname{F}^\mathrm M(P,Q) $. This concludes the proof.

\end{proof}

\begin{theorembox}
    \begin{theorem}[Path 10, restated from Theorem~\ref{Thm:Path10}]\label{App:ThmPath10}
        Let $P, Q \in \mathbb P_d$ and let $R = \gamma_{P ^{-1} Q ^{-1} }^\mathrm{Euc}(t)$ for any $t \in [0,1]$. Then 
        \begin{equation}
             \operatorname{F}_R(P,Q) = \operatorname{F}^\mathrm M(P,Q). 
        \end{equation}
    \end{theorem}
\end{theorembox}

\begin{proof}
By Lemma~\ref{Lem:GenFidMatFidSuffCondition}, it suffices to show that $P R Q = Q R P$. Recall that $R$ takes the form
$ R  = (1-t) P ^{-1} + t Q ^{-1} $ for any $t \in [0,1]$. Use this form of $R$ to write
\begin{equation}
\begin{aligned}
    P R Q &= P \left((1-t) P^{-1} + t Q^{-1}\right) Q = (1-t)Q + t P \\
 &= Q \left((1-t) P^{-1} + t Q^{-1}\right) P = Q R P.
\end{aligned}
\end{equation}
 This completes the proof.
\end{proof}

\begin{theorembox}
    \begin{theorem}[Path 11, restated from Theorem~\ref{Thm:Path11}]\label{App:ThmPath11}
       Let $P, Q \in \mathbb P_d$ and let $R = [\gamma_{PQ}^\mathrm{Euc}(t)]^{-1} $ for any $t \in [0,1]$. Then 
        \begin{equation}
             \operatorname{F}_R(P,Q) = \operatorname{F}^\mathrm M(P,Q). 
        \end{equation}    
    \end{theorem}
\end{theorembox}
\begin{proof}
Here too we will prove the sufficient condition $P R Q = Q R P$, which is equivalent to showing $Q^{-1} R^{-1} P^{-1} = P^{-1} R^{-1} Q^{-1}$. Noting that along the path $[\gamma_{PQ}^\mathrm{Euc}(t)]^{-1}$, we have $R = \left((1-t) P + t Q\right)^{-1}$ for $t \in [0,1]$.  Thus we have
\begin{equation}
\begin{aligned}
    P^{-1} R^{-1} Q^{-1} &= P^{-1} \left((1-t) P + t Q\right) Q^{-1} = (1-t) Q^{-1} + t P^{-1} \\
&= Q^{-1} \left((1-t) P + t Q\right) P^{-1} = Q^{-1} R^{-1} P^{-1},
\end{aligned}
\end{equation}
which implies $P R Q = Q R P$. By Lemma~\ref{Lem:GenFidMatFidSuffCondition}, this completes the proof.
\end{proof}

\end{document}